\documentclass[fleqn,usenatbib]{mnras}

\usepackage{newtxtext,newtxmath}

\usepackage[T1]{fontenc}

\DeclareRobustCommand{\VAN}[3]{#2}
\let\VANthebibliography\thebibliography
\def\thebibliography{\DeclareRobustCommand{\VAN}[3]{##3}\VANthebibliography}

\usepackage{graphicx}	
\usepackage{amsmath}	
\usepackage{booktabs}
\usepackage{multirow}

\usepackage{comment}

\title[Radio Source Detection and Classification]{Transformer-Based Source Detection and Morphological Classification in LOFAR Deep-Field Continuum Images}

\author[Chen et al.]{
Guangwen Chen$^{1,2,3}$\thanks{E-mail: guangwen.chen@physics.ox.ac.uk}, 
Kristian Z. Adami$^{1,4}$\thanks{E-mail: kristian.zarbadami@physics.ox.ac.uk}, 
John Abela$^{5}$, 
Caijuan Yue$^{6}$,
Weibin Sun$^{7,8}$,
Fujia Li$^{7,8}$, \newauthor 
Zhaoting Chen$^{9}$,
Daniel Magro$^{4}$,
Yogesh Wadadekar$^{10}$,
Leah K. Morabito$^{11,12}$ \\
$^{1}$Sub-department of Astrophysics, Department of Physics, University of Oxford, Keble Road, Oxford OX1 3RH, UK \\
$^{2}$Instituto de Astrof\'isica de Canarias, calle Vía L\'actea s/n, E-38205 La Laguna, Tenerife, Spain \\
$^{3}$Departamento de Astrof\'isica, Universidad de La Laguna, Avenida Astrof\'isico Francisco S\'anchez s/n, E-38206 La Laguna, Spain \\
$^{4}$Institute of Space Sciences and Astronomy, University of Malta, Msida MSD2080, Malta \\
$^{5}$Department of Information Systems, Faculty of ICT, University of Malta, Msida MSD2080, Malta \\
$^{6}$School of Physics and Astronomy, Anqing Normal University, Anqing 246133, China \\
$^{7}$Department of Astronomy, University of Science and Technology of China, Hefei 230026, China\\
$^{8}$School of Astronomy and Space Sciences, University of Science and Technology of China, Hefei 230026, China\\
$^{9}$Institute for Astronomy, School of Physics and Astronomy, The University of Edinburgh, Royal Observatory, Edinburgh EH9 3HJ, UK \\
$^{10}$National Centre for Radio Astrophysics, TIFR, Post Bag 3, Ganeshkhind, Pune 411007, India  \\
$^{11}$Centre for Extragalactic Astronomy, Department of Physics, Durham University, South Road, Durham DH1 3LE, UK \\
$^{12}$Institute for Computational Cosmology, Department of Physics, Durham University, Durham DH1 3LE, UK
}

\date{Accepted XXX. Received YYY; in original form ZZZ}

\pubyear{\the\year{}}

\begin{document}
\label{firstpage}
\pagerange{\pageref{firstpage}--\pageref{lastpage}}
\maketitle

\begin{abstract}

Radio source detection and morphological classification are fundamental for exploiting the scientific potential of modern radio continuum surveys. However, the rapidly increasing data volumes and the wide diversity of radio morphologies make traditional visual inspection infeasible and pose significant challenges for automated source finding. We apply a transformer-based set-prediction detector (RF-DETR) to 150\,MHz continuum images from the LOFAR Deep Fields for instance-level source detection and morphological classification. The method is adapted to multi-frequency-synthesis images of interferometric data and trained with a morphology-driven scheme using five mutually exclusive classes. The model is trained on the ELAIS-N1 Deep Field, where it achieves high detection and classification performance ($\mathrm{F1}\simeq 91$\,per\,cent), and is then applied without retraining to the other three LOFAR Deep Fields. Across all four fields, the model yields consistent catalogues with modest field-to-field differences arising from survey depth and calibration. Compared with widely used PyBDSF catalogues, RF-DETR recovers the majority of PyBDSF sources while representing classical multi-component radio galaxies as single source-level detections rather than fragmented Gaussian components. Artefact-affected and spurious detections are identified as explicit classes, allowing these detections to be distinguished from general astrophysical sources in the resulting catalogues. As external validation, RF-DETR recovers the majority of visually identified extended and giant radio galaxies in the LOFAR Deep Fields and assigns them predominantly to extended morphological classes. These results indicate that transformer-based detectors provide a practical, scalable, morphology-aware approach to source finding in deep radio surveys, with clear relevance for forthcoming facilities such as SKA-Low.

\end{abstract}

\begin{keywords}
radio continuum: galaxies — methods: data analysis — methods: statistical
\end{keywords}



\section{Introduction}

Radio observations provide dust-unbiased views of star formation and nuclear activity in galaxies \citep[e.g.,][]{Best2012,Bendo2015,Bendo2016,Hardcastle2020,Cochrane2023,Chen2023,Chen2024}. Large-area, deep surveys with Square Kilometre Array (SKA) pathfinders, including the LOw Frequency ARray (LOFAR; \citealt{vanHaarlem2013_LOFAR}), the Murchison Widefield Array (MWA; \citealt{Tingay2013_MWA}) and the Australian SKA Pathfinder (ASKAP; \citealt{Hotan2021_ASKAP}), are transforming how we discover and characterise the extragalactic radio sky. 

Accurate detection and classification are prerequisites for two complementary outcomes. On the one hand, complete, well-characterised catalogues enable number counts, angular clustering and cross-correlations that link galaxy populations to their dark-matter environments and test structure growth on large scales \citep[e.g.,][]{Jarvis2015,Alonso2021,Hale2024,Hale2025,Tanidis2025}. On the other hand, morphological labels, in combination with multi-wavelength observations, provide indicators of the dominant emission mechanisms and facilitate the separation of star-formation-dominated systems from jetted active galactic nuclei (AGN), allowing interpretable inferences about AGN feedback, galaxy evolution, and environments \citep[e.g.,][]{Mingo2019,Mingo2022,Chen2020,Chen2022,Cochrane2023,Morabito2025,Horton2025}. Within this landscape, the LOFAR Two-metre Sky Survey (LoTSS) delivers arcsecond-scale imaging with sub-0.1\,mJy\,beam$^{-1}$ sensitivity at $120$–$168$\,MHz, enabling the detection of complex, low-surface-brightness structures over tens of square degrees \citep{Shimwell2017_LOFAR_description,Shimwell2019_LoTSS_DR1,Tasse2021_LOTSS_DeepField,Shimwell2022_LOTSS_DR2,Shimwell2025}. At the same time, the growing heterogeneity of radio-galaxy morphologies, spanning from compact cores to extended multi-component systems, complicates catalogue production at survey scale \citep{Rudnick2021,Ndung2023_Review}. Looking ahead, SKA Phase~1 will increase sensitivities and survey speed and confront the community with exascale data volumes, making automation essential \citep[e.g.,][]{Braun2019_SKA1,Bonaldi2021,Bonaldi2025}.

Classical radio source finders—such as the Python Blob Detection and Source Finder \citep[PyBDSF;][]{Mohan2015_PyBDSF} and AEGEAN \citep{Hancock2012_Aegean,Hancock2018_Aegean}—have been the workhorses of continuum imaging for more than a decade. They model emission via thresholding and Gaussian decomposition (PyBDSF) or Laplacian-of-Gaussian clustering (AEGEAN), returning catalogues of compact/extended components with fitted shapes and fluxes. Despite their maturity, challenging regimes persist: artefacts around bright sources, diffuse or filamentary emission \citep[e.g.,][]{Mohan2015_PyBDSF,Riggi2021}, and physically connected multi-component systems that are split into separate islands and require a subsequent association step before cross-matching \citep[e.g.,][]{Magliocchetti1998,Williams2019}. These difficulties motivate methods that learn morphological priors directly from the data and provide richer, instance-level outputs beyond simple Gaussian components.

Motivated by these limitations, deep learning source finders—particularly those based on convolutional architectures—have evolved from early point-source detectors into multi-class, instance-level systems that jointly localise and label sources. \texttt{DeepSource} benchmarked point-source detection and reported completeness–purity gains over PyBDSF on simulations at fixed thresholds \citep{Vafaei2019_DeepSource}. Region-based models were then explored for radio maps, such as \texttt{ClaRAN}, a Faster R–CNN \citep{Ren2017_FastRCNN} variant that detects and assigns morphology labels using radio and infrared inputs \citep{Wu2019_CLARAN}, and \texttt{ConvoSource} for semantic segmentation trained on SKA Science Data Challenge~1–style skies \citep{Lukic2019_ConvoSource}. More recently, instance-segmentation frameworks based on Mask R–CNN \citep{He2017_MaskRCNN} have been adapted to radio images to jointly detect and classify compact and extended sources as well as spurious detections, enabling pixel-level masks and downstream morphological analyses; examples include the \texttt{caesar–mrcnn} line and related applications to ASKAP and the Meer Karoo Array Telescope (MeerKAT) data \citep{Riggi2023,Riggi2024}. In parallel, one-stage detectors in the YOLO family have also been adapted to radio data; for example, \texttt{YOLO–CIANNA} introduces astronomy-specific adaptations and reports strong performance on the SKAO SDC1 simulations \citep{Cornu2024}. Hybrid designs that blend convolutional backbones with Transformer blocks report improved handling of extended and diffuse emission via stronger context modelling \citep[e.g., HeTu–v2;][]{Lao2023_HeTu}. Recent work has also begun to explore the applicability of Vision Transformers (ViTs) and foundation models to astrophysical data, demonstrating strong performance in optical classification and competitive results in detection settings, while revealing significant limitations for radio galaxy morphology due to domain shift and task-specific challenges \citep{Lastufka2025}.

Complementary to detectors, supervised image-level classifiers operating on pre-centred cutouts—together with increasingly adopted self- and semi-supervised representation learning—aid morphology studies and improve label efficiency \citep[e.g.,][]{Alhassan2018,Lukic2019,Samudre2022,Brand2023,Ndung2024,Perez2025}, yet they presuppose existing candidates and provide no localisation, and therefore cannot substitute for instance-level detection in catalogue production. In addition, practical constraints of current models persist at survey scale: anchor design and non-maximum suppression may yield duplicate or fragmented detections near bright, multi-component systems; dense-mask architectures increase annotation burden and memory footprint; and performance can degrade under domain shift across fields/instruments and when running tiled inference on large images \citep[e.g.,][]{Ndung2023_Review,Sortino2023,Cornu2024}. 

Beyond convolutional pipelines, transformer-based detectors offer a unified route to localisation and labelling. The DEtection TRansformer (DETR; \citealt{Carion2020_DETR}) casts detection as set prediction with bipartite (Hungarian) matching, which removes anchors and, by construction, duplicate assignments, while self-attention provides long-range context that is well suited to extended, multi-component radio morphology. For survey-grade cataloguing, such properties are attractive because they reduce reliance on hand-tuned anchors and non-maximum suppression and allow reasoning over diffuse structures within large tiles. A recent radio benchmark that evaluates DETR alongside Mask R-CNN and YOLO across cutouts from ASKAP, the Australia Telescope Compact Array (ATCA), the Karl G. Jansky Very Large Array (VLA), and the Radio Galaxy Zoo (RGZ; \citealt{Banfield2015_RGZ}) project finds transformer detectors competitive yet still comparatively underused in this domain, indicating both potential and room for further development \citep{Sortino2023}. 

Building on the original DETR framework, successors such as Deformable DETR, Conditional DETR and Lightweight DETR further improve convergence and small-object sensitivity \citep{Zhu2021_DeformableDETR,Meng2021_ConditionalDETR,Chen2024_LWDETR}. Among these, RF-DETR attains a favourable accuracy–throughput trade-off on the Common Objects in Context (COCO) benchmark by coupling a ViT encoder with a shallow DETR decoder and training-oriented refinements \citep{Robinson_rf-detr_ICLR}. RF-DETR has already been applied beyond natural-image benchmarks \citep[e.g.,][]{Sapkota2025,Dahiya2025}, yet it has not been systematically evaluated on radio astronomical data.

In this work, we investigate whether a DETR-family, set-prediction transformer with strong self-supervised pretraining can be effectively repurposed for LOFAR deep-field continuum imaging. Specifically, we adapt RF-DETR to deliver unified, instance-level source detection with five mutually exclusive morphological classes, and assess its performance not only in terms of detection accuracy but also in terms of catalogue-level scientific usability. Our aim is to determine whether such a framework can produce scalable and morphology-aware source catalogues across deep fields with differing depths and calibration characteristics, providing a practical reference for forthcoming SKA-Low surveys.

This paper is organised as follows. In Section~\ref{sec:dataset}, we describe the LoTSS Deep Fields data and the source classification scheme, together with the construction of the training and validation datasets. Section~\ref{sec:methods} introduces the RF-DETR transformer-based detector and outlines the methods adopted in this work. The model evaluation, the source catalogues, and the comparison with PyBDSF are presented in Section~\ref{sec:results}. A brief summary of the main results is given in Section~\ref{sec:summary}.

\section{Dataset} \label{sec:dataset}



\subsection{LoTSS Deep Fields DR1} \label{subsec:LOFAR}


The LoTSS Deep Fields data release 1 (DR1)\footnote{\url{https://lofar-surveys.org/deepfields.html}} provide ultra-deep 120$-$168\,MHz imaging in four extragalactic regions: European Large-Area ISO Survey-North 1 (ELAIS-N1), Lockman Hole, Bo\"otes, and the Euclid Deep Field North (EDFN), reaching root mean square (rms) depths in the central regions down to $\sim$ 20, 22, 32 and $32~\mu\mathrm{Jy\,beam}^{-1}$ at central frequencies $\sim 150~\mathrm{MHz}$, respectively \citep{Tasse2021_LOTSS_DeepField, Sabater2021_ELAISN1,Bondi2024_LOFAR_EDFN}. These depths are comparable to those anticipated for SKA1-Low ($\sim 26~\mu\mathrm{Jy\,beam}^{-1}$ at $\sim 110$\,MHz for a 1-hour continuum observation; \citealt{Braun2019_SKA1}). Each pointing covers approximately $20~\mathrm{deg}^2$, with an angular resolution of $\simeq$6 arcsec. The Deep Fields also benefit from extensive multi-wavelength coverage from X-ray to far-infrared over a combined cross-matched area of $\sim 26~\mathrm{deg}^2$ \citep{Kondapally2021_LOTSS_VAC}, which we use for verification and contextual interpretation of radio sources. 

It is worth noting that a DR2 release for ELAIS–N1 is now available, but DR2 products for the other Deep Fields have not yet been released. To maintain uniformity across training and evaluation, we adopt DR1 throughout and will use DR2 for future analysis. Specifically, we train and validate on the ELAIS–N1 DR1 continuum images, and assess generalisation by applying the trained model to the Lockman Hole, Bo\"otes, and EDFN DR1 continuum images.

\begin{figure*}
    \centering
    \includegraphics[width=1.\textwidth]{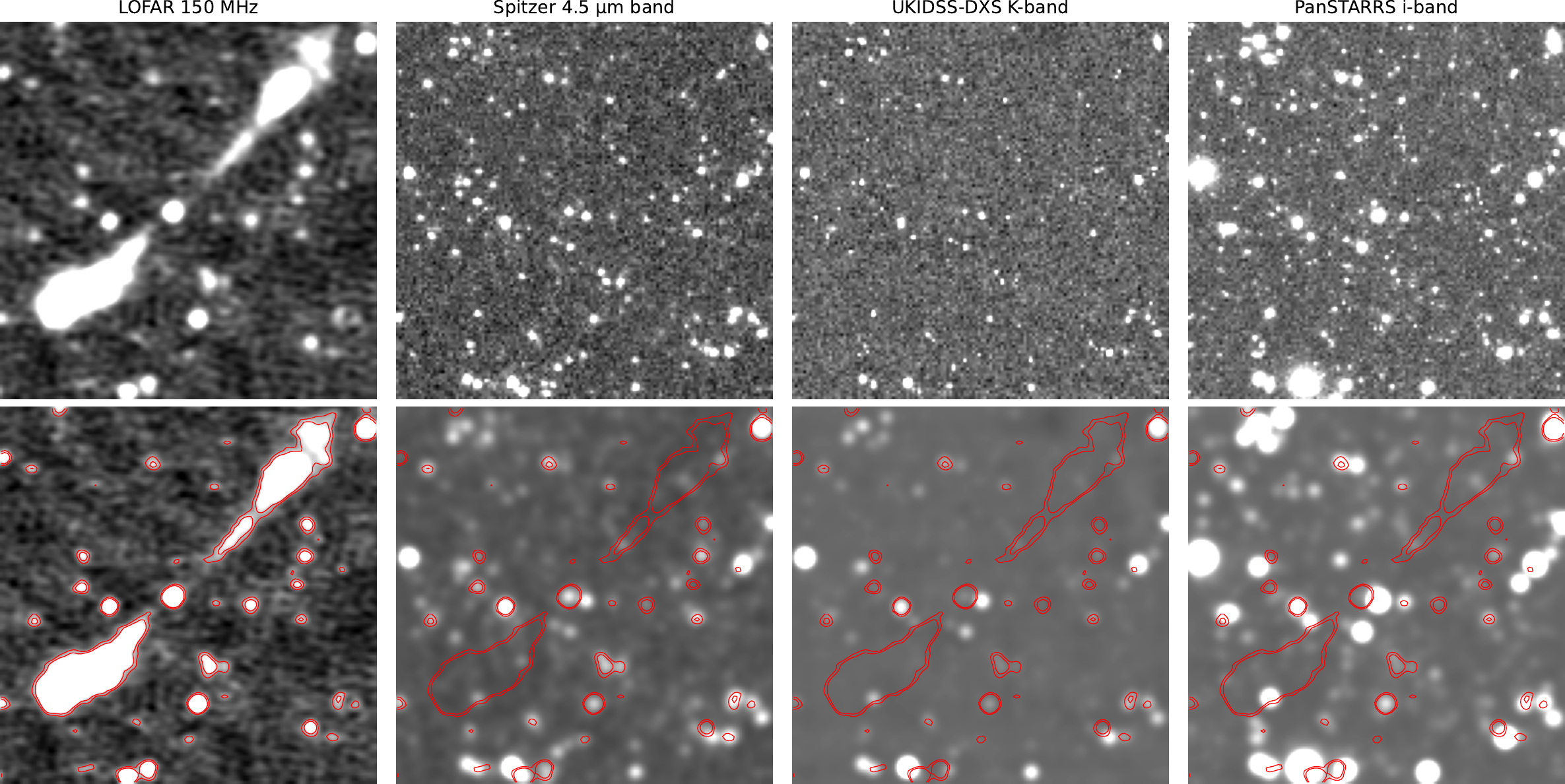} 
    \caption{Examples of LOFAR 150\,MHz and ancillary imaging for a $132\times132$ pixel cutout ($3.3\times3.3$ arcmin at 1.5 arcsec pixel$^{-1}$), displayed with a ZScale stretch (contrast $=0.2$). Left to right: LOFAR 150\,MHz, \emph{Spitzer} 4.5\,$\mu$m, UKIDSS--DXS $K$, and Pan-STARRS $i$. Upper panels: native ancillary images (no PSF convolution). Lower panels: ancillary images PSF-matched to the LOFAR synthesized beam (FWHM $\simeq 6^{\prime\prime}$) to enable direct morphological comparison. Red contours trace LOFAR surface brightness at $3\sigma$ and $5\sigma$ (with $\sigma$ estimated from the local rms noise). Ancillary images are used solely for visual verification and are not included in the training set.}
    \label{fig:multipanels}
\end{figure*}

\subsection{Source classification scheme}
\label{subsec:class-label-schema}

We adopt a morphology–driven taxonomy of five \emph{mutually exclusive} classes, designed for automated source finding in next–generation radio surveys and conceptually aligned with \citet{Riggi2023}. Throughout, an island denotes a contiguous region of emission in the radio image as identified by standard source finders.

\begin{enumerate}
\item \texttt{Compact}---single–island system with point-like or only slightly resolved structure, broadly consistent with the synthesized beam. Blending of one or more unresolved components within the same island is allowed.

\item \texttt{Single-Extended}---single–island system exhibiting clear, intrinsic resolved structure beyond that expected from ionospheric smearing or beam broadening (e.g.,\ lobes, jets, or extended haloes), possibly comprising several blended components within one connected island.

\item \texttt{Multi-Extended}---physically associated system comprising two or more disconnected islands judged to belong to the same astrophysical source (e.g., classical double-lobed radio galaxies, a core with lobes, or multi–knot jets). For such systems, a single source-level annotation is defined as a rectangular bounding box drawn to enclose the full spatial extent of the system, covering the combined area spanned by all constituent islands.

\item \texttt{Flagged}---single–island system with compact or extended morphology whose measurements are unreliable because of severe imaging artefacts or dynamic-range limitations near bright objects. Such sources are kept for completeness in statistics but are explicitly flagged in science catalogues since derived quantities (flux density, size, morphology) are not trustworthy.

\item \texttt{Spurious}---false detection produced by different types of artefacts (e.g.,\ sidelobes, ripples, striping) introduced by calibration and/or imaging. We focus on the sidelobes, which can mimic compact or elongated sources and frequently exceed standard $5\sigma$ thresholds, thereby compromising catalogue reliability.

\end{enumerate}

This scheme is intentionally conservative and oriented toward source detection rather than detailed astrophysical taxonomy. It nonetheless covers nearly all kinds of sources in radio observations: \texttt{Compact} captures most ordinary radio sources; \texttt{Single-Extended} together with \texttt{Multi-Extended} encompass the majority of classical Fanaroff–Riley~I/II galaxies; and \texttt{Flagged} plus \texttt{Spurious} capture artefact-related detections, supporting quality control and reliability assessment.

Definitions are based on the LOFAR 150\,MHz continuum image from the ELAIS-N1 field. 
Ancillary images, including 4.5\,$\mu$m data from the \texttt{Spitzer} Wide-area Infrared Extragalactic survey (SWIRE; \citealt{Lonsdale2003}), $K$-band imaging from the UK Infrared Deep Sky Survey (UKIDSS) Deep Extragalactic Survey (DXS) DR10 \citep{Lawrence2007}, and $i$-band data from the Panoramic Survey Telescope and Rapid Response System (Pan-STARRS; \citealt{Kaiser2010}), are used exclusively during the manual annotation process. 
They provide complementary morphological information to assist in judging source association and in identifying artefact-affected and spurious detections, particularly for \texttt{Multi-Extended}, \texttt{Flagged}, and \texttt{Spurious} cases. These ancillary images are not used as inputs to the model during training or inference and serve no role beyond human-guided annotation and visual verification (Figure~\ref{fig:multipanels}).

\subsection{Dataset preparation} 
\label{subsec:data-prep} 

\begin{figure*}
  \centering
  {\bfseries Manual Annotations}\\[0.5mm]
    \includegraphics[height=0.33\textwidth]{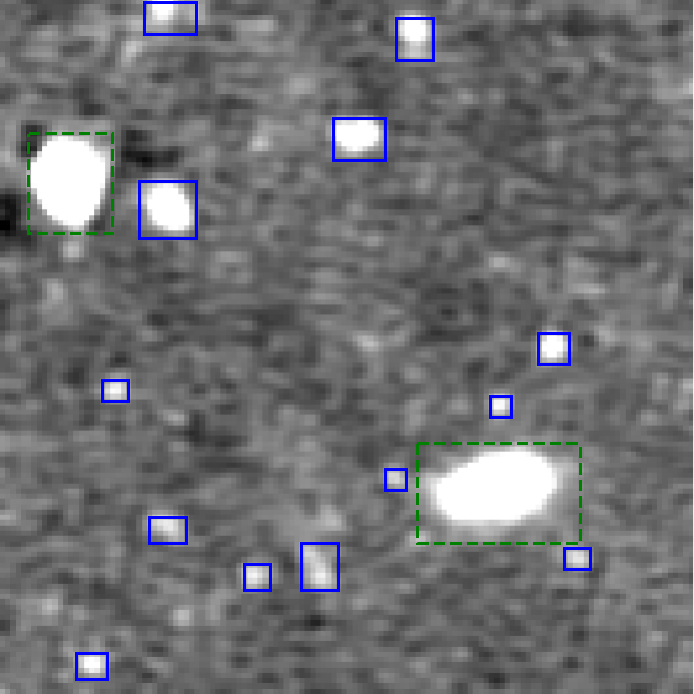}   
    \includegraphics[height=0.33\textwidth]{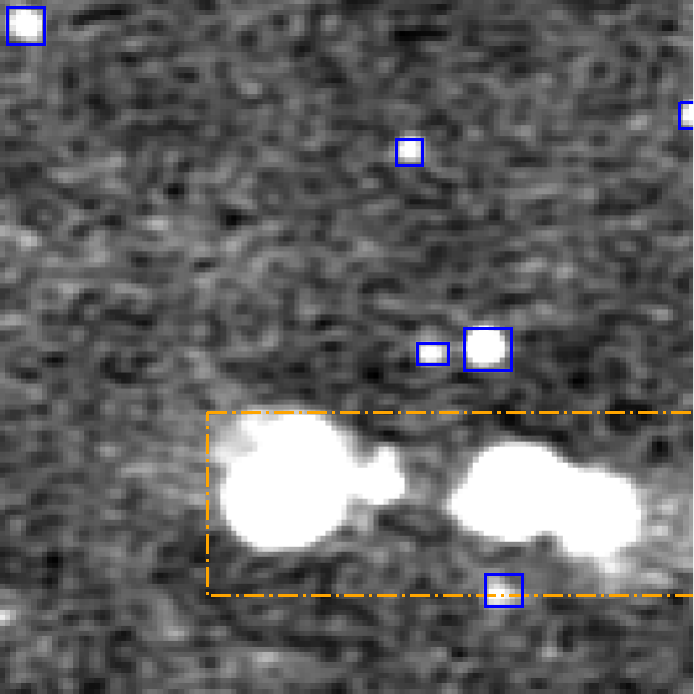} 
    \includegraphics[height=0.33\textwidth]{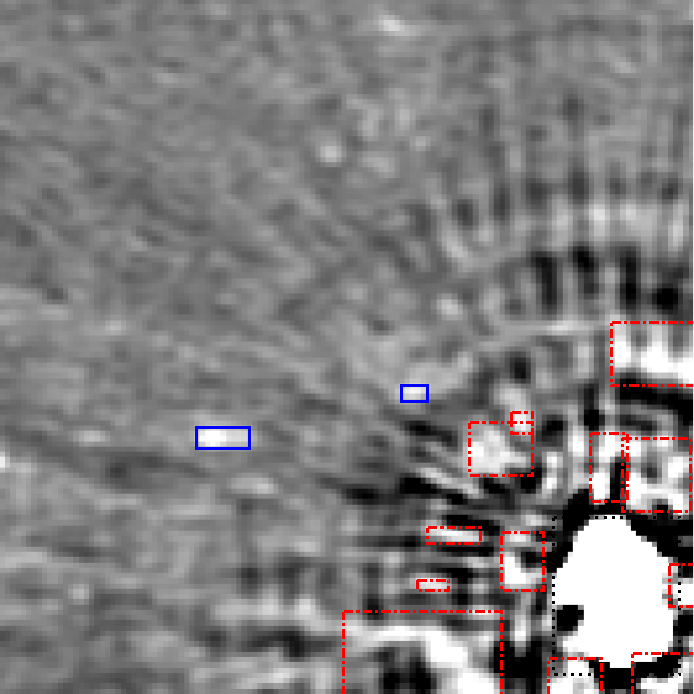} 
    \caption{Examples of the annotation classes shown on $132\times132$ pixel cutouts from the LOFAR 150\,MHz continuum images, with a ZScale stretch (contrast $=0.2$). Bounding boxes are distinguished using both colour and line style: \texttt{Compact} (blue, solid), \texttt{Single-Extended} (green, dashed), \texttt{Multi-Extended} (orange, dash--dot), \texttt{Flagged} (black, dotted), and \texttt{Spurious} (red, dash--dot--dotted).
    \emph{Left:} predominantly \texttt{Compact} sources and two \texttt{Single-Extended} sources. 
    \emph{Middle:} a \texttt{Multi-Extended} system together with nearby \texttt{Compact} sources. 
    \emph{Right:} a bright source affected by severe imaging artefacts is labelled \texttt{Flagged}, and surrounding artefact--induced false detections are marked as \texttt{Spurious}.
    All annotations are axis-aligned rectangular bounding boxes drawn by annotators.}
  \label{fig:source_scheme}
\end{figure*}

Annotations were created by a team of trained annotators within a private Zooniverse project. Annotators performed direct visual inspection of the LOFAR 150\,MHz deep-field continuum images and were not provided with any pre-existing source catalogues or automated source-finding outputs. Radio sources were identified solely on the basis of visual assessment of the continuum images, following the morphological definitions described in Section~\ref{subsec:class-label-schema}. For each identified source, annotators manually drew an axis-aligned rectangular bounding box enclosing the perceived radio emission and assigned one of the five mutually exclusive classes. Each bounding box together with its assigned class constitutes one annotation. Sources near the edges of the cutouts were treated in the same manner as interior sources and were annotated whenever clear radio emission was visible within the cutout. 

Illustrative examples of all classes and the associated colour key are shown in Figure~\ref{fig:source_scheme}. For \texttt{Multi-Extended} systems, these bounding boxes are intended to represent source-level envelopes enclosing all physically associated islands, rather than precise morphological outlines. While more general inclined rectangles could in principle provide a tighter geometric description, we adopt axis-aligned bounding boxes throughout to ensure consistency with the representation required by DETR-style detectors used in this work. The set-prediction framework of RF-DETR allows multiple, potentially overlapping bounding boxes to be predicted, enabling compact or single-extended sources embedded within larger \texttt{Multi-Extended} envelopes to be detected and classified without ambiguity.

To minimise confusion between genuine radio sources and artefact-induced responses around bright objects, annotators considered not only the morphology in the LOFAR 150\,MHz images but also the spatial context and symmetry of emission, as well as information from ancillary infrared and optical images, which were used to identify likely sidelobes or calibration artefacts that lack plausible multi-wavelength counterparts. Ambiguous cases in the vicinity of bright sources were conservatively labelled as \texttt{Flagged} or \texttt{Spurious}, rather than as astrophysical sources.

As a further quality-control measure, a subset of identical or overlapping cutouts was independently inspected by more than one annotator, and the resulting labels were cross-checked prior to model training to ensure consistent application of the morphological definitions. Following this quality-control step, all annotated cutouts were retained for training and validation, including those with partial spatial overlap.

Using this taxonomy, we curated $1{,}057$ $132\times132$ pixel cutouts from the ELAIS-N1 150\,MHz mosaic (1.5 arcsec per pixel), yielding $13{,}821$ labelled instances distributed as:
\texttt{Compact} $=12{,}535$,
\texttt{Single-Extended} $=520$,
\texttt{Multi-Extended} $=112$,
\texttt{Flagged} $=90$,
\texttt{Spurious} $=564$.
This corresponds to an average surface density of $\approx13.1$ sources per cutout, i.e., $\approx$ 1.21 $~\mathrm{arcmin}^{-2}$. For comparison, \citet{Riggi2023} trained Mask R-CNN source detectors on data with a surface density of $\approx$ 0.22 $\mathrm{arcmin}^{-2}$ (with 0.15 $~\mathrm{arcmin}^{-2}$ from VLA, 0.42 $~\mathrm{arcmin}^{-2}$ from ASKAP, and 0.33 $~\mathrm{arcmin}^{-2}$ from ATCA, respectively). In addition, \citet{Lao2023_HeTu} trained the HeTu-v2 source detector on the VLA FIRST survey at $\approx$ 0.10$~\mathrm{arcmin}^{-2}$. By contrast, based on deeper observations, our dataset operates at $\sim5.5\times$ and $\sim12\times$ higher source densities, respectively, than these previous studies. While the increased depth of the LoTSS Deep Fields implies higher SNRs for many detected sources, which can aid the identification of individual radio components, the corresponding rise in source density also leads to substantially more crowded fields. This introduces additional challenges for source-level morphology classification, including increased blending, ambiguous association of neighbouring emission, and the coexistence of compact and extended structures within small angular scales. Consequently, classification in deep LOFAR images constitutes a particularly stringent test of a model’s ability to reason over complex spatial context at survey depth.


\begin{table}
\centering
\caption{Class counts in our original training and validation datasets.}
\label{tab:class-stats}
\begin{tabular}{llll}
\hline
Class & Total & Training & Validation \\
\hline
\texttt{Compact} & 12{,}535 & 10{,}199 & 2{,}336\\
\texttt{Single-Extended} & 520 & 409 & 111 \\
\texttt{Multi-Extended} & 112 & 94 & 18 \\
\texttt{Flagged} & 90 & 62 & 28 \\
\texttt{Spurious} & 564 & 444 & 120 \\
\hline
\end{tabular}
\end{table}

We split the dataset $\approx80{:}20$ by cutout into training (850) and validation (207) subsets to avoid leakage. Because the cutouts are partially overlapping, a small fraction of validation cutouts also overlap spatially with those in the training set. In practice, such overlap is mostly confined to sources located near cutout boundaries, and involves only a small number of instances ($\approx 1.8\%$ of the validation annotations). This limited overlap does not materially affect the statistical independence of the validation set. The details of the subsets are summarised in Table~\ref{tab:class-stats}. No explicit rebalancing or resampling was applied to the class composition. Training and evaluation under such survey-like conditions allows us to assess the performance of RF-DETR in a realistic catalogue production setting, rather than optimising detection or classification metrics under artificially balanced class proportions. In this context, our primary aim is to evaluate whether RF-DETR can reliably detect and classify the full range of radio source morphologies encountered in LOFAR deep-field data, including rarer extended and multi-component systems.

To improve robustness to source orientation and image contrast variations, we apply a uniform data augmentation scheme to the training subset.  Each training cutout is randomly rotated by multiples of $90^\circ$ with equal probability and horizontally flipped with a probability of 0.6, so that extended and elongated sources are seen under different orientations without altering their intrinsic morphology. In addition, a ZScale-style contrast remapping is applied with probability 0.6, with the contrast parameter independently sampled per cutout in the range 0.1–0.6, to account for variations in background level and dynamic range across the images. The original training instances are retained, and multiple augmented views are generated per cutout. Taken together with the probabilities above, this results in approximately 15\% of training instances remaining unchanged in both geometry and contrast. All geometric transformations are applied consistently to the corresponding bounding boxes and class labels, and the augmented data are converted into the model-specific input formats described in Section~\ref{sec:methods}.

\section{Methods} \label{sec:methods}

\subsection{Transformer detector: RF-DETR}
\label{subsec:rfdetr}

\begin{figure*}
  \centering
   \includegraphics[width=1.0\textwidth]{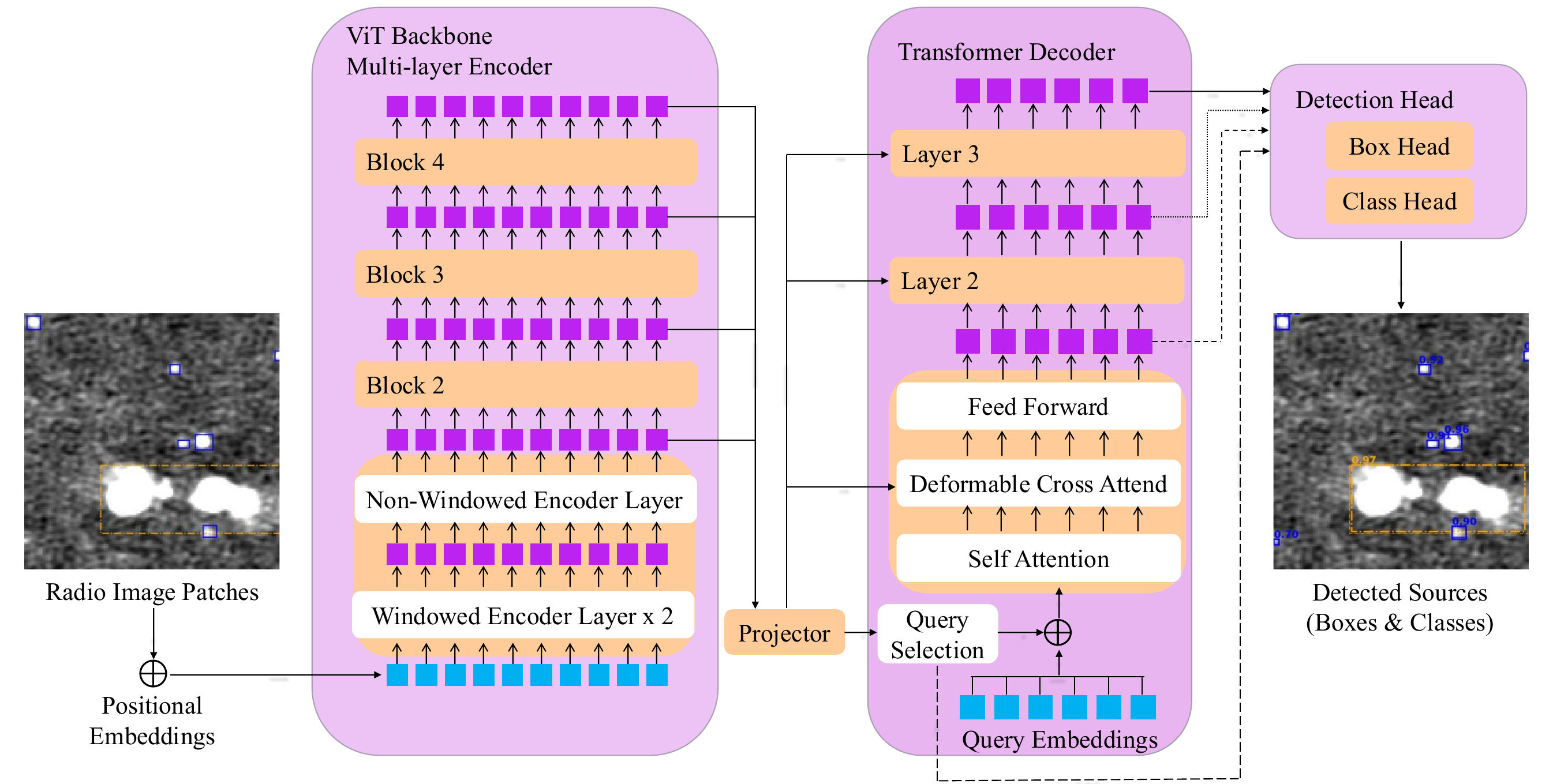} ~ ~
    \caption{Schematic overview of the RF-DETR architecture as used in this work. A LOFAR 150 MHz radio image cutout is processed by a ViT-based backbone with interleaved windowed and non-windowed encoder layers to extract image features, which are then passed to a Transformer encoder–decoder. A fixed set of object queries interacts with the encoded features and produces a set of predictions through the detection head, yielding source-level bounding boxes and corresponding class labels. This formulation enables one-to-one matching between predictions and radio sources, providing a unified framework for instance-level source detection and morphological classification.}
    \label{fig:rfdetr-architecture}
\end{figure*}

We adopt RF-DETR v1.3.0\footnote{\url{https://github.com/roboflow/rf-detr}} \citep{Robinson_rf-detr_ICLR} as our deep-learning detector for LOFAR 150 MHz continuum images. The public implementation follows the DETR set--prediction paradigm \citep{Carion2020_DETR}, in which a fixed set of object queries is matched one-to-one to ground-truth instances via the Hungarian algorithm. A vision backbone together with a Transformer encoder--decoder maps the queries to class logits and normalised bounding-box parameters. A schematic overview of the RF-DETR architecture as used in this work is shown in Figure~\ref{fig:rfdetr-architecture}.

RF-DETR builds on LW–DETR by coupling a ViT backbone with a real-time transformer decoder and introducing a number of training-effective refinements \citep{Chen2024_LWDETR}. The backbone is initialised with DINOv2 self-supervised features to provide strong general-purpose representations across domains \citep{Oquab2023_DINOv2}, while decoder conditioning with learned reference points follows the Conditional-DETR formulation \citep{Meng2021_ConditionalDETR}. A key component of RF-DETR is multi-scale deformable attention \citep{Zhu2021_DeformableDETR}. Rather than attending densely over all spatial positions, each object query samples a small number of locations around learned reference points across multiple feature scales. This sparse attention mechanism reduces the computational cost associated with high-resolution feature maps while focusing attention on informative regions of the image. In radio continuum data, where faint compact sources occupy only a few pixels and extended systems span a wide range of angular scales, this combination of multi-scale representation and spatially focused attention provides an effective framework for capturing both compact and complex morphologies.

Given the optimal Hungarian matching, the loss is computed on the matched prediction--target pairs. For each decoder output, the objective is
\begin{equation}
\mathcal{L} = \lambda_{\mathrm{cls}} \, \mathcal{L}_{\mathrm{cls}} 
+ \lambda_{\ell_1} \, \mathcal{L}_{\ell_1} 
+ \lambda_{\mathrm{giou}} \, \mathcal{L}_{\mathrm{giou}} ,
\end{equation}
where $\mathcal{L}_{\mathrm{cls}}$ is a sigmoid focal classification loss, and $\mathcal{L}_{\ell_1}$ and $\mathcal{L}_{\mathrm{giou}}$ are the standard $L_1$ and generalised-intersection-over-union (GIoU) losses for box regression. The focal formulation down-weights easy negative predictions and mitigates the strong class imbalance arising from the large number of unmatched queries. The same weighted loss is applied to intermediate decoder layers as auxiliary supervision, while the final layer provides the main predictions. A cardinality term is evaluated for monitoring purposes but is not included in the optimisation.

For training, the annotations described in Section~\ref{subsec:data-prep} were converted to the COCO format using axis-aligned bounding boxes parameterised as $[x, y, w, h]$, where $(x, y)$ denotes the coordinates of the upper-left corner of the box and $w$ and $h$ its width and height, respectively. All geometric augmentations were applied consistently to the bounding boxes. Single-channel LOFAR cutouts were replicated to three channels to match the RGB input expected by the pre-trained encoder. We adopted the \texttt{RFDETRMedium} variant with a DINOv2 encoder (specified as \texttt{dinov2\_windowed\_small}), initialised from a COCO-pretrained checkpoint. The network was trained for 200 epochs using mixed-precision training and synchronized batch normalization. Optimisation employed AdamW with a base learning rate of $5\times10^{-5}$ (encoder learning rate $2\times10^{-5}$), a weight decay of 0.02, and a cosine learning-rate schedule with a 10-epoch linear warm-up and a final minimum learning-rate factor of 0.05. Gradient accumulation was used (effective batch size $4\times4$) to stabilise optimisation, and exponential moving average (EMA) tracking was enabled with a decay factor of 0.9998. Prediction--target assignment followed DETR's bipartite (Hungarian) matching. We retained the default matching costs and loss coefficients for classification, \(L_1\), and GIoU, giving the usual stronger weight to the \(L_1\) box term relative to GIoU, following standard DETR formulation \citep{Carion2020_DETR}. Transformer drop-path regularisation was left at the model default for this run. Checkpoints were selected by the highest validation mean Average Precision from per-epoch COCO evaluation. The main parameters are summarised in Table~\ref{tab:rfdetr-config}. 

\begin{table}
\centering
\caption{Main parameters of RF-DETR used in this work.}
\label{tab:rfdetr-config}
\begin{tabular}{ll}
\hline
Component & Value \\
\hline
\multicolumn{2}{l}{\emph{Architecture \& Initialisation}} \\
Model variant                  & \texttt{RFDETRMedium} \\
Backbone                       & \texttt{dinov2 windowed} \\
Pretrained checkpoint          & \texttt{rf-detr-medium} \\ 
Number of classes              & 5 \\ 
Number of decoding layers      & 3 \\ 
Number of query slots          & 300 \\
\hline
\multicolumn{2}{l}{\emph{Optimisation \& Schedule}} \\
Optimizer                      & AdamW \\
Base learning rate             & $5\times10^{-5}$ \\
Backbone learning rate          & $2\times10^{-5}$ \\
LR schedule                    & cosine \\
Warm-up epochs                 & 10 \\
Weight decay                   & 0.02 \\
\hline                          
\multicolumn{2}{l}{\emph{Batching \& Train length}} \\
Multi-scale training           & on \\
Batch size                     & 4 \\
Gradient accumulation          & 4 \\
Max epochs                     & 200 \\
\hline
\multicolumn{2}{l}{\emph{Stability}} \\
EMA                            & on \\
EMA decay                      & 0.9998 \\
\hline
\multicolumn{2}{l}{\emph{Matching \& Loss}} \\
Matching cost (class)          & 2 \\
Matching cost ($L_1$)          & 5 \\
Matching cost (GIoU)           & 2 \\
Loss coefficient (cls)         & 2 \\
Loss coefficient (bbox $L_1$)  & 5 \\
Loss coefficient (GIoU)        & 2 \\
\hline
\end{tabular}
\end{table}

\subsection{Evaluation protocol and metrics}\label{sec:eval}

We evaluate RF-DETR on the validation set under a strict one-to-one matching between ground-truth (GT) sources and model detections (DTs). For each image, we retain detections with score $\ge 0.5$ and deem a GT--DT pair valid when their axis-aligned bounding boxes have intersection-over-union (IoU) $\ge 0.5$. All valid pairs are sorted by IoU and greedily matched to enforce that each GT and each DT appears in at most one pair; ties are broken by higher IoU and then by higher detection score. This protocol prevents double counting and therefore yields more conservative estimates than more permissive matching schemes.

We adopt metric definitions consistent with prior machine learning studies on radio data \citep[e.g.,][]{Riggi2023,Riggi2024}. For the detection metrics, we compute 
\begin{equation}
C=\frac{|\mathcal{A}_{1\text{--}1}|}{|G|},\qquad
R=\frac{|\mathcal{A}_{1\text{--}1}|}{|D|},\qquad
\mathrm{F1}_{\text{det}}=\frac{2CR}{C+R}.
\end{equation}
where $G$ is the set of GT objects, $D$ is the set of filtered detections, and $\mathcal{A}_{1\text{--}1}\subseteq G\times D$ is  the one-to-one matched-pair set defined above. Here, \emph{completeness} ($C$) is the fraction of true sources recovered by a matched detection, \emph{reliability} ($R$) is the fraction of detections corresponding to true sources, and $\mathrm{F1}_{\text{det}}$ is their harmonic mean.

Classification is assessed only on the matched pairs $\mathcal{A}_{1\text{--}1}$. For each class, a GT–DT pair counts as a true positive (TP) if the GT label and the predicted label are the same; otherwise it contributes a false negative (FN) for the GT class and a false positive (FP) for the predicted class. We then compute
\begin{equation}
\mathcal{R}=\frac{\mathrm{TP}}{\mathrm{TP}+\mathrm{FN}},\qquad
\mathcal{P}=\frac{\mathrm{TP}}{\mathrm{TP}+\mathrm{FP}},\qquad
\mathrm{F1}_{\text{cls}}=\frac{2\,\mathcal{P}\,\mathcal{R}}{\mathcal{P}+\mathcal{R}}.
\end{equation}
Here, \emph{recall} ($\mathcal{R}$) is the fraction of matched GT objects of a class that are correctly classified; \emph{precision} ($\mathcal{P}$) is the fraction of matched detections predicted as that class that are correct; and $\mathrm{F1}_{\text{cls}}$ is their harmonic mean. Both the detection metrics ($\mathrm{F1}_{\text{det}}$) and the classification metrics ($\mathrm{F1}_{\text{cls}}$) are computed separately for each class. The overall classification metrics referred to as ``All'' are obtained by pooling matched pairs across classes (micro average). By construction, classification metrics are conditional on successful detection and matching; end-to-end performance is bounded by detection completeness.

All detection and classification metrics are computed in the same manner for all five classes. However, the interpretation of these metrics differs between genuine source classes and the \texttt{Spurious} category. The \texttt{Compact}, \texttt{Single-Extended}, \texttt{Multi-Extended}, and \texttt{Flagged} classes all correspond to real sources, with \texttt{Flagged} identifying systems whose measurements are unreliable owing to severe imaging artefacts or dynamic-range limitations. By contrast, the \texttt{Spurious} class represents false detections induced by imaging artefacts rather than genuine sources. 
For this reason, the detection completeness or reliability of the \texttt{Spurious} class is not a primary determinant of catalogue quality. The purpose of introducing an explicit \texttt{Spurious} class is to distinguish artefact-induced detections from real sources. Accordingly, the primary concern is whether genuine astrophysical sources are incorrectly classified as \texttt{Spurious}, which would lead to their exclusion from subsequent analyses.

\subsection{Inference on LoTSS Deep Field continuum maps}\label{subsec:inference}

We run RF-DETR on each LoTSS Deep Field mosaic using a padded sliding window and a two-stage consolidation of per tile detections. Windows of $132\times132$ pixels with a stride of 66 pixels in both directions are laid out with end compensation so that the right and bottom edges are fully covered. Each window is extracted with a symmetric halo of $4\times\mathrm{FWHM}$ (in 16 pixels) using reflection padding outside the image. RF-DETR is applied to the haloed tile; we then perform per tile non-maximum suppression (NMS) and retain only boxes whose centroids fall inside the inner $132\times132$ region to avoid halo artefacts. Surviving detections are mapped back to mosaic coordinates and concatenated.

A class-aware consolidation step encodes radio-specific priors consistent with the annotation scheme used for training and evaluation. First, we retain detections with score $\ge 0.5$ and remove boxes with width or height $<3$\,pixels that are consistent with noise fragments. Next, a global per-class suppression step resolves duplicate detections while applying class-dependent overlap rules that preserve \texttt{Multi-Extended} envelopes, which are intended to represent physically associated multi-component radio systems. Finally, containment and same class overlap rules prune nested or near duplicate boxes that arise from tile boundaries or closely spaced detections, keeping the highest confidence instance so that each source is represented once. These consolidation rules are designed to reflect the same source-level assumptions encoded in the training annotations. The resulting catalogue provides world coordinate system (WCS) positions, pixel space boxes, and corresponding classes as well as their fluxes. 
Details on the computational resources and runtimes associated with training and inference are provided in Appendix~\ref{subsec:computation}.

\section{Results}
\label{sec:results}

\subsection{Model performance on the validation set} 
\label{subsec:evaluation}

\begin{figure*}
  \centering
   \includegraphics[width=0.975\textwidth]{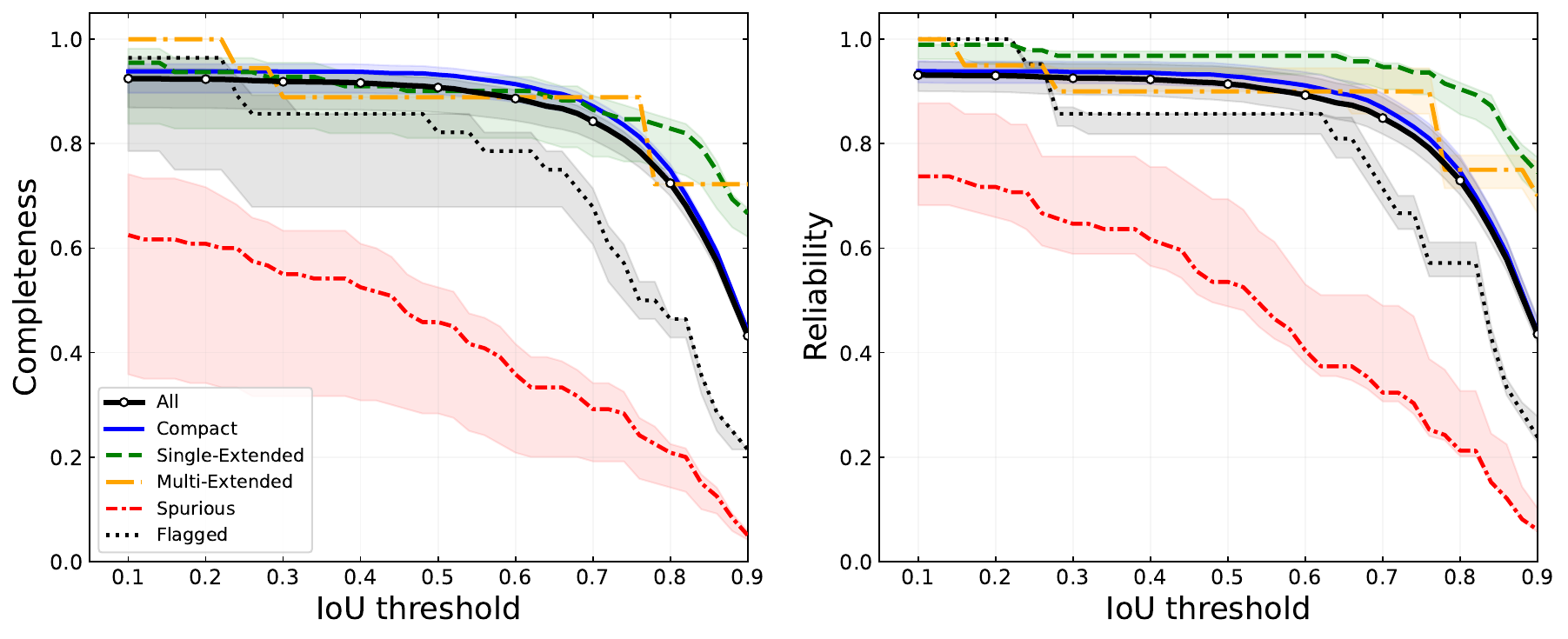} 
   \caption{Detection completeness (left) and reliability (right) as a function of the IoU threshold for the full sample (All) and the five morphological classes. The curves correspond to the reference score threshold of 0.5, while shaded regions indicate the variation of score thresholds in the range 0.25–0.75.}
   \label{fig:cr-curves}
\end{figure*}

\begin{figure*}
  \centering
  {\bfseries AI Model Predictions}\\[0.5mm]
   \includegraphics[height=0.33\textwidth]{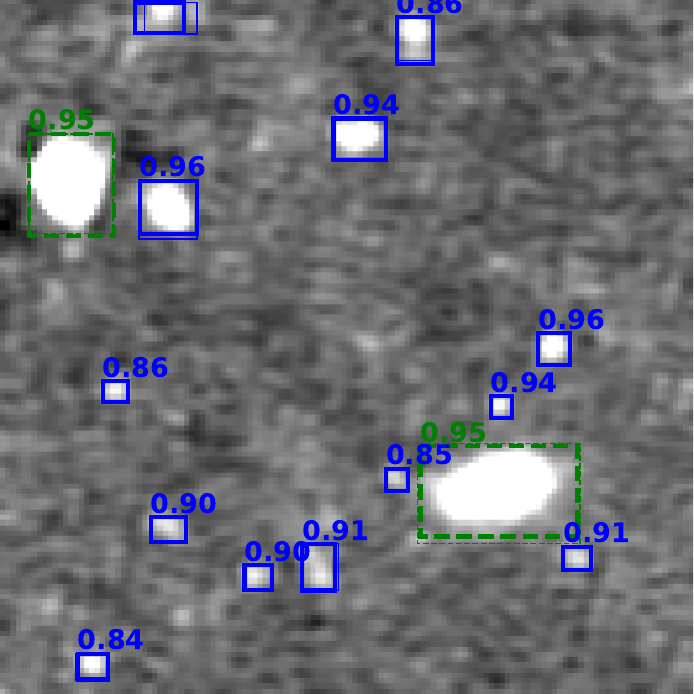} 
   \includegraphics[height=0.33\textwidth]{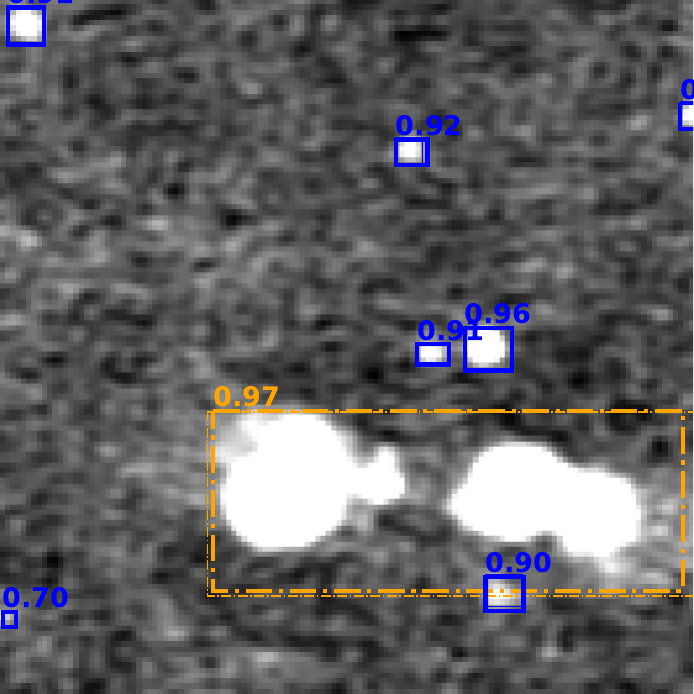}  
   \includegraphics[height=0.33\textwidth]{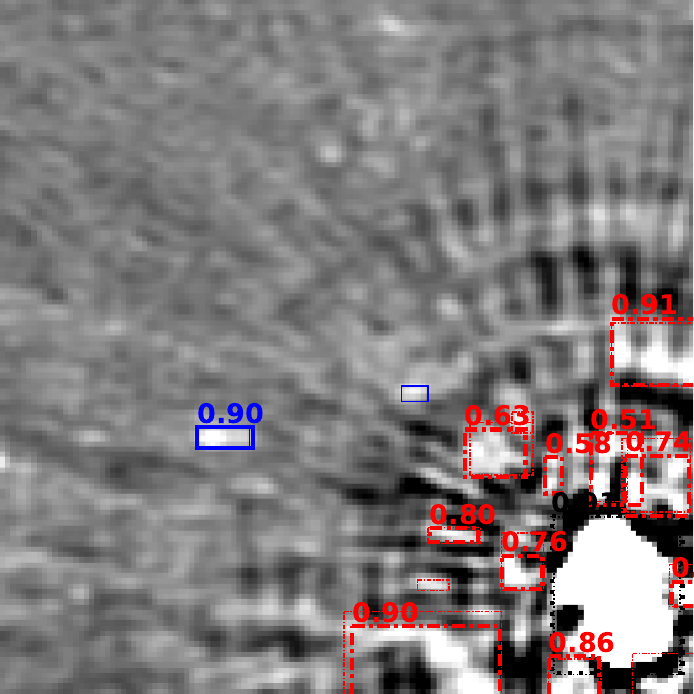} 
   \caption{RF-DETR predictions on the validation set for the same image cutouts shown in Figure~\ref{fig:source_scheme}, with bounding boxes following the same colour and line-style scheme. Predicted source instances are shown with thicker bounding boxes and annotated with detector confidence scores, while ground-truth annotations are shown with thinner boxes. Only predictions with score $\ge 0.5$ are displayed. The left panel shows \texttt{Compact} and \texttt{Single-Extended} predictions, the middle panel shows a \texttt{Multi-Extended} system together with nearby \texttt{Compact} sources, and the right panel contains \texttt{Spurious} detections around a \texttt{Flagged} source.}
    \label{fig:rfdetr-examples}
\end{figure*}

\begin{table*}
\centering
\caption{Validation results for RF--DETR on the validation set using one-to-one matching at IoU$\ge 0.5$ and score$\ge 0.5$.} 
\label{tab:main-quant}
\begin{tabular}{lccccccc}
\toprule
\multirow{2}{*}{Class} & \multicolumn{3}{c}{Detection} & \multicolumn{3}{c}{Classification} \\
\cmidrule(lr){2-4}\cmidrule(lr){5-7}
 & Completeness (\%) & Reliability (\%) & F1 score (\%) & Recall (\%) & Precision (\%) & F1 score (\%) \\
\midrule
All                      & 90.7 & 91.4 & 91.1 & 98.7 & 98.7 & 98.7 \\
\texttt{Compact}         & 93.2 & 92.9 & 93.1 & 99.9 & 99.1 & 99.5 \\
\texttt{Single-Extended} & 90.1 & 96.8 & 93.3 & 85.0 & 95.5 & 89.9 \\
\texttt{Multi-Extended}  & 88.9 & 90.0 & 89.4 & 100 & 88.9 & 94.1 \\
\texttt{Flagged}         & 82.1 & 85.7 & 83.9 & 73.9 & 94.4 & 82.9 \\
\texttt{Spurious}        & 45.8 & 53.5 & 49.4 & 89.1 & 92.5 & 90.7 \\
\bottomrule
\end{tabular}\\
\end{table*}


We trained RF-DETR on the LOFAR images with the main settings listed in Table~\ref{tab:rfdetr-config}. 
Figure~\ref{fig:rfdetr-examples} illustrates representative RF-DETR evaluation examples on the validation set for the same image cutouts shown in Figure~\ref{fig:source_scheme}. In fields dominated by \texttt{Compact} sources with occasional \texttt{Single-Extended} objects (\emph{left}), detections are dense and labels are correct. In the presence of large, multi-component systems (\emph{middle}), the model assigns the \texttt{Multi-Extended} label to the aggregate structure while preserving nearby \texttt{Compact} sources; this behaviour is desirable for radio surveys, as accurate identification of extended and especially \texttt{Multi-Extended} morphologies facilitates the discovery of nearby Fanaroff–Riley~I/II galaxies and/or giant radio galaxies. Around bright \texttt{Flagged} regions (\emph{right}), the detector also returns \texttt{Flagged} and \texttt{Spurious} predictions; however, predicted bounding boxes can differ from the human annotations, and the model occasionally produces multiple detections associated with ambiguous artefacts, which locally reduces completeness and reliability. This is expected because artefact morphology depends on bright-source sidelobes, $u\!v$ coverage, and calibration residuals, and thus lacks a stable pattern. Nevertheless, the detector seldom confuses real sources with \texttt{Spurious} detections, so the uncertainty in these classes---being imaging artefacts rather than astrophysical objects---does not bias detections of real galaxies.

To assess the robustness of the detection metrics with respect to the IoU threshold, we examine completeness–reliability curves (as defined in Section~\ref{sec:eval}) as a function of the IoU threshold, shown in Figure~\ref{fig:cr-curves}. For the three main morphological classes (\texttt{Compact}, \texttt{Single-Extended}, and \texttt{Multi-Extended}), both completeness and reliability remain nearly unchanged over the range \(0.4 \lesssim \mathrm{IoU} \lesssim 0.6\), indicating that the detection performance is stable within this interval. At higher IoU thresholds (\(\gtrsim 0.7\)), both metrics decrease, as expected from the increasingly stringent matching criterion, which amplifies small spatial mismatches between predicted and reference bounding boxes. This effect is more pronounced for diffuse or extended radio emission, where axis-aligned bounding boxes act as source-level envelopes rather than precise geometric descriptions. The shaded regions indicate the variation across detection score thresholds, demonstrating robustness to both IoU and score thresholds. As a sensitivity check, adopting more permissive IoU thresholds (e.g. IoU $\geq 0.3$) leads to a modest increase in completeness for the \texttt{Single-Extended} and \texttt{Multi-Extended} classes, without altering the relative ordering of classes or the overall conclusions. We therefore adopt IoU and score thresholds both set to 0.5 throughout this work as a conservative and uniform operating point. Table~\ref{tab:main-quant} reports the corresponding overall and per-class detection and classification metrics evaluated under this configuration.

Under this configuration, RF-DETR achieves strong performance on the ELAIS-N1 validation set, with an overall detection F1 score of $\simeq91$\,\% (completeness $C\simeq91$\,\%, reliability $R\simeq91$\,\%). Classification performance after one-to-one source matching is excellent, with both recall and precision exceeding $98$\,\%, yielding a classification F1 of $\simeq99$\,\%. We emphasise that these classification metrics are conditional on successful detection and matching; the overall catalogue-level performance is accordingly bounded by the detection completeness. For the dominant \texttt{Compact} population, the model performs exceptionally well in both detection and classification (F1$_{\mathrm{det}}\simeq93$\,\%; F1$_{\mathrm{cls}}\simeq99$\,\%), reflecting the relative morphological simplicity and abundance of these sources. \texttt{Single-Extended} systems are detected reliably (F1$_{\mathrm{det}}\simeq93$\,\%), while classification is slightly more challenging due to their structural diversity, with recall $\simeq85$\,\%, precision $\simeq96$\,\%, and F1$_{\mathrm{cls}}\simeq90$\,\%.  \texttt{Multi-Extended} sources exhibit moderately lower completeness ($\simeq89$\,\%) but high reliability ($\simeq90$\,\%), consistent with occasional partial overlaps between predicted and reference bounding regions for complex, multi-component systems. Importantly, their classification is robust, achieving perfect recall and a classification F1 of $\simeq94$\,\%, demonstrating that physically associated multi-island systems are rarely misidentified once detected. The \texttt{Flagged} class shows moderate detection performance (F1$_{\mathrm{det}}\simeq84$\,\%) and lower but still reasonable classification accuracy (F1$_{\mathrm{cls}}\simeq83$\,\%), reflecting the heterogeneous morphologies and limited training support for this category.  For the \texttt{Spurious} category, the reduced detection completeness is primarily driven by ground-truth \texttt{Spurious} instances that are not detected at all, rather than by systematic confusion with astrophysical source classes. These unmatched cases correspond to artefact-like structures in the radio images that do not give rise to sufficiently confident detections. This behaviour is consistent with the high classification precision achieved for the \texttt{Spurious} class (precision $\simeq 93$~per~cent), indicating that genuine radio sources are rarely assigned a \texttt{Spurious} label once detected. The spatial association of \texttt{Spurious} detections with regions affected by strong imaging artefacts, as reflected by the \texttt{Flagged} class, further supports this interpretation. Other morphological classes do not exhibit comparable clustering around such regions. Consequently, the \texttt{Spurious} class predominantly captures artefact-related features, and does not significantly impact the identification of genuine radio sources.

Recently, \citet{Riggi2023} developed \texttt{caesar-mrcnn} from Mask R--CNN and successfully applied it to VLA/ASKAP/ATCA datasets. With IoU and score thresholds of 0.6 and 0.5, they report overall \(C\simeq88\)~\% and \(R\simeq64\)~\%, with a typical class ordering: \texttt{Compact} showing the highest in completeness, \texttt{Extended} classes trading completeness for reliability, and \texttt{Spurious} difficult to detect (\(C\simeq46\)~\%, \(R\simeq37\)~\%). Our results follow the same qualitative trend while operating at, on average, \(\simeq5.5\times\) higher source density driven by deeper LOFAR imaging (Section~\ref{subsec:data-prep}). To provide a controlled in-domain benchmark, we also trained the latest \texttt{caesar-mrcnn-tf2} on our training set and evaluated it on our LOFAR validation set at the same thresholds (see Table~\ref{tab:mrcnn-evaluation} in Appendix \ref{subsec:mrcnn}). It remains competitive on \texttt{Compact} sources (F1$_{\text{det}}\simeq88$\,\%), but shows larger gaps on the complex classes relative to RF-DETR, indicating that RF-DETR handles complex morphology better in our denser and class-imbalanced dataset.
 

\subsection{RF-DETR source catalogues in the LoTSS Deep Fields} \label{subsec:catalogues}

\begin{figure*}
  \centering
   \includegraphics[height=0.47\textwidth]{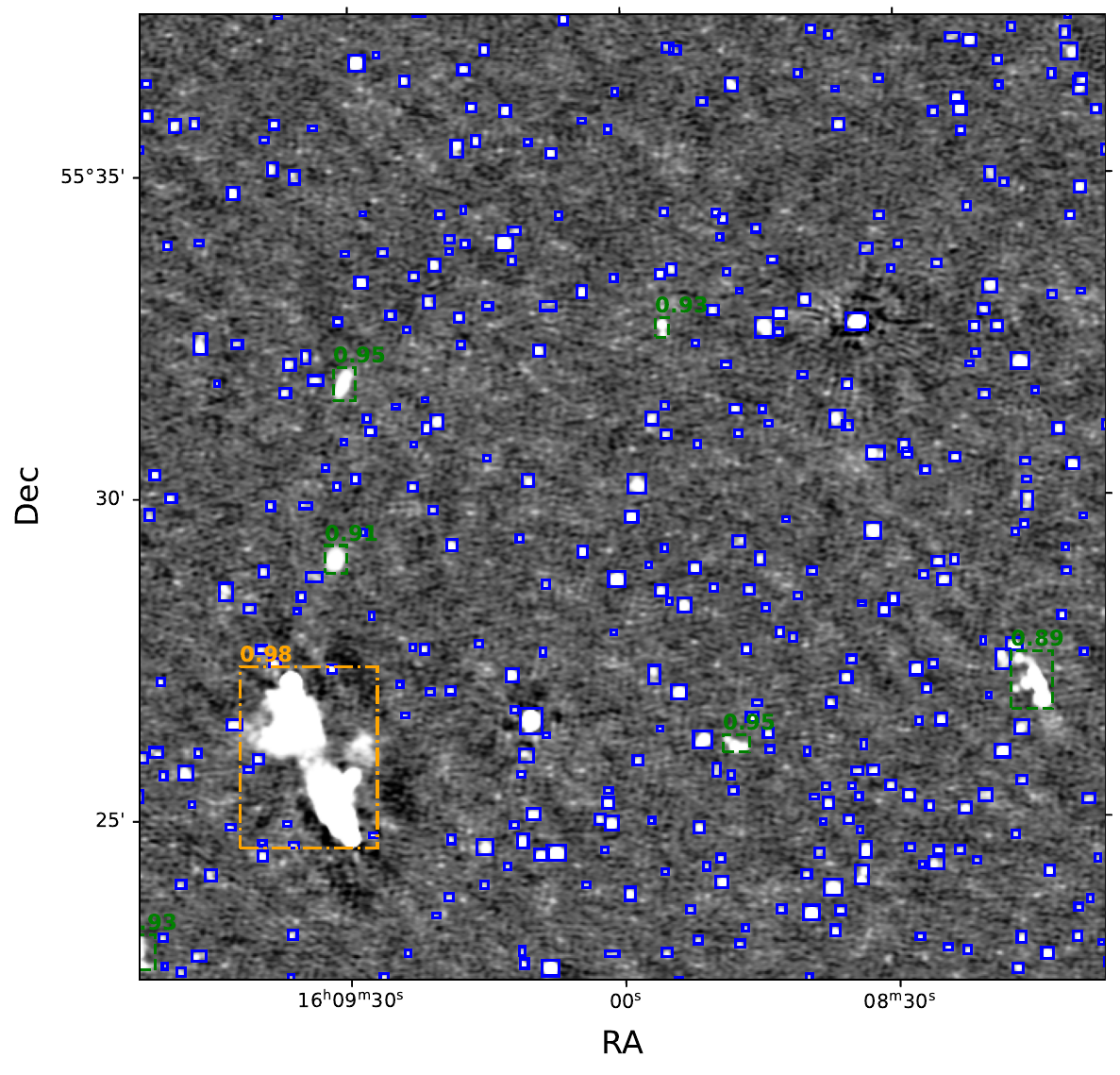} ~ ~  
   \includegraphics[height=0.47\textwidth]{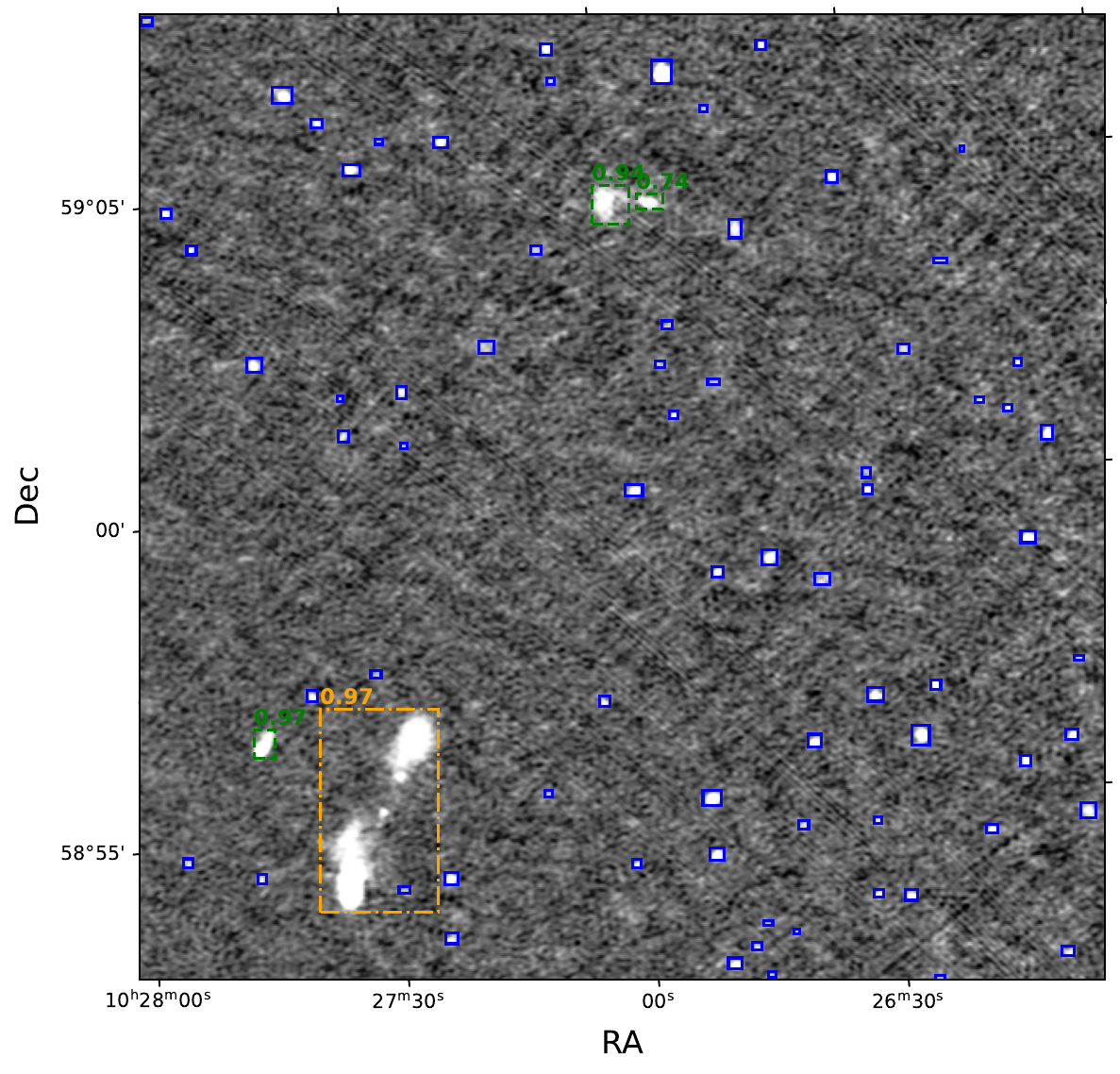} ~ ~  
   \includegraphics[height=0.47\textwidth]{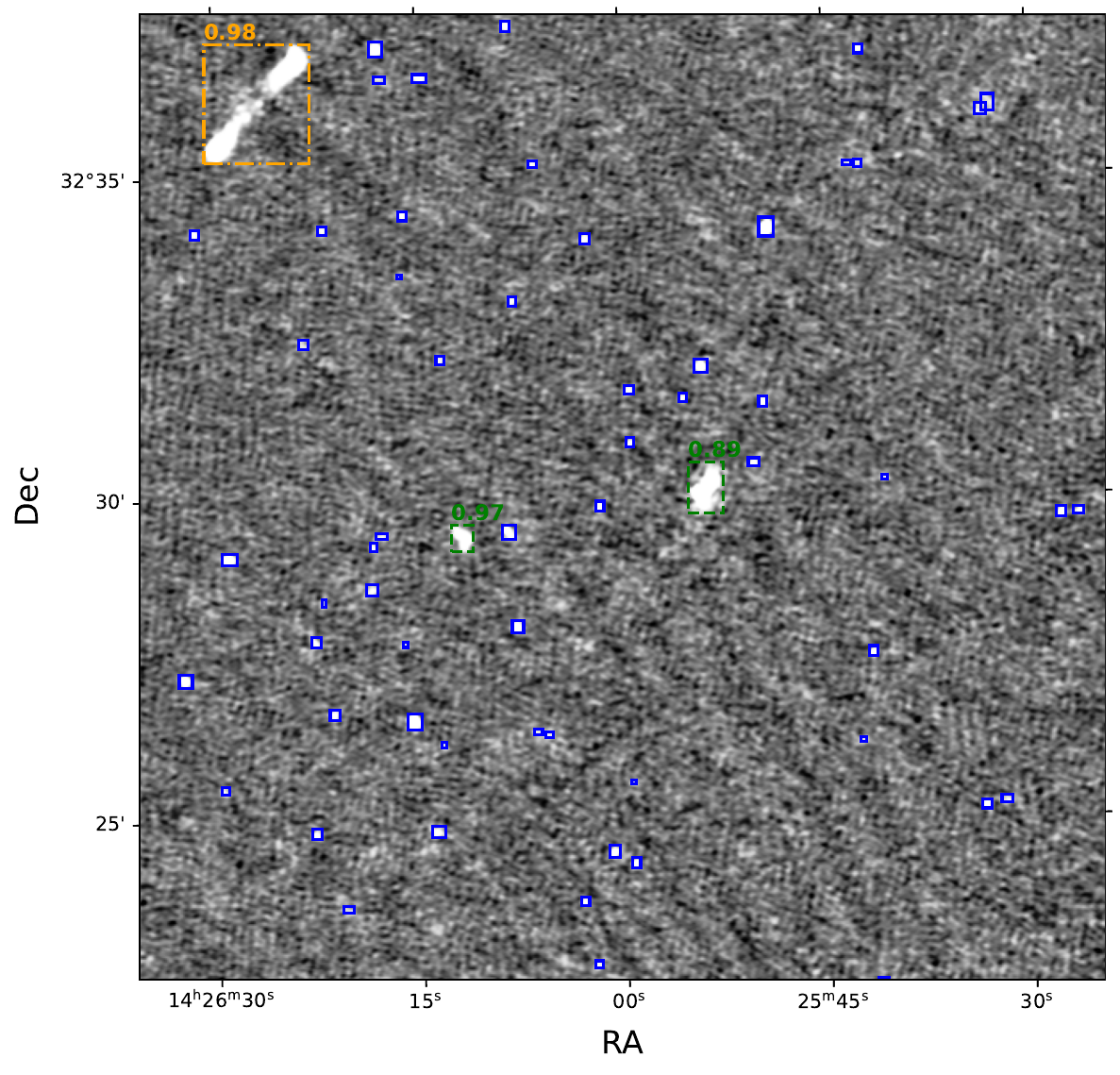} ~ ~  
   \includegraphics[height=0.47\textwidth]{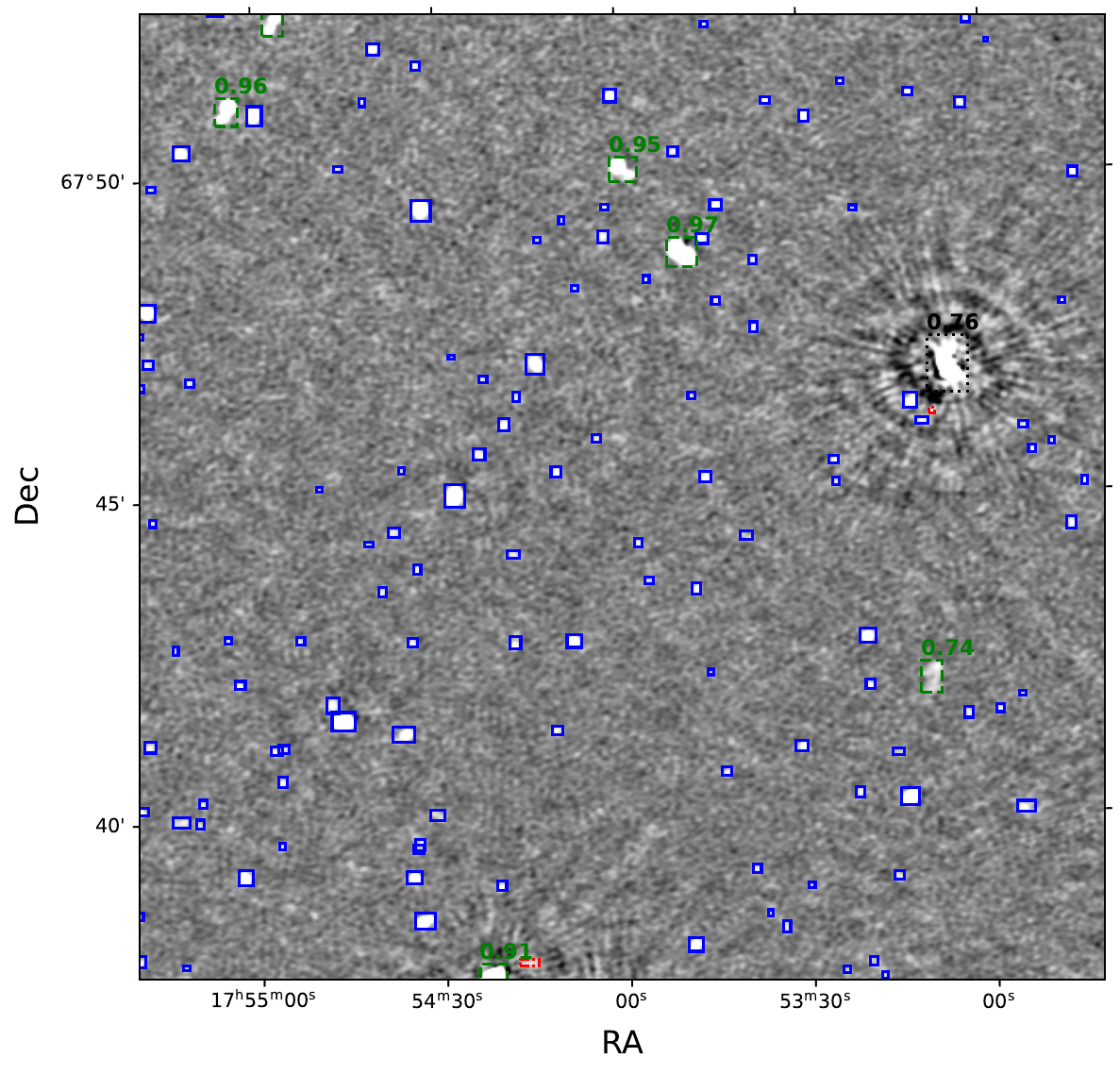} ~ ~  
   \caption{RF-DETR detections in randomly selected $15\arcmin\times15\arcmin$ regions from the four LoTSS Deep Fields (ELAIS-N1, Lockman Hole, Bo\"otes, and EDFN). Predicted bounding boxes follow the same colour and line-style scheme as defined in Figure~\ref{fig:source_scheme}. Detector confidence scores are shown only for \texttt{Single-Extended}, \texttt{Multi-Extended} and \texttt{Flagged} for clarity.}
  \label{fig:rf-detection}
\end{figure*}

\begin{table*}
\centering
\caption{RF-DETR detections in the LoTSS Deep Fields.}
\label{tab:rfdetr-catalog}
\begin{tabular}{lrrrrrr}
\toprule
Field & All & \texttt{Compact} & \texttt{Single-Extended} & \texttt{Multi-Extended} & \texttt{Flagged} & \texttt{Spurious} \\
\midrule
ELAIS-N1     & 73,539 & 69,441 (94.4\%) & 1,434 (2.0\%) & 130 (0.18\%) & 246 (0.3\%) & 2,288 (3.1\%) \\
Lockman Hole & 75,609 & 69,865 (92.4\%) & 1,855 (2.4\%) & 196 (0.26\%) & 390 (0.5\%) & 3,303 (4.4\%) \\
Boötes        & 43,805 & 41,187 (94.0\%) & 870 (2.0\%)   & 91 (0.21\%)  & 184 (0.4\%) & 1,473 (3.4\%) \\
EDFN          & 56,969 & 53,029 (93.1\%) & 1,306 (2.3\%)   & 142 (0.25\%)  & 266 (0.5\%) & 2,226 (3.9\%) \\
\bottomrule
\end{tabular}
\end{table*}

We apply the trained RF-DETR model to the full continuum maps of all four LoTSS Deep Fields (ELAIS-N1, Lockman Hole, Bo\"otes, and EDFN) using the tiled inference and consolidation procedure described in Section~\ref{subsec:inference}. Although the model is trained exclusively on ELAIS-N1, the same inference configuration and operating point are adopted for all fields. The four Deep Fields cover comparable sky areas of order $\sim$ 20~deg$^{2}$, but differ in achieved depth and calibration history, providing a realistic test of cross-field transfer under heterogeneous imaging conditions.

Table~\ref{tab:rfdetr-catalog} summarises the RF-DETR detections in each field, reporting the total number of sources and their breakdown by morphological class. As expected, fields reaching lower central rms levels, such as ELAIS-N1 and Lockman Hole, yield systematically larger detection counts than the shallower Bo\"otes and EDFN fields. In all cases, however, the resulting catalogues are strongly dominated by \texttt{Compact} sources, which account for $\simeq92$--95\,\% of all detections, in line with expectations for deep low-frequency radio surveys. The relative contributions of \texttt{Single-Extended} and \texttt{Multi-Extended} sources are small and remarkably stable across the four Deep Fields, together comprising only $\simeq2$--3\,\% of the catalogues. This consistency indicates that the RF-DETR detector generalises well beyond the training field, producing comparable morphological populations despite differences in depth, sky coverage, and calibration strategy. Representative cutouts from each field, selected to illustrate the range of detected morphologies, are shown in Figure~\ref{fig:rf-detection}.
Minor field-to-field variations are primarily confined to the \texttt{Flagged} and \texttt{Spurious} categories. Such behaviour is consistent with differences in imaging characteristics, including stronger sidelobes, elevated or spatially varying noise, and direction-dependent calibration residuals, rather than with intrinsic changes in the underlying source population \citep{Tasse2021_LOTSS_DeepField,Sabater2021_ELAISN1,Bondi2024_LOFAR_EDFN}. Overall, these variations do not affect the stability of the primary source populations and have no impact on the scientific interpretation of the resulting catalogues.

\subsection{Comparison with PyBDSF catalogues} \label{sec:rf-vs-pybdsf}

\begin{figure*}
  \centering
   \includegraphics[width=0.4\textwidth]{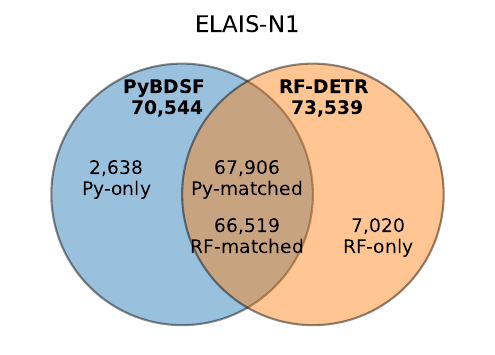}  
   \includegraphics[width=0.4\textwidth]{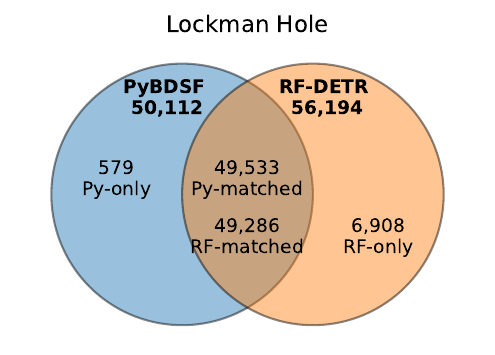}  
   \includegraphics[width=0.4\textwidth]{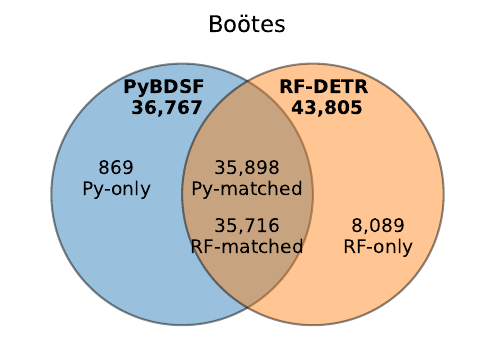}  
   \includegraphics[width=0.4\textwidth]{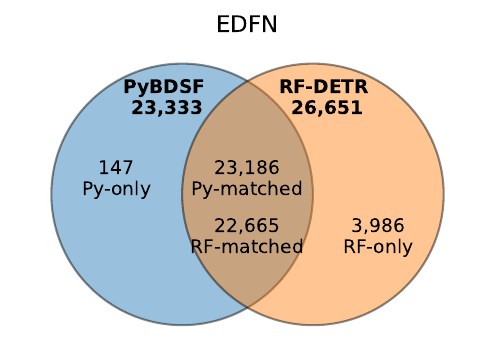}  
   \caption{Venn diagrams illustrating the catalogue overlap between PyBDSF (blue) and RF-DETR (orange) in the four LoTSS Deep Fields. Numbers inside each circle indicate the number of detections in the corresponding catalogue. The overlap region shows the number of Py-matched sources, defined as PyBDSF sources that are associated with at least one RF-DETR bounding box, as well as the corresponding number of RF-matched detections. Py-only sources correspond to PyBDSF detections without an RF-DETR counterpart, while RF-only detections denote RF-DETR bounding boxes that are not matched to any PyBDSF source.
   Specifically, for the Lockman Hole and EDFN fields, both RF-DETR and PyBDSF counts are restricted to the central regions covered by the publicly available PyBDSF catalogues, matching the sky coverage used for the comparison. See more details in text.}
  \label{fig:rf-vs-pybdsf-venn}
\end{figure*}

Having characterised the RF-DETR catalogues in the LoTSS Deep Fields, we now compare them with the widely used component-based catalogues produced by PyBDSF on the same continuum maps so as to assess catalogue-level completeness, overlap, and structural differences between the two approaches. The PyBDSF catalogues used in this comparison are the publicly released LoTSS Deep Fields DR1 products (e.g.\ \citealt{Sabater2021_ELAISN1, Tasse2021_LOTSS_DeepField}). 
To establish correspondences between the two catalogues, we adopt a PyBDSF-centric containment-based matching scheme. Specifically, a PyBDSF source is considered matched if its catalogue position (RA, Dec) lies within the area of a predicted RF-DETR bounding box. In cases where a PyBDSF source falls within more than one RF-DETR bounding box, the match is assigned to the RF-DETR detection whose bounding-box centre is closest to the PyBDSF position. Because multiple PyBDSF sources can be associated with a single RF-DETR bounding box, the number of Py-matched sources is systematically larger than the number of RF-DETR detections that contain at least one PyBDSF source.

Figure~\ref{fig:rf-vs-pybdsf-venn} illustrates the overlap between RF-DETR and PyBDSF catalogues across all four LoTSS Deep Fields, showing the total number of detections, as well as the Py-only, RF-only, and matched populations. The majority of PyBDSF sources are successfully recovered by RF-DETR. In the ELAIS-N1 field, approximately 96\% of PyBDSF sources (67,906 out of 70,544) are matched to at least one RF-DETR detection, with similarly high recovery fractions ($\gtrsim 97$\,\%) observed in the other fields. In addition to this high level of overlap, RF-DETR yields a slightly larger number of catalogue entries than PyBDSF. This difference reflects the distinct detection strategies adopted by the two methods. At the operating point used in Section~\ref{sec:methods}, RF-DETR applies a single confidence threshold to the detector’s class probabilities, without a subsequent island-growth stage. In contrast, PyBDSF first identifies emission islands via pixel-thresholding and then fits one or more Gaussians; sources that do not seed an island are therefore not included in the catalogue. As a result, spatially coherent but low-surface-brightness sources that fall below per-pixel thresholds may be recovered by RF-DETR while remaining absent from PyBDSF.
For the Lockman Hole and EDFN fields, the publicly available PyBDSF catalogues are restricted to the central regions of the continuum maps and therefore contain fewer sources overall (50,112 and 23,333, respectively). Accordingly, the RF-DETR counts reported in Figure~\ref{fig:rf-vs-pybdsf-venn} for these two fields are restricted to the same sky areas covered by the PyBDSF catalogues. In view of the reduced sky coverage of the available catalogues for the Lockman Hole and EDFN fields, and to avoid introducing additional field-dependent systematics associated with variations in noise properties and survey depth, we therefore focus the detailed comparison between RF-DETR and PyBDSF on the full ELAIS-N1 map, which serves as the deepest and most uniformly sensitive LoTSS Deep Field and thus provides a natural reference for catalogue-level comparisons \citep[e.g.][]{Sabater2021_ELAISN1}.

\begin{figure*}
  \centering
   \includegraphics[width=0.475\textwidth]{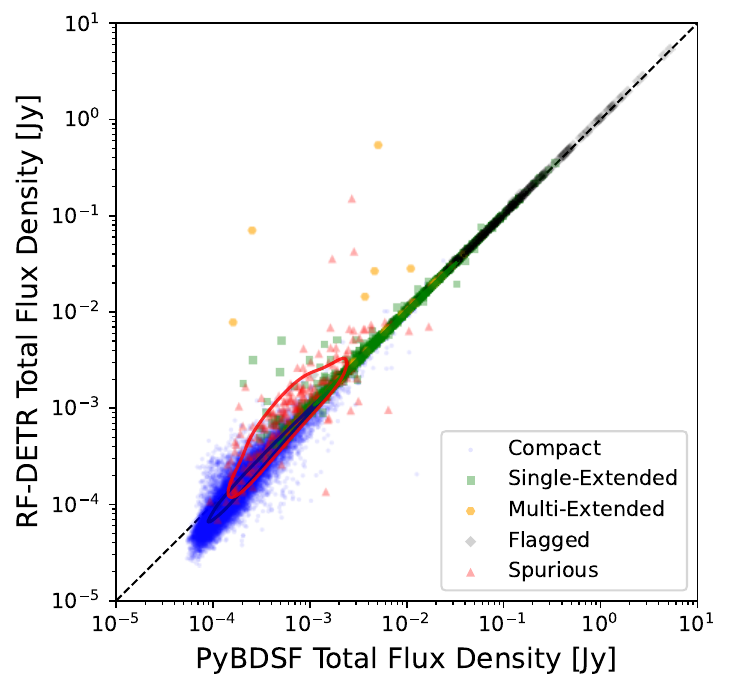}   
   \includegraphics[width=0.475\textwidth]{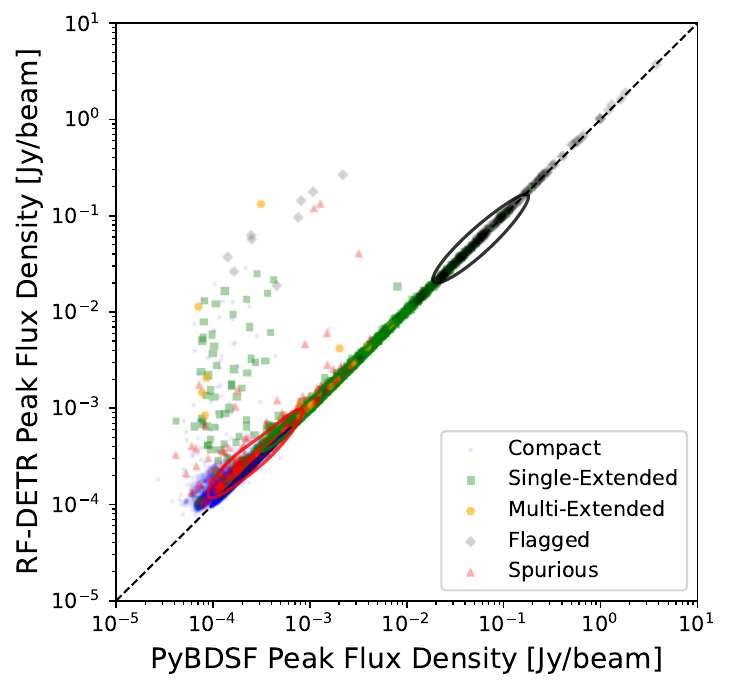} 
   \caption{Comparison of RF-DETR and PyBDSF flux densities for matched sources in the ELAIS–N1 field. 
   Left panel: comparison of total flux densities obtained via direct pixel summation within RF-DETR bounding boxes versus PyBDSF islands.
   Right panel: comparison of peak flux densities, where RF-DETR values correspond to the maximum pixel intensity within the bounding boxes, while PyBDSF measurements are derived from Gaussian component fitting within the associated islands. Data points are shown using different colours and marker shapes according to the RF-DETR morphological class. For classes with more than 30 matched sources, kernel-density contours are overlaid to enclose the densest 68.3\% of the points.}
  \label{fig:rf-vs-pybdsf-1to1}
\end{figure*}

Figure~\ref{fig:rf-vs-pybdsf-1to1} compares total (left) and peak (right) flux densities for matched RF-DETR and PyBDSF sources in the ELAIS-N1 field. Total flux densities are obtained by summation within each RF-DETR bounding box and within the associated PyBDSF island, while peak flux densities correspond to the maximum pixel intensity inside the RF-DETR bounding box and the fitted Gaussian peak for PyBDSF. For bright sources, including those classified as \texttt{Flagged}, the flux densities measured by RF-DETR and PyBDSF closely follow the one-to-one relation in both total and peak flux, down to approximately $0.01$~Jy and $0.01$~Jy~beam$^{-1}$, respectively. \texttt{Flagged} sources are identified as real radio sources rather than false detections; this class reflects limitations in flux measurements caused by dynamic-range and calibration artefacts around bright emission, rather than uncertainty in source identification (Section \ref{subsec:class-label-schema}). Accordingly, these sources are retained in the catalogue for completeness and source association, while their flux densities are best refined through dedicated re-imaging or improved direction-dependent calibration when used for quantitative analyses.

Deviations from the one-to-one relation reflect methodological differences: RF-DETR often reports higher total fluxes for \texttt{Multi-Extended} systems because a single RF-DETR bounding box can enclose spatially separated components that PyBDSF treats as multiple islands.  A systematic excess is also seen for a subset of RF-DETR \texttt{Single-Extended} sources, particularly in the peak flux comparison. This may reflect differences in how PyBDSF peak fluxes are derived from Gaussian component fitting and can be suppressed for structured extended emission, while RF-DETR values are measured directly from pixel-level intensities and are more sensitive to localized brightness enhancements.
At lower flux densities, \texttt{Spurious} detections exhibit substantial scatter, and since these detections are produced by different types of artefacts, their flux density measurements are not expected to be reliable and should not be used for quantitative analyses. \texttt{Compact} sources display increased scatter towards the faint end, and a subset of these sources show systematically lower RF-DETR total flux densities relative to PyBDSF. 
This behaviour is consistent with differences in how low-surface-brightness emission near the detection threshold is treated by the two approaches, particularly for faint \texttt{Compact} sources whose extended emission can be partially excluded by RF-DETR bounding boxes, and this effect is further quantified below.

\begin{figure*}
  \centering
   \includegraphics[width=0.33\textwidth]{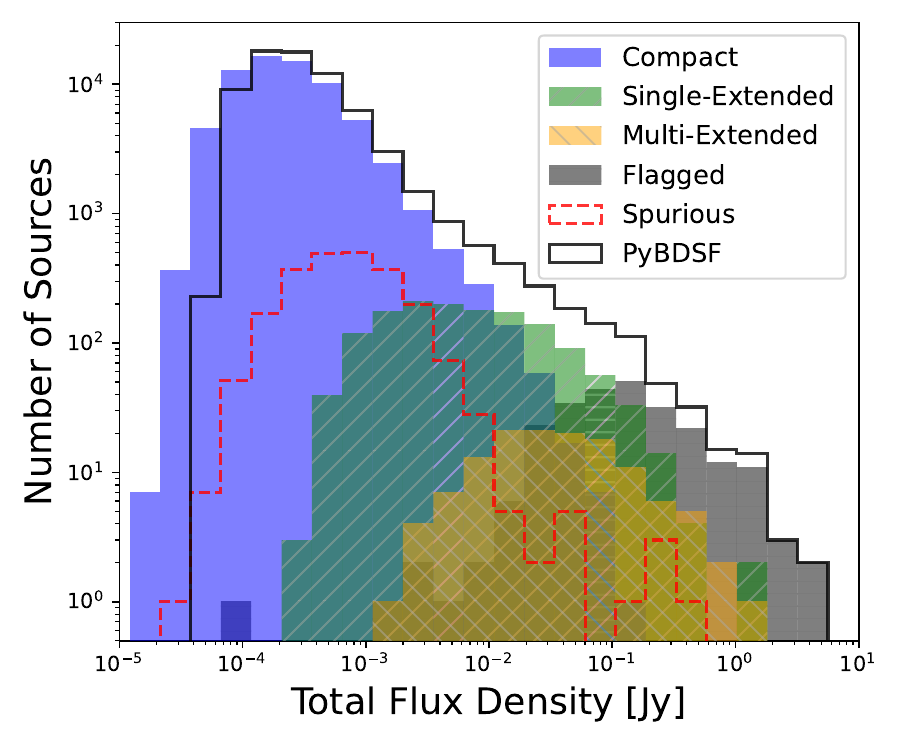}   
   \includegraphics[width=0.33\textwidth]{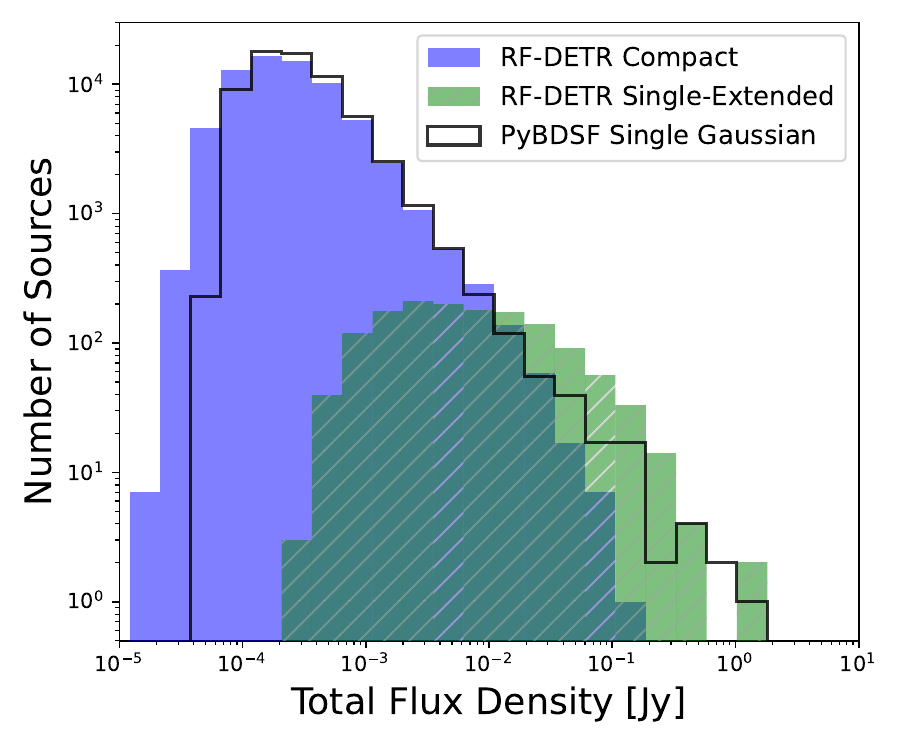}  
   \includegraphics[width=0.33\textwidth]{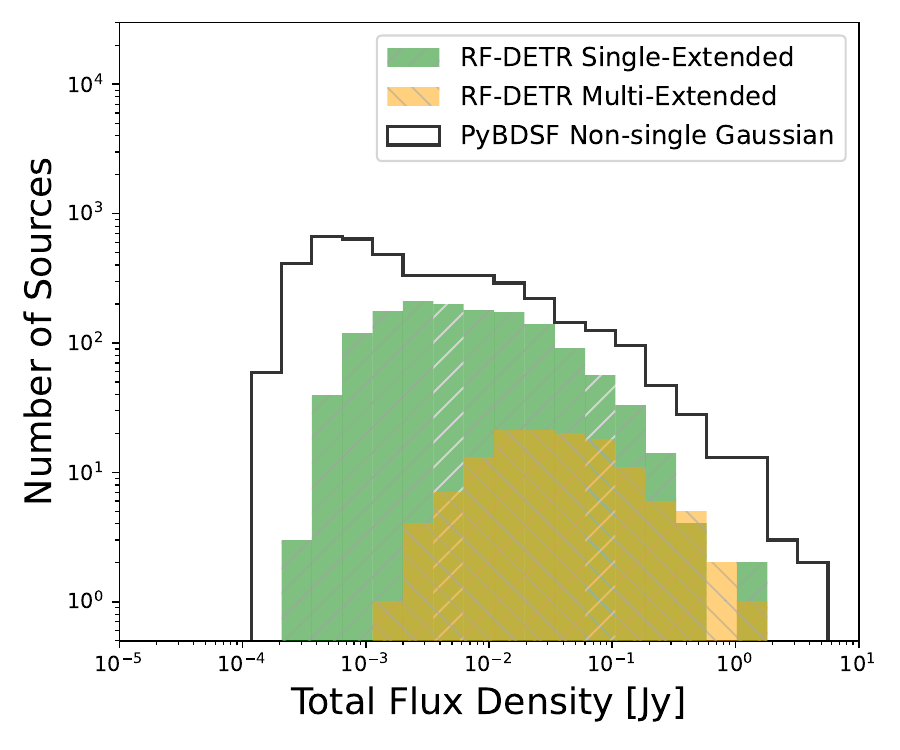}  
   \caption{Total flux distributions for sources detected in the ELAIS-N1 field.
   Left: RF-DETR catalogue decomposed by morphological class (histograms with different colours and hatch patterns: \texttt{Compact}, \texttt{Single-Extended}, \texttt{Multi-Extended}, \texttt{Flagged} and \texttt{Spurious}), overlaid with the PyBDSF total flux distribution (black open histograms).
   Middle: Comparison between RF-DETR \texttt{Compact}, \texttt{Single-Extended} and PyBDSF single-Gaussian sources.
   Right: Comparison between RF-DETR extended classes and PyBDSF component groupings (\texttt{M} and \texttt{C}).}
  \label{fig:rf-vs-pybdsf-dis}
\end{figure*}

Figure~\ref{fig:rf-vs-pybdsf-dis} places these source-level differences into a population context by comparing total flux distributions decomposed by morphological class. The overall RF-DETR flux distribution closely tracks PyBDSF at intermediate and high fluxes, including \texttt{Single-Extended}, \texttt{Multi-Extended} and \texttt{Flagged} classes. 
Notably, RF-DETR shows a modest excess of faint detections relative to PyBDSF that is dominated by \texttt{Compact} sources. To interpret this excess, we examine the detection behaviour as a function of signal-to-noise ratio (SNR), focusing on \texttt{Compact} sources where discrepancies are most pronounced.  The peak SNR is defined as the ratio of the peak flux density to the local background rms. Using the matching method described above, approximately 7,000 RF-DETR detections are not matched to any PyBDSF source, of which about 5,000 are classified as \texttt{Compact} and about 2,000 as \texttt{Spurious}.  These RF-DETR-only \texttt{Compact} detections are predominantly low-SNR candidates; approximately half of these sources have peak SNR below 5, with a median peak SNR of $\simeq 5.1$, compared to a median peak SNR of 9.0 for \texttt{Compact} sources matched between RF-DETR and PyBDSF.  In addition, these RF-DETR-only \texttt{Compact} detections are not spatially clustered around regions classified as \texttt{Flagged} or \texttt{Spurious}; none lie within 10~arcsec of such regions, and fewer than 5~per~cent are located within 60~arcsec. This indicates that the majority of the additional low-SNR compact detections returned by RF-DETR are not trivially attributable to obvious imaging artefacts, but instead represent a broader population of low-SNR compact source candidates near the detection threshold. Conversely, approximately 2,600 PyBDSF sources are not matched to any RF-DETR detection. These unmatched PyBDSF sources are systematically faint, with a median total flux density of $2.5\times10^{-4}$~Jy, compared to $3.4\times10^{-4}$~Jy for PyBDSF sources recovered by RF-DETR. A Kolmogorov--Smirnov test confirms that the detected and missing PyBDSF populations are statistically distinct ($p \ll 0.01$), indicating that incompleteness in both catalogues is concentrated toward the low-SNR, faint source regime. These results indicate that RF-DETR tends to return a larger population of low-SNR compact candidates, whereas PyBDSF is more conservative in this regime but may recover some faint Gaussian components that RF-DETR does not. 

Nevertheless, the modest increase in RF-DETR faint detections does not lead to a higher catalogue-level integrated flux density; summing the total flux densities over all detected sources in the ELAIS-N1 field yields 150~Jy for RF-DETR, compared to 157~Jy for the PyBDSF catalogue, corresponding to a relative difference of $-4.6\%$.  In particular, RF-DETR total fluxes are obtained by direct pixel summation within annotation-driven bounding boxes, which can be conservative around low-surface-brightness emission, while PyBDSF measurements integrate emission over the contiguous islands, including flux associated with fitted Gaussian wings \citep{Mohan2015_PyBDSF}. As a result, although RF-DETR returns a slightly larger number of faint detections, their individual flux contributions are typically small and do not offset the systematically larger per-source apertures adopted by PyBDSF, leading to a modestly lower total integrated flux density for RF-DETR at the catalogue level. In this sense, RF-DETR can be regarded as a generator of faint compact source candidates, for which SNR-based post-processing or additional validation should be applied to balance completeness and reliability according to specific scientific objectives.

\begin{figure*}
  \centering
   \includegraphics[height=0.47\textwidth]{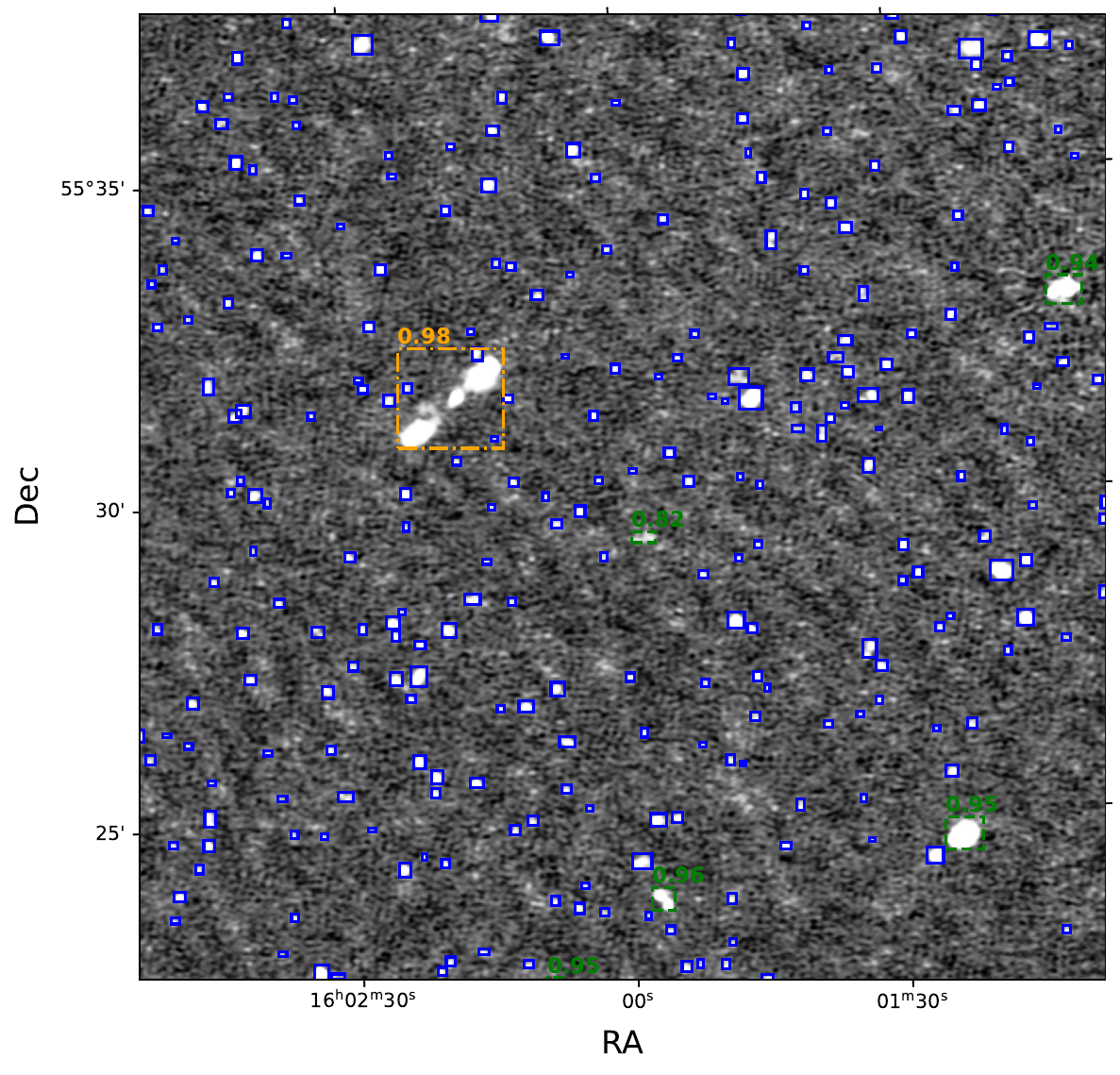} ~ ~  
   \includegraphics[height=0.47\textwidth]{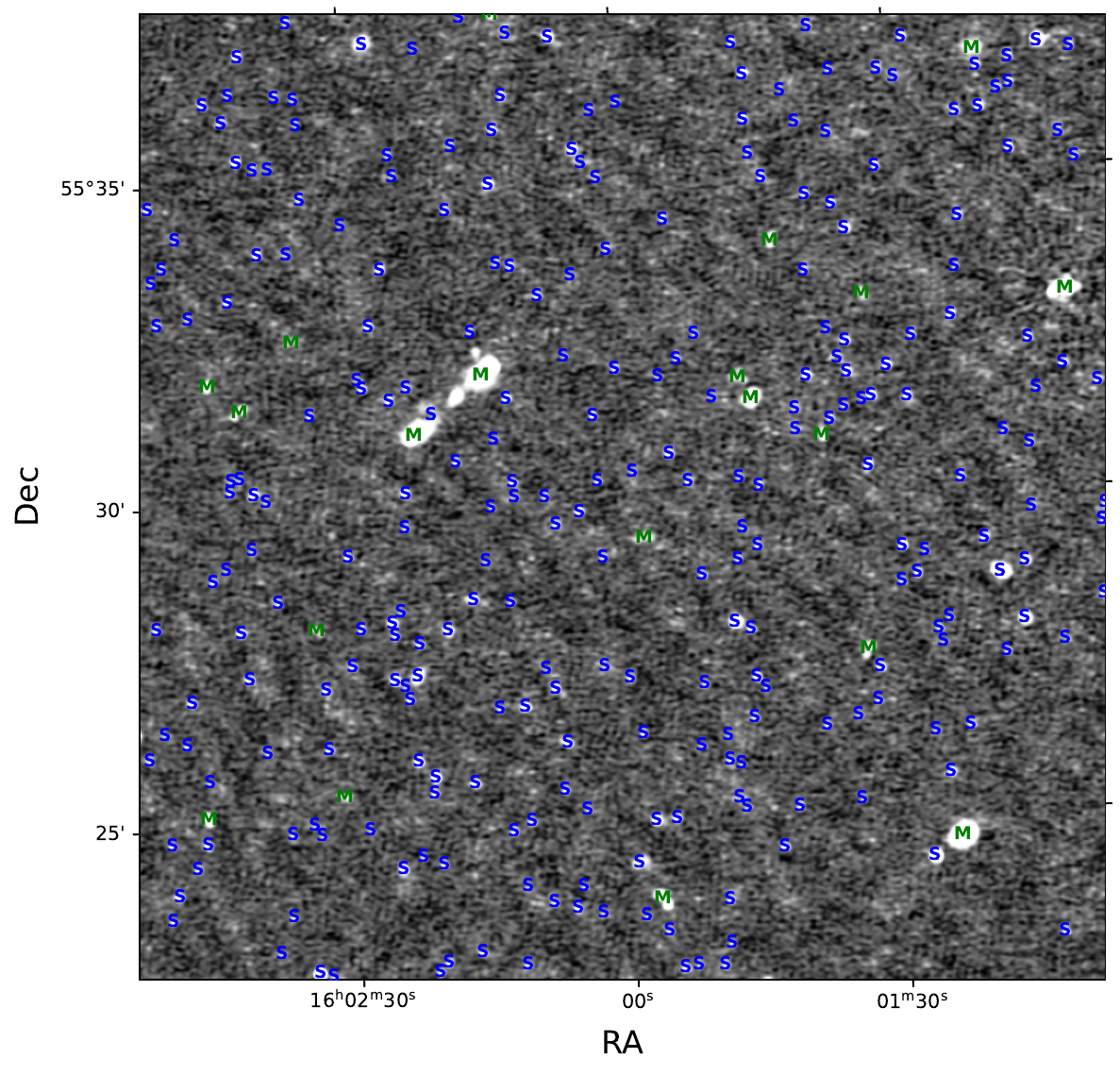} 
   \includegraphics[height=0.47\textwidth]{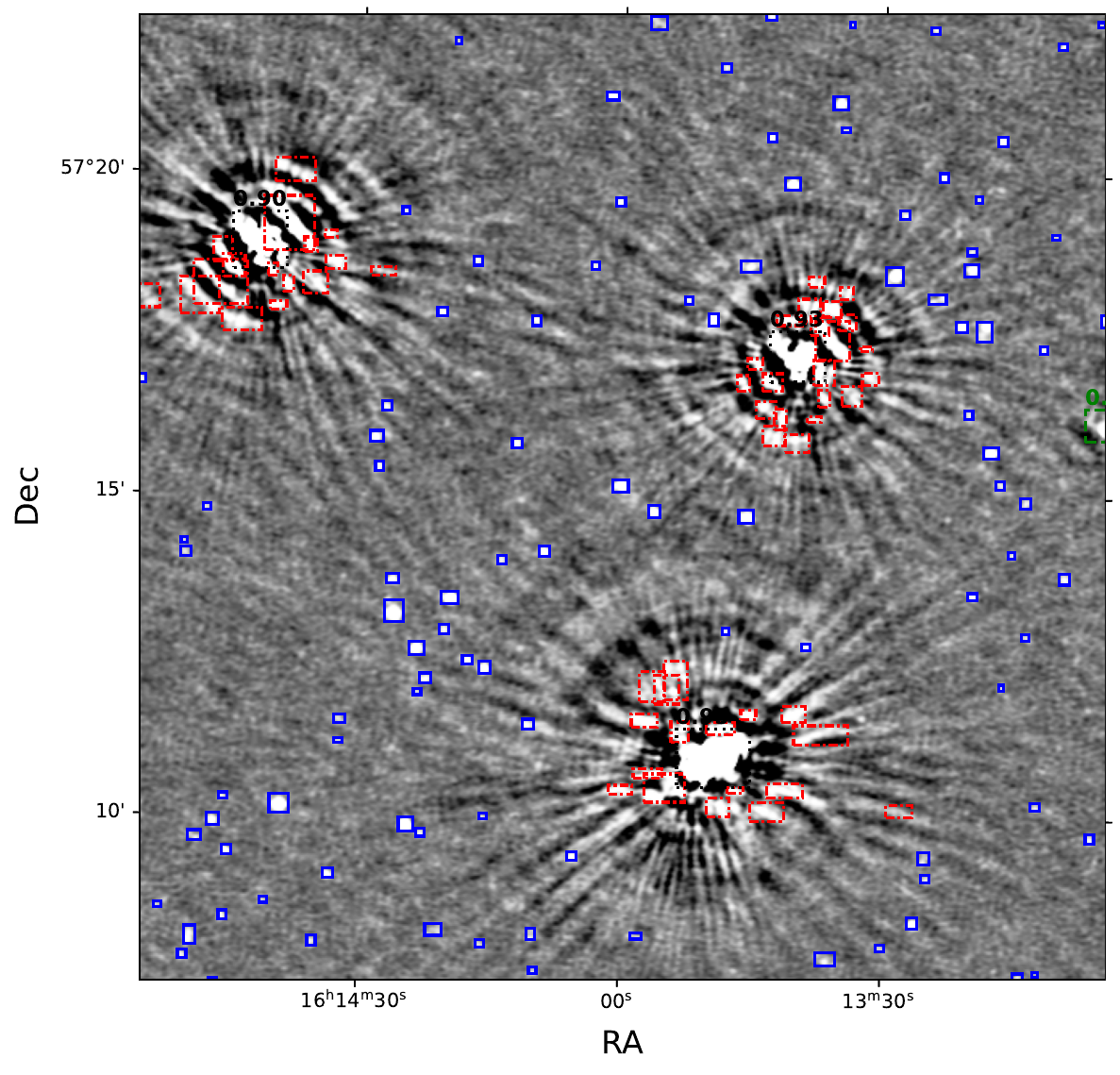} ~ ~   
   \includegraphics[height=0.47\textwidth]{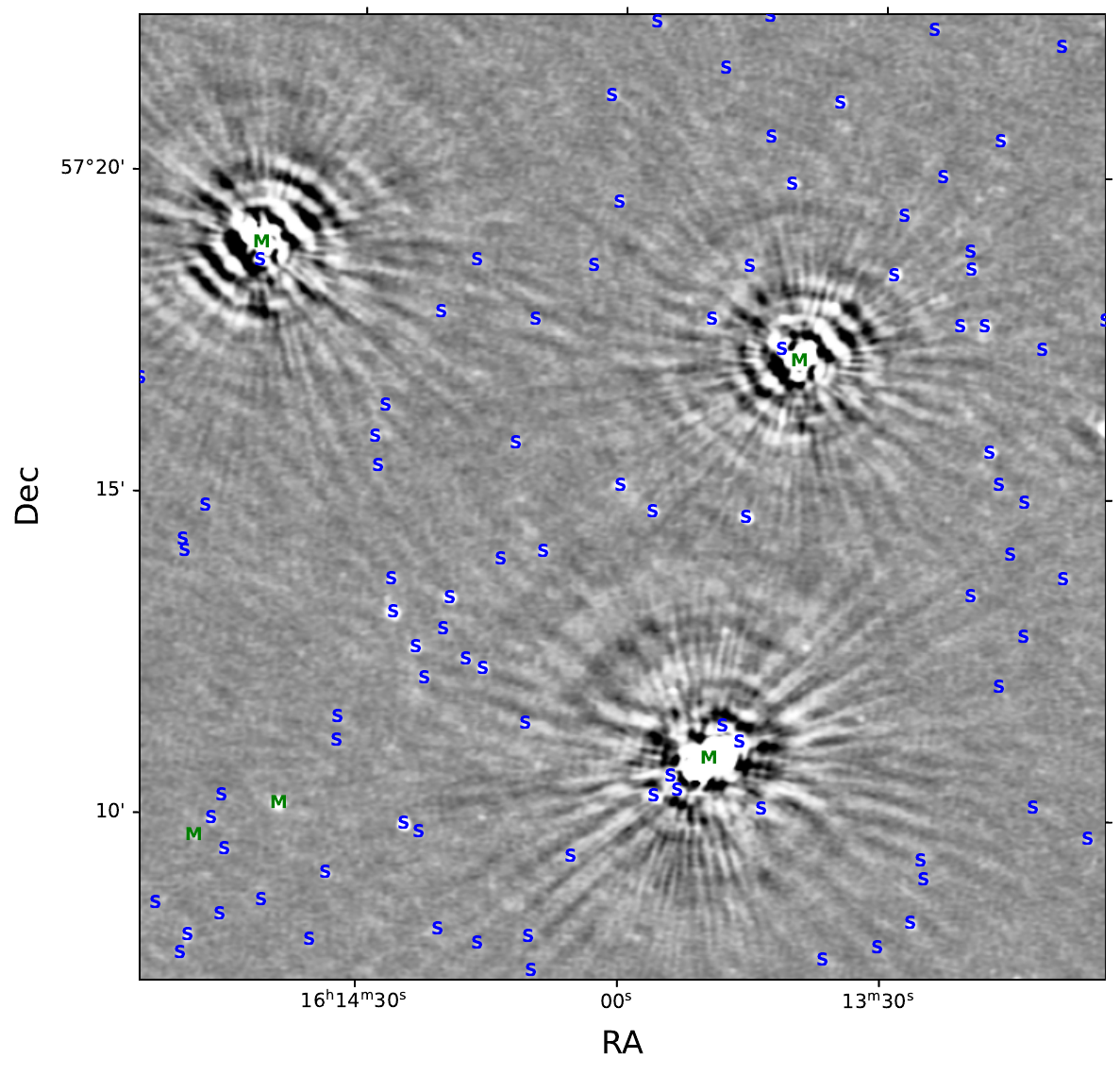} 
   \caption{Left: RF-DETR detections overlaid on the $15\arcmin\times15\arcmin$ cutouts of ELAIS-N1 mosaic. Right: PyBDSF catalogue entries on the same regions. Blue and green labels represent single Gaussian (\texttt{S}) and multiple Gaussians (\texttt{M}) sources, respectively.}
  \label{fig:rf-vs-pybdsf-example}
\end{figure*}

Figure~\ref{fig:rf-vs-pybdsf-example} presents a qualitative comparison between RF-DETR detections (left panels) and PyBDSF catalogue entries (right panels) over representative $15\arcmin\times15\arcmin$ regions of the ELAIS-N1 field. The upper panels illustrate a region containing a classical double-lobed radio galaxy embedded in a dense field of compact sources. RF-DETR assigns a single \texttt{Multi-Extended} detection whose bounding box encloses both lobes and the core, corresponding to the astrophysical interpretation of a single radio source. In contrast, the PyBDSF catalogue decomposes the same system into multiple Gaussian components, listed as separate catalogue entries. It is important to note that the PyBDSF structural flag encodes the composition of the Gaussian model (\texttt{S}: single Gaussian; \texttt{M}: multiple Gaussians; \texttt{C}: single Gaussian within an island containing other sources), rather than a source-level morphological classification \citep{Mohan2015_PyBDSF}. As a result, physically connected multi-component systems appear as multiple rows in the PyBDSF catalogue and require an additional association step before they can be treated as a single astrophysical source. By construction, RF-DETR instead provides a unified, source-level entry, which is advantageous for cross-identification and catalogue compactness.

The lower panels show a region dominated by three bright sources with pronounced sidelobes and deconvolution artefacts. RF-DETR correctly identifies the bright artefact-affected systems as \texttt{Flagged} and labels a small number of nearby responses as \texttt{Spurious}, while simultaneously preserving numerous genuine faint \texttt{Compact} detections in the surrounding area. PyBDSF relies on fitting Gaussian components to emission islands identified above user-defined thresholds and does not provide explicit labels for artefacts. While PyBDSF supports adjustable parameters such as rms estimation boxes and island thresholds that can be tuned to mitigate the inclusion of sidelobe and other instrumental artefacts, these settings do not categorically exclude artefact emission. In practice, under conditions of elevated and spatially varying noise due to bright sources, artefact-dominated islands can still be fitted with Gaussian components that enter the PyBDSF catalogue. The explicit \texttt{Flagged} and \texttt{Spurious} labels produced by RF-DETR thus enable artefact-aware catalogue filtering without image-specific parameter tuning, while retaining genuine sources in the vicinity of bright objects.

\subsection{Recovery of extended and giant radio galaxies}

To further assess the practical performance of our RF-DETR–based source finder, we cross-matched our LOFAR deep-field catalogues with the sample of extended radio galaxies (ERGs) compiled by \citet{Simonte2024}, which was constructed through careful visual inspection of the ELAIS-N1, Lockman Hole and Bo\"otes deep fields. Their study provides a stringent external benchmark, as it targets large, morphologically complex radio galaxies that are known to be challenging for automated source-finding algorithms.

Restricting the comparison to ERGs whose host positions lie within the LOFAR image footprints, we find that RF-DETR successfully recovers the vast majority of these sources across all three fields. Overall, approximately 94\% (1,512 out of 1,609) of the ERGs are detected by RF-DETR, while the recovery fraction for giant radio galaxies (GRGs; defined as systems with projected linear sizes $\mathrm{LLS} \ge 0.7\,\mathrm{Mpc}$) remains high at $\simeq$86\%. In this analysis, a source is considered recovered if the host position reported by \citet{Simonte2024} falls within at least one RF-DETR bounding box, which constitutes a conservative matching criterion, particularly for very large, lobe-dominated systems. More than 83\% of recovered ERGs and GRGs are assigned to extended morphological classes by RF-DETR, either \texttt{Single-Extended} or \texttt{Multi-Extended}. The latter class is frequently associated with classical double-lobed and multi-component systems, consistent with the physical nature of large radio galaxies. Only a small fraction of ERGs are classified as \texttt{Flagged} or \texttt{Compact}, indicating that RF-DETR does not confuse genuine extended emission with artefacts or noise features.

Beyond recovering the majority of visually identified ERGs, RF-DETR assigns a substantially larger number of sources in the LoTSS Deep Fields to extended morphological classes than those included in the \citet{Simonte2024} catalogue. While not all sources classified as \texttt{Single-Extended} or \texttt{Multi-Extended} by RF-DETR necessarily correspond to genuine ERGs as defined by projected linear size, this indicates the presence of additional extended radio galaxy candidates beyond those identified in visual catalogues. Given that compact sources constitute more than 90\% of detections in LoTSS Deep Fields, starting from the RF-DETR extended classes provides a highly efficient and physically motivated pre-selection for subsequent ERG and GRG searches. These results demonstrate that RF-DETR not only achieves high completeness on a visually curated sample of extended radio galaxies, but also produces morphologically consistent classifications in a fully automated and scalable manner, underscoring its potential for large-scale, systematic searches for extended and giant radio galaxies in forthcoming wide-area radio surveys.

\section{Conclusions} \label{sec:summary}

We have demonstrated that a transformer-based, set-prediction detector, RF-DETR, can be successfully adapted for instance-level source detection and morphology classification in deep, low-frequency radio continuum surveys. Our approach involves a balance between survey-scale source detection and capturing the complexity of radio morphologies, and adopts a conservative morphology scheme designed for automated source finding. Applied to the LoTSS Deep Fields at 150\,MHz, the model delivers unified, source-level catalogues that simultaneously localise radio emission and assign physically motivated morphological labels. 

Trained on the ELAIS-N1 Deep Field and applied without retraining to three additional fields, RF-DETR achieves robust and stable performance under realistic deep-field conditions. For the main source populations, including \texttt{Compact}, \texttt{Single-Extended}, and \texttt{Multi-Extended} systems, the model attains detection and classification performance at the $\gtrsim 90$\,per\,cent level on a held-out validation set. Importantly, physically associated multi-lobed systems are preferentially detected as single instances, reflecting the astrophysical concept of a radio source rather than a collection of individual components. 

At the catalogue level, RF-DETR recovers the vast majority of sources identified by the widely used PyBDSF source finder and maintains good overall agreement in flux measurements, while exhibiting distinct detection behaviour in specific regimes. In particular, RF-DETR provides source-level representations of classical multi-component radio galaxies as unified entries rather than fragmented Gaussian components, and returns a broader set of compact source candidates near the detection threshold. Beyond conventional morphology classes, the explicit identification of \texttt{Flagged} and \texttt{Spurious} detections enables artefact-aware filtering and facilitates reliable catalogue construction in the vicinity of bright sources.

As an external validation, we compared the RF-DETR catalogues with a visually curated sample of extended and giant radio galaxies. The detector successfully recovers the majority of these large, morphologically complex systems and assigns them predominantly to extended morphological classes. This demonstrates its effectiveness on physically associated, multi-component radio galaxies that are known to challenge traditional automated source-finding approaches.

These results demonstrate that modern transformer-based detectors can provide a practical and scalable framework for morphology-aware source finding in deep radio continuum surveys. The RF-DETR approach is well suited for source detection and population-level morphological studies, with clear applicability to forthcoming large-area surveys with facilities such as SKA-Low.

\section*{Acknowledgements}

This work is funded by the International Science Partnerships Fund (ISPF) and this grant is awarded by the STFC on behalf of UK Research and Innovation (UKRI).
GC acknowledges financial support provided by the Spanish Ministerio de Ciencia, Innovación y Universidades (MICIU) through the project PID2023-153342NB-I00 / 10.13039/501100011033. 
LKM is grateful for support from ISPF via STFC [ST/Y004159/1]. 
YW acknowledges the support of the Department of Atomic Energy, Government of India, under project no. 12-R\&D-TFR5.02-0700. 
ZC is funded by a UKRI Future Leaders Fellowship [grant MR/X005399/1; PI: Alkistis Pourtsidou].

This publication also uses data generated via the \href{https://www.zooniverse.org}{Zooniverse.org} platform, development of which is funded by generous support, including a Global Impact Award from Google, and by a grant from the Alfred P. Sloan Foundation.

\section*{Data Availability}
The LoTSS Deep Fields DR1 continuum images are publicly available through the LoTSS data releases, and ancillary imaging data (SWIRE, UKIDSS--DXS, Pan--STARRS) are distributed by their respective surveys.
The trained RF-DETR model weights and the resulting source catalogues for the LoTSS Deep Fields will be made publicly available upon publication. 
The annotation data underlying this work are not publicly released due to data-use constraints associated with the original annotation platform, but may be made available upon reasonable request to the corresponding authors.



\bibliographystyle{mnras}
\bibliography{AI4Astro} 

@ARTICLE{Chen2023,
       author = {{Chen}, Guangwen and {Bendo}, George J. and {Fuller}, Gary A. and {Henkel}, Christian and {Kong}, Xu},
        title = "{Star formation in the centre of NGC 1808 as observed by ALMA}",
      journal = {\mnras},
         year = 2023,
        month = nov,
       volume = {525},
       number = {3},
        pages = {3645-3661},
          doi = {10.1093/mnras/stad2450},
archivePrefix = {arXiv},
       eprint = {2308.08598},
 primaryClass = {astro-ph.GA},
       adsurl = {https://ui.adsabs.harvard.edu/abs/2023MNRAS.525.3645C}
}

@ARTICLE{Chen2024,
       author = {{Chen}, Guangwen and {Bendo}, George J. and {Fuller}, Gary A. and {Zhang}, Hong-Xin and {Kong}, Xu},
        title = "{Radio-to-submillimetre spectral energy distributions of NGC 1365}",
      journal = {\mnras},
         year = 2024,
        month = may,
       volume = {530},
       number = {1},
        pages = {819-835},
          doi = {10.1093/mnras/stae859},
archivePrefix = {arXiv},
       eprint = {2403.15620},
 primaryClass = {astro-ph.GA},
       adsurl = {https://ui.adsabs.harvard.edu/abs/2024MNRAS.530..819C}
}

@ARTICLE{Chen2020,
       author = {{Chen}, Guangwen and {Zhang}, Hong-Xin and {Kong}, Xu and {Lin}, Zesen and {Liang}, Zhixiong and {Chen}, Xinkai and {Chen}, Zuyi and {Song}, Zhiyuan},
        title = "{The Most Predictive Physical Properties for the Stellar Population Radial Profiles of Nearby Galaxies}",
      journal = {\apj},
         year = 2020,
        month = jun,
       volume = {895},
       number = {2},
          eid = {146},
        pages = {146},
          doi = {10.3847/1538-4357/ab8cc2},
archivePrefix = {arXiv},
       eprint = {2004.13044},
 primaryClass = {astro-ph.GA},
       adsurl = {https://ui.adsabs.harvard.edu/abs/2020ApJ...895..146C}
}

@ARTICLE{Chen2022,
       author = {{Chen}, Guangwen and {Zhang}, Hong-Xin and {Kong}, Xu and {Lin}, Zesen and {Liang}, Zhixiong and {Chen}, Zuyi and {Tang}, Yimeng and {Chen}, Xinkai},
        title = "{Discovery of a Bimodal Environmental Distribution of Compact Ellipticals in the Local Universe}",
      journal = {\apjl},
         year = 2022,
        month = aug,
       volume = {934},
       number = {2},
          eid = {L35},
        pages = {L35},
          doi = {10.3847/2041-8213/ac8354},
archivePrefix = {arXiv},
       eprint = {2207.12422},
 primaryClass = {astro-ph.GA},
       adsurl = {https://ui.adsabs.harvard.edu/abs/2022ApJ...934L..35C}
}

@ARTICLE{Bendo2015,
       author = {{Bendo}, G.~J. and {Beswick}, R.~J. and {D'Cruze}, M.~J. and {Dickinson}, C. and {Fuller}, G.~A. and {Muxlow}, T.~W.~B.},
        title = "{ALMA observations of 99 GHz free-free and H40{\ensuremath{\alpha}} line emission from star formation in the centre of NGC 253}",
      journal = {\mnras},
         year = 2015,
        month = jun,
       volume = {450},
       number = {1},
        pages = {L80-L84},
          doi = {10.1093/mnrasl/slv053},
archivePrefix = {arXiv},
       eprint = {1504.02142},
 primaryClass = {astro-ph.GA},
       adsurl = {https://ui.adsabs.harvard.edu/abs/2015MNRAS.450L..80B}
}

@ARTICLE{Bendo2016,
       author = {{Bendo}, G.~J. and {Henkel}, C. and {D'Cruze}, M.~J. and {Dickinson}, C. and {Fuller}, G.~A. and {Karim}, A.},
        title = "{Free-free and H42{\ensuremath{\alpha}} emission from the dusty starburst within NGC 4945 as observed by ALMA}",
      journal = {\mnras},
         year = 2016,
        month = nov,
       volume = {463},
       number = {1},
        pages = {252-269},
          doi = {10.1093/mnras/stw1659},
archivePrefix = {arXiv},
       eprint = {1607.02304},
 primaryClass = {astro-ph.GA},
       adsurl = {https://ui.adsabs.harvard.edu/abs/2016MNRAS.463..252B}
}

@inproceedings{
    Robinson_rf-detr_ICLR,
    title={{RF}-{DETR}: Neural Architecture Search for Real-Time Detection Transformers},
    author={Isaac Robinson and Peter Robicheaux and Matvei Popov and Deva Ramanan and Neehar Peri},
    booktitle={International Conference on Learning Representations (ICLR)},
    year={2026},
}

@INPROCEEDINGS{He2017_MaskRCNN,
  author={He, Kaiming and Gkioxari, Georgia and Dollár, Piotr and Girshick, Ross},
  booktitle={2017 IEEE International Conference on Computer Vision (ICCV)}, 
  title={Mask R-CNN}, 
  year={2017},
  volume={},
  number={},
  pages={2980-2988},
  doi={10.1109/ICCV.2017.322}}

@ARTICLE{Riggi2021,
       author = {{Riggi}, S. and {Umana}, G. and {Trigilio}, C. and {Cavallaro}, F. and {Ingallinera}, A. and {Leto}, P. and {Bufano}, F. and {Norris}, R.~P. and {Hopkins}, A.~M. and {Filipovi{\'c}}, M.~D. and {Andernach}, H. and {van Loon}, J. Th and {Micha{\l}owski}, M.~J. and {Bordiu}, C. and {An}, T. and {Buemi}, C. and {Carretti}, E. and {Collier}, J.~D. and {Joseph}, T. and {Koribalski}, B.~S. and {Kothes}, R. and {Loru}, S. and {McConnell}, D. and {Pommier}, M. and {Sciacca}, E. and {Schillir{\`o}}, F. and {Vitello}, F. and {Warhurst}, K. and {Whiting}, M.},
        title = "{Evolutionary map of the Universe (EMU): Compact radio sources in the SCORPIO field towards the galactic plane}",
      journal = {\mnras},
         year = 2021,
        month = mar,
       volume = {502},
       number = {1},
        pages = {60-79},
          doi = {10.1093/mnras/stab028},
archivePrefix = {arXiv},
       eprint = {2101.03843},
 primaryClass = {astro-ph.GA},
       adsurl = {https://ui.adsabs.harvard.edu/abs/2021MNRAS.502...60R}
}

@ARTICLE{Riggi2023,
       author = {{Riggi}, S. and {Magro}, D. and {Sortino}, R. and {De Marco}, A. and {Bordiu}, C. and {Cecconello}, T. and {Hopkins}, A.~M. and {Marvil}, J. and {Umana}, G. and {Sciacca}, E. and {Vitello}, F. and {Bufano}, F. and {Ingallinera}, A. and {Fiameni}, G. and {Spampinato}, C. and {Zarb Adami}, K.},
        title = "{Astronomical source detection in radio continuum maps with deep neural networks}",
      journal = {Astronomy and Computing},
         year = 2023,
        month = jan,
       volume = {42},
          eid = {100682},
        pages = {100682},
          doi = {10.1016/j.ascom.2022.100682},
archivePrefix = {arXiv},
       eprint = {2212.02538},
 primaryClass = {astro-ph.IM},
       adsurl = {https://ui.adsabs.harvard.edu/abs/2023A&C....4200682R}
}

@ARTICLE{Riggi2024,
       author = {{Riggi}, S. and {Cecconello}, T. and {Palazzo}, S. and {Hopkins}, A.~M. and {Gupta}, N. and {Bordiu}, C. and {Ingallinera}, A. and {Buemi}, C. and {Bufano}, F. and {Cavallaro}, F. and {Filipovi{\'c}}, M.~D. and {Leto}, P. and {Loru}, S. and {Ruggeri}, A.~C. and {Trigilio}, C. and {Umana}, G. and {Vitello}, F.},
        title = "{Self-supervised contrastive learning of radio data for source detection, classification and peculiar object discovery}",
      journal = {\pasa},
         year = 2024,
        month = nov,
       volume = {41},
          eid = {e085},
        pages = {e085},
          doi = {10.1017/pasa.2024.84},
archivePrefix = {arXiv},
       eprint = {2404.18462},
 primaryClass = {astro-ph.IM},
       adsurl = {https://ui.adsabs.harvard.edu/abs/2024PASA...41...85R}
}

@ARTICLE{Rudnick2021,
       author = {{Rudnick}, Lawrence},
        title = "{Radio Galaxy Classification: \#Tags, Not Boxes}",
      journal = {Galaxies},
         year = 2021,
        month = oct,
       volume = {9},
       number = {4},
          eid = {85},
        pages = {85},
          doi = {10.3390/galaxies9040085},
archivePrefix = {arXiv},
       eprint = {2110.13733},
 primaryClass = {astro-ph.GA},
       adsurl = {https://ui.adsabs.harvard.edu/abs/2021Galax...9...85R}
}

@ARTICLE{Vafaei2019_DeepSource,
       author = {{Vafaei Sadr}, A. and {Vos}, Etienne E. and {Bassett}, Bruce A. and {Hosenie}, Zafiirah and {Oozeer}, N. and {Lochner}, Michelle},
        title = "{DEEPSOURCE: point source detection using deep learning}",
      journal = {\mnras},
         year = 2019,
        month = apr,
       volume = {484},
       number = {2},
        pages = {2793-2806},
          doi = {10.1093/mnras/stz131},
archivePrefix = {arXiv},
       eprint = {1807.02701},
 primaryClass = {astro-ph.IM},
       adsurl = {https://ui.adsabs.harvard.edu/abs/2019MNRAS.484.2793V}
}

@ARTICLE{Wu2019_CLARAN,
       author = {{Wu}, Chen and {Wong}, Oiwei Ivy and {Rudnick}, Lawrence and {Shabala}, Stanislav S. and {Alger}, Matthew J. and {Banfield}, Julie K. and {Ong}, Cheng Soon and {White}, Sarah V. and {Garon}, Avery F. and {Norris}, Ray P. and {Andernach}, Heinz and {Tate}, Jean and {Lukic}, Vesna and {Tang}, Hongming and {Schawinski}, Kevin and {Diakogiannis}, Foivos I.},
        title = "{Radio Galaxy Zoo: CLARAN - a deep learning classifier for radio morphologies}",
      journal = {\mnras},
         year = 2019,
        month = jan,
       volume = {482},
       number = {1},
        pages = {1211-1230},
          doi = {10.1093/mnras/sty2646},
archivePrefix = {arXiv},
       eprint = {1805.12008},
 primaryClass = {astro-ph.IM},
       adsurl = {https://ui.adsabs.harvard.edu/abs/2019MNRAS.482.1211W}
}

@ARTICLE{Ren2017_FastRCNN,
  author={Ren, Shaoqing and He, Kaiming and Girshick, Ross and Sun, Jian},
  journal={IEEE Transactions on Pattern Analysis and Machine Intelligence}, 
  title={Faster R-CNN: Towards Real-Time Object Detection with Region Proposal Networks}, 
  year={2017},
  volume={39},
  number={6},
  pages={1137-1149},
  doi={10.1109/TPAMI.2016.2577031}
}

@ARTICLE{Lukic2019_ConvoSource,
       author = {{Lukic}, Vesna and {de Gasperin}, Francesco and {Br{\"u}ggen}, Marcus},
        title = "{ConvoSource: Radio-Astronomical Source-Finding with Convolutional Neural Networks}",
      journal = {Galaxies},
         year = 2019,
        month = dec,
       volume = {8},
       number = {1},
          eid = {3},
        pages = {3},
          doi = {10.3390/galaxies8010003},
archivePrefix = {arXiv},
       eprint = {1910.03631},
 primaryClass = {astro-ph.IM},
       adsurl = {https://ui.adsabs.harvard.edu/abs/2019Galax...8....3L}
}

@ARTICLE{Lukic2019,
       author = {{Lukic}, V. and {Br{\"u}ggen}, M. and {Mingo}, B. and {Croston}, J.~H. and {Kasieczka}, G. and {Best}, P.~N.},
        title = "{Morphological classification of radio galaxies: capsule networks versus convolutional neural networks}",
      journal = {\mnras},
         year = 2019,
        month = aug,
       volume = {487},
       number = {2},
        pages = {1729-1744},
          doi = {10.1093/mnras/stz1289},
archivePrefix = {arXiv},
       eprint = {1905.03274},
 primaryClass = {astro-ph.IM},
       adsurl = {https://ui.adsabs.harvard.edu/abs/2019MNRAS.487.1729L}
}

@ARTICLE{Brand2023,
       author = {{Brand}, Kevin and {Grobler}, Trienko L. and {Kleynhans}, Waldo and {Vaccari}, Mattia and {Prescott}, Matthew and {Becker}, Burger},
        title = "{Feature guided training and rotational standardization for the morphological classification of radio galaxies}",
      journal = {\mnras},
         year = 2023,
        month = jun,
       volume = {522},
       number = {1},
        pages = {292-311},
          doi = {10.1093/mnras/stad989},
archivePrefix = {arXiv},
       eprint = {2304.05095},
 primaryClass = {astro-ph.IM},
       adsurl = {https://ui.adsabs.harvard.edu/abs/2023MNRAS.522..292B}
}

@ARTICLE{Ndung2023_Review,
       author = {{Ndung'u}, Steven and {Grobler}, Trienko and {Wijnholds}, Stefan J. and {Karastoyanova}, Dimka and {Azzopardi}, George},
        title = "{Advances on the morphological classification of radio galaxies: A review}",
      journal = {\nar},
         year = 2023,
        month = dec,
       volume = {97},
          eid = {101685},
        pages = {101685},
          doi = {10.1016/j.newar.2023.101685},
       adsurl = {https://ui.adsabs.harvard.edu/abs/2023NewAR..9701685N}
}

@ARTICLE{Ndung2024,
       author = {{Ndung'u}, Steven and {Grobler}, Trienko and {Wijnholds}, Stefan J. and {Karastoyanova}, Dimka and {Azzopardi}, George},
        title = "{Classification of radio galaxies with trainable COSFIRE filters}",
      journal = {\mnras},
         year = 2024,
        month = may,
       volume = {530},
       number = {1},
        pages = {783-794},
          doi = {10.1093/mnras/stae821},
archivePrefix = {arXiv},
       eprint = {2311.11286},
 primaryClass = {astro-ph.IM},
       adsurl = {https://ui.adsabs.harvard.edu/abs/2024MNRAS.530..783N}
}

@ARTICLE{Perez2025,
       author = {{Baron Perez}, Nicolas and {Br{\"u}ggen}, Marcus and {Kasieczka}, Gregor and {Lucie-Smith}, Luisa},
        title = "{Classification of radio sources through self-supervised learning}",
      journal = {\aap},
         year = 2025,
        month = jul,
       volume = {699},
          eid = {A302},
        pages = {A302},
          doi = {10.1051/0004-6361/202554735},
archivePrefix = {arXiv},
       eprint = {2503.19111},
 primaryClass = {astro-ph.IM},
       adsurl = {https://ui.adsabs.harvard.edu/abs/2025A&A...699A.302B}
}

@ARTICLE{Sortino2023,
       author = {{Sortino}, Renato and {Magro}, Daniel and {Fiameni}, Giuseppe and {Sciacca}, Eva and {Riggi}, Simone and {DeMarco}, Andrea and {Spampinato}, Concetto and {Hopkins}, Andrew M. and {Bufano}, Filomena and {Schillir{\`o}}, Francesco and {Bordiu}, Cristobal and {Pino}, Carmelo},
        title = "{Radio astronomical images object detection and segmentation: a benchmark on deep learning methods}",
      journal = {Experimental Astronomy},
         year = 2023,
        month = aug,
       volume = {56},
       number = {1},
        pages = {293-331},
          doi = {10.1007/s10686-023-09893-w},
archivePrefix = {arXiv},
       eprint = {2303.04506},
 primaryClass = {cs.CV},
       adsurl = {https://ui.adsabs.harvard.edu/abs/2023ExA....56..293S}
}

@ARTICLE{Cornu2024,
       author = {{Cornu}, D. and {Salom{\'e}}, P. and {Semelin}, B. and {Marchal}, A. and {Freundlich}, J. and {Aicardi}, S. and {Lu}, X. and {Sainton}, G. and {Mertens}, F. and {Combes}, F. and {Tasse}, C.},
        title = "{YOLO-CIANNA: Galaxy detection with deep learning in radio data: I. A new YOLO-inspired source detection method applied to the SKAO SDC1}",
      journal = {\aap},
         year = 2024,
        month = oct,
       volume = {690},
          eid = {A211},
        pages = {A211},
          doi = {10.1051/0004-6361/202449548},
archivePrefix = {arXiv},
       eprint = {2402.05925},
 primaryClass = {astro-ph.IM},
       adsurl = {https://ui.adsabs.harvard.edu/abs/2024A&A...690A.211C}
}

@ARTICLE{Alhassan2018,
       author = {{Alhassan}, Wathela and {Taylor}, A.~R. and {Vaccari}, Mattia},
        title = "{The FIRST Classifier: compact and extended radio galaxy classification using deep Convolutional Neural Networks}",
      journal = {\mnras},
         year = 2018,
        month = oct,
       volume = {480},
       number = {2},
        pages = {2085-2093},
          doi = {10.1093/mnras/sty2038},
archivePrefix = {arXiv},
       eprint = {1807.10380},
 primaryClass = {astro-ph.GA},
       adsurl = {https://ui.adsabs.harvard.edu/abs/2018MNRAS.480.2085A}
}

@ARTICLE{Samudre2022,
       author = {{Samudre}, Ashwin and {George}, Lijo T. and {Bansal}, Mahak and {Wadadekar}, Yogesh},
        title = "{Data-efficient classification of radio galaxies}",
      journal = {\mnras},
         year = 2022,
        month = jan,
       volume = {509},
       number = {2},
        pages = {2269-2280},
          doi = {10.1093/mnras/stab3144},
archivePrefix = {arXiv},
       eprint = {2011.13311},
 primaryClass = {astro-ph.IM},
       adsurl = {https://ui.adsabs.harvard.edu/abs/2022MNRAS.509.2269S}
}

@ARTICLE{Lastufka2025,
       author = {{Lastufka}, E. and {Bait}, O. and {Drozdova}, M. and {Kinakh}, V. and {Piras}, D. and {Audard}, M. and {Dessauges-Zavadsky}, M. and {Holotyak}, T. and {Schaerer}, D. and {Voloshynovskiy}, S.},
        title = "{Examining vision foundation models for classification and detection in optical and radio astronomy}",
      journal = {\aap},
         year = 2025,
        month = nov,
       volume = {703},
          eid = {A217},
        pages = {A217},
          doi = {10.1051/0004-6361/202553691},
archivePrefix = {arXiv},
       eprint = {2409.11175},
 primaryClass = {astro-ph.IM},
       adsurl = {https://ui.adsabs.harvard.edu/abs/2025A&A...703A.217L}
}

@ARTICLE{vanHaarlem2013_LOFAR,
       author = {{van Haarlem}, M.~P. and {Wise}, M.~W. and {Gunst}, A.~W. and {Heald}, G. and {McKean}, J.~P. and {Hessels}, J.~W.~T. and {de Bruyn}, A.~G. and {Nijboer}, R. and {Swinbank}, J. and {Fallows}, R. and {Brentjens}, M. and {Nelles}, A. and {Beck}, R. and {Falcke}, H. and {Fender}, R. and {H{\"o}randel}, J. and {Koopmans}, L.~V.~E. and {Mann}, G. and {Miley}, G. and {R{\"o}ttgering}, H. and {Stappers}, B.~W. and {Wijers}, R.~A.~M.~J. and {Zaroubi}, S. and {van den Akker}, M. and {Alexov}, A. and {Anderson}, J. and {Anderson}, K. and {van Ardenne}, A. and {Arts}, M. and {Asgekar}, A. and {Avruch}, I.~M. and {Batejat}, F. and {B{\"a}hren}, L. and {Bell}, M.~E. and {Bell}, M.~R. and {van Bemmel}, I. and {Bennema}, P. and {Bentum}, M.~J. and {Bernardi}, G. and {Best}, P. and {B{\^\i}rzan}, L. and {Bonafede}, A. and {Boonstra}, A.-J. and {Braun}, R. and {Bregman}, J. and {Breitling}, F. and {van de Brink}, R.~H. and {Broderick}, J. and {Broekema}, P.~C. and {Brouw}, W.~N. and {Br{\"u}ggen}, M. and {Butcher}, H.~R. and {van Cappellen}, W. and {Ciardi}, B. and {Coenen}, T. and {Conway}, J. and {Coolen}, A. and {Corstanje}, A. and {Damstra}, S. and {Davies}, O. and {Deller}, A.~T. and {Dettmar}, R.-J. and {van Diepen}, G. and {Dijkstra}, K. and {Donker}, P. and {Doorduin}, A. and {Dromer}, J. and {Drost}, M. and {van Duin}, A. and {Eisl{\"o}ffel}, J. and {van Enst}, J. and {Ferrari}, C. and {Frieswijk}, W. and {Gankema}, H. and {Garrett}, M.~A. and {de Gasperin}, F. and {Gerbers}, M. and {de Geus}, E. and {Grie{\ss}meier}, J.-M. and {Grit}, T. and {Gruppen}, P. and {Hamaker}, J.~P. and {Hassall}, T. and {Hoeft}, M. and {Holties}, H.~A. and {Horneffer}, A. and {van der Horst}, A. and {van Houwelingen}, A. and {Huijgen}, A. and {Iacobelli}, M. and {Intema}, H. and {Jackson}, N. and {Jelic}, V. and {de Jong}, A. and {Juette}, E. and {Kant}, D. and {Karastergiou}, A. and {Koers}, A. and {Kollen}, H. and {Kondratiev}, V.~I. and {Kooistra}, E. and {Koopman}, Y. and {Koster}, A. and {Kuniyoshi}, M. and {Kramer}, M. and {Kuper}, G. and {Lambropoulos}, P. and {Law}, C. and {van Leeuwen}, J. and {Lemaitre}, J. and {Loose}, M. and {Maat}, P. and {Macario}, G. and {Markoff}, S. and {Masters}, J. and {McFadden}, R.~A. and {McKay-Bukowski}, D. and {Meijering}, H. and {Meulman}, H. and {Mevius}, M. and {Middelberg}, E. and {Millenaar}, R. and {Miller-Jones}, J.~C.~A. and {Mohan}, R.~N. and {Mol}, J.~D. and {Morawietz}, J. and {Morganti}, R. and {Mulcahy}, D.~D. and {Mulder}, E. and {Munk}, H. and {Nieuwenhuis}, L. and {van Nieuwpoort}, R. and {Noordam}, J.~E. and {Norden}, M. and {Noutsos}, A. and {Offringa}, A.~R. and {Olofsson}, H. and {Omar}, A. and {Orr{\'u}}, E. and {Overeem}, R. and {Paas}, H. and {Pandey-Pommier}, M. and {Pandey}, V.~N. and {Pizzo}, R. and {Polatidis}, A. and {Rafferty}, D. and {Rawlings}, S. and {Reich}, W. and {de Reijer}, J.-P. and {Reitsma}, J. and {Renting}, G.~A. and {Riemers}, P. and {Rol}, E. and {Romein}, J.~W. and {Roosjen}, J. and {Ruiter}, M. and {Scaife}, A. and {van der Schaaf}, K. and {Scheers}, B. and {Schellart}, P. and {Schoenmakers}, A. and {Schoonderbeek}, G. and {Serylak}, M. and {Shulevski}, A. and {Sluman}, J. and {Smirnov}, O. and {Sobey}, C. and {Spreeuw}, H. and {Steinmetz}, M. and {Sterks}, C.~G.~M. and {Stiepel}, H.-J. and {Stuurwold}, K. and {Tagger}, M. and {Tang}, Y. and {Tasse}, C. and {Thomas}, I. and {Thoudam}, S. and {Toribio}, M.~C. and {van der Tol}, B. and {Usov}, O. and {van Veelen}, M. and {van der Veen}, A.-J. and {ter Veen}, S. and {Verbiest}, J.~P.~W. and {Vermeulen}, R. and {Vermaas}, N. and {Vocks}, C. and {Vogt}, C. and {de Vos}, M. and {van der Wal}, E. and {van Weeren}, R. and {Weggemans}, H. and {Weltevrede}, P. and {White}, S. and {Wijnholds}, S.~J. and {Wilhelmsson}, T. and {Wucknitz}, O. and {Yatawatta}, S. and {Zarka}, P. and {Zensus}, A.},
        title = "{LOFAR: The LOw-Frequency ARray}",
      journal = {\aap},
         year = 2013,
        month = aug,
       volume = {556},
          eid = {A2},
        pages = {A2},
          doi = {10.1051/0004-6361/201220873},
archivePrefix = {arXiv},
       eprint = {1305.3550},
 primaryClass = {astro-ph.IM},
       adsurl = {https://ui.adsabs.harvard.edu/abs/2013A&A...556A...2V}
}

@ARTICLE{Sabater2021_ELAISN1,
       author = {{Sabater}, J. and {Best}, P.~N. and {Tasse}, C. and {Hardcastle}, M.~J. and {Shimwell}, T.~W. and {Nisbet}, D. and {Jelic}, V. and {Callingham}, J.~R. and {R{\"o}ttgering}, H.~J.~A. and {Bonato}, M. and {Bondi}, M. and {Ciardi}, B. and {Cochrane}, R.~K. and {Jarvis}, M.~J. and {Kondapally}, R. and {Koopmans}, L.~V.~E. and {O'Sullivan}, S.~P. and {Prandoni}, I. and {Schwarz}, D.~J. and {Smith}, D.~J.~B. and {Wang}, L. and {Williams}, W.~L. and {Zaroubi}, S.},
        title = "{The LOFAR Two-meter Sky Survey: Deep Fields Data Release 1. II. The ELAIS-N1 LOFAR deep field}",
      journal = {\aap},
         year = 2021,
        month = apr,
       volume = {648},
          eid = {A2},
        pages = {A2},
          doi = {10.1051/0004-6361/202038828},
archivePrefix = {arXiv},
       eprint = {2011.08211},
 primaryClass = {astro-ph.GA},
       adsurl = {https://ui.adsabs.harvard.edu/abs/2021A&A...648A...2S}
}

@ARTICLE{Tasse2021_LOTSS_DeepField,
       author = {{Tasse}, C. and {Shimwell}, T. and {Hardcastle}, M.~J. and {O'Sullivan}, S.~P. and {van Weeren}, R. and {Best}, P.~N. and {Bester}, L. and {Hugo}, B. and {Smirnov}, O. and {Sabater}, J. and {Calistro-Rivera}, G. and {de Gasperin}, F. and {Morabito}, L.~K. and {R{\"o}ttgering}, H. and {Williams}, W.~L. and {Bonato}, M. and {Bondi}, M. and {Botteon}, A. and {Br{\"u}ggen}, M. and {Brunetti}, G. and {Chy{\.z}y}, K.~T. and {Garrett}, M.~A. and {G{\"u}rkan}, G. and {Jarvis}, M.~J. and {Kondapally}, R. and {Mandal}, S. and {Prandoni}, I. and {Repetti}, A. and {Retana-Montenegro}, E. and {Schwarz}, D.~J. and {Shulevski}, A. and {Wiaux}, Y.},
        title = "{The LOFAR Two-meter Sky Survey: Deep Fields Data Release 1. I. Direction-dependent calibration and imaging}",
      journal = {\aap},
         year = 2021,
        month = apr,
       volume = {648},
          eid = {A1},
        pages = {A1},
          doi = {10.1051/0004-6361/202038804},
archivePrefix = {arXiv},
       eprint = {2011.08328},
 primaryClass = {astro-ph.IM},
       adsurl = {https://ui.adsabs.harvard.edu/abs/2021A&A...648A...1T}
}

@ARTICLE{Kondapally2021_LOTSS_VAC,
       author = {{Kondapally}, R. and {Best}, P.~N. and {Hardcastle}, M.~J. and {Nisbet}, D. and {Bonato}, M. and {Sabater}, J. and {Duncan}, K.~J. and {McCheyne}, I. and {Cochrane}, R.~K. and {Bowler}, R.~A.~A. and {Williams}, W.~L. and {Shimwell}, T.~W. and {Tasse}, C. and {Croston}, J.~H. and {Goyal}, A. and {Jamrozy}, M. and {Jarvis}, M.~J. and {Mahatma}, V.~H. and {R{\"o}ttgering}, H.~J.~A. and {Smith}, D.~J.~B. and {Wo{\l}owska}, A. and {Bondi}, M. and {Brienza}, M. and {Brown}, M.~J.~I. and {Br{\"u}ggen}, M. and {Chambers}, K. and {Garrett}, M.~A. and {G{\"u}rkan}, G. and {Huber}, M. and {Kunert-Bajraszewska}, M. and {Magnier}, E. and {Mingo}, B. and {Mostert}, R. and {Nikiel-Wroczy{\'n}ski}, B. and {O'Sullivan}, S.~P. and {Paladino}, R. and {Ploeckinger}, T. and {Prandoni}, I. and {Rosenthal}, M.~J. and {Schwarz}, D.~J. and {Shulevski}, A. and {Wagenveld}, J.~D. and {Wang}, L.},
        title = "{The LOFAR Two-meter Sky Survey: Deep Fields Data Release 1. III. Host-galaxy identifications and value added catalogues}",
      journal = {\aap},
         year = 2021,
        month = apr,
       volume = {648},
          eid = {A3},
        pages = {A3},
          doi = {10.1051/0004-6361/202038813},
archivePrefix = {arXiv},
       eprint = {2011.08201},
 primaryClass = {astro-ph.GA},
       adsurl = {https://ui.adsabs.harvard.edu/abs/2021A&A...648A...3K}
}

@ARTICLE{Bondi2024_LOFAR_EDFN,
       author = {{Bondi}, M. and {Scaramella}, R. and {Zamorani}, G. and {Ciliegi}, P. and {Vitello}, F. and {Arias}, M. and {Best}, P.~N. and {Bonato}, M. and {Botteon}, A. and {Brienza}, M. and {Brunetti}, G. and {Hardcastle}, M.~J. and {Magliocchetti}, M. and {Massaro}, F. and {Morabito}, L.~K. and {Pentericci}, L. and {Prandoni}, I. and {R{\"o}ttgering}, H.~J.~A. and {Shimwell}, T.~W. and {Tasse}, C. and {van Weeren}, R.~J. and {White}, G.~J.},
        title = "{LOFAR HBA observations of the Euclid Deep Field North (EDFN)}",
      journal = {\aap},
         year = 2024,
        month = mar,
       volume = {683},
          eid = {A179},
        pages = {A179},
          doi = {10.1051/0004-6361/202348333},
archivePrefix = {arXiv},
       eprint = {2312.06247},
 primaryClass = {astro-ph.GA},
       adsurl = {https://ui.adsabs.harvard.edu/abs/2024A&A...683A.179B}
}

@ARTICLE{Shimwell2025,
       author = {{Shimwell}, T.~W. and {Hale}, C.~L. and {Best}, P.~N. and {Botteon}, A. and {Drabent}, A. and {Hardcastle}, M.~J. and {Jeli{\'c}}, V. and {de Jong}, J.~M.~G.~H.~J. and {Kondapally}, R. and {R{\"o}ttgering}, H.~J.~A. and {Tasse}, C. and {van Weeren}, R.~J. and {Williams}, W.~L. and {Bonafede}, A. and {Bondi}, M. and {Br{\"u}ggen}, M. and {Brunetti}, G. and {Callingham}, J.~R. and {De Gasperin}, F. and {Duncan}, K.~J. and {Horellou}, C. and {Iyer}, S. and {de Ruiter}, I. and {Ma{\l}ek}, K. and {Nair}, D.~G. and {Morabito}, L.~K. and {Prandoni}, I. and {Rowlinson}, A. and {Sabater}, J. and {Shulevski}, A. and {Smith}, D.~J.~B. and {Sweijen}, F.},
        title = "{The LOFAR Two-metre Sky Survey: Deep Fields Data Release 2: I. The ELAIS-N1 field}",
      journal = {\aap},
         year = 2025,
        month = mar,
       volume = {695},
          eid = {A80},
        pages = {A80},
          doi = {10.1051/0004-6361/202452930},
archivePrefix = {arXiv},
       eprint = {2501.04093},
 primaryClass = {astro-ph.CO},
       adsurl = {https://ui.adsabs.harvard.edu/abs/2025A&A...695A..80S}
}

@ARTICLE{Williams2019,
       author = {{Williams}, W.~L. and {Hardcastle}, M.~J. and {Best}, P.~N. and {Sabater}, J. and {Croston}, J.~H. and {Duncan}, K.~J. and {Shimwell}, T.~W. and {R{\"o}ttgering}, H.~J.~A. and {Nisbet}, D. and {G{\"u}rkan}, G. and {Alegre}, L. and {Cochrane}, R.~K. and {Goyal}, A. and {Hale}, C.~L. and {Jackson}, N. and {Jamrozy}, M. and {Kondapally}, R. and {Kunert-Bajraszewska}, M. and {Mahatma}, V.~H. and {Mingo}, B. and {Morabito}, L.~K. and {Prandoni}, I. and {Roskowinski}, C. and {Shulevski}, A. and {Smith}, D.~J.~B. and {Tasse}, C. and {Urquhart}, S. and {Webster}, B. and {White}, G.~J. and {Beswick}, R.~J. and {Callingham}, J.~R. and {Chy{\.z}y}, K.~T. and {de Gasperin}, F. and {Harwood}, J.~J. and {Hoeft}, M. and {Iacobelli}, M. and {McKean},R J.~P. and {Mechev}, A.~P. and {Miley}, G.~K. and {Schwarz}, D.~J. and {van Weeren}, R.~J.},
        title = "{The LOFAR Two-metre Sky Survey. III. First data release: Optical/infrared identifications and value-added catalogue}",
      journal = {\aap},
         year = 2019,
        month = feb,
       volume = {622},
          eid = {A2},
        pages = {A2},
          doi = {10.1051/0004-6361/201833564},
archivePrefix = {arXiv},
       eprint = {1811.07927},
 primaryClass = {astro-ph.GA},
       adsurl = {https://ui.adsabs.harvard.edu/abs/2019A&A...622A...2W}
}

@ARTICLE{Lonsdale2003,
       author = {{Lonsdale}, Carol J. and {Smith}, Harding E. and {Rowan-Robinson}, Michael and {Surace}, Jason and {Shupe}, David and {Xu}, Cong and {Oliver}, Sebastian and {Padgett}, Deborah and {Fang}, Fan and {Conrow}, Tim and {Franceschini}, Alberto and {Gautier}, Nick and {Griffin}, Matt and {Hacking}, Perry and {Masci}, Frank and {Morrison}, Glenn and {O'Linger}, Joanne and {Owen}, Frazer and {P{\'e}rez-Fournon}, Ismael and {Pierre}, Marguerite and {Puetter}, Rick and {Stacey}, Gordon and {Castro}, Sandra and {Polletta}, Maria del Carmen and {Farrah}, Duncan and {Jarrett}, Tom and {Frayer}, Dave and {Siana}, Brian and {Babbedge}, Tom and {Dye}, Simon and {Fox}, Matt and {Gonzalez-Solares}, Eduardo and {Salaman}, Malcolm and {Berta}, Stefano and {Condon}, Jim J. and {Dole}, Herv{\'e} and {Serjeant}, Steve},
        title = "{SWIRE: The SIRTF Wide-Area Infrared Extragalactic Survey}",
      journal = {\pasp},
         year = 2003,
        month = aug,
       volume = {115},
       number = {810},
        pages = {897-927},
          doi = {10.1086/376850},
archivePrefix = {arXiv},
       eprint = {astro-ph/0305375},
 primaryClass = {astro-ph},
       adsurl = {https://ui.adsabs.harvard.edu/abs/2003PASP..115..897L}
}

@ARTICLE{Lawrence2007,
       author = {{Lawrence}, A. and {Warren}, S.~J. and {Almaini}, O. and {Edge}, A.~C. and {Hambly}, N.~C. and {Jameson}, R.~F. and {Lucas}, P. and {Casali}, M. and {Adamson}, A. and {Dye}, S. and {Emerson}, J.~P. and {Foucaud}, S. and {Hewett}, P. and {Hirst}, P. and {Hodgkin}, S.~T. and {Irwin}, M.~J. and {Lodieu}, N. and {McMahon}, R.~G. and {Simpson}, C. and {Smail}, I. and {Mortlock}, D. and {Folger}, M.},
        title = "{The UKIRT Infrared Deep Sky Survey (UKIDSS)}",
      journal = {\mnras},
         year = 2007,
        month = aug,
       volume = {379},
       number = {4},
        pages = {1599-1617},
          doi = {10.1111/j.1365-2966.2007.12040.x},
archivePrefix = {arXiv},
       eprint = {astro-ph/0604426},
 primaryClass = {astro-ph},
       adsurl = {https://ui.adsabs.harvard.edu/abs/2007MNRAS.379.1599L}
}

@INPROCEEDINGS{Kaiser2010,
       author = {{Kaiser}, Nick and {Burgett}, William and {Chambers}, Ken and {Denneau}, Larry and {Heasley}, Jim and {Jedicke}, Robert and {Magnier}, Eugene and {Morgan}, Jeff and {Onaka}, Peter and {Tonry}, John},
        title = "{The Pan-STARRS wide-field optical/NIR imaging survey}",
    booktitle = {Ground-based and Airborne Telescopes III},
         year = 2010,
       editor = {{Stepp}, Larry M. and {Gilmozzi}, Roberto and {Hall}, Helen J.},
       series = {Society of Photo-Optical Instrumentation Engineers (SPIE) Conference Series},
       volume = {7733},
        month = jul,
          eid = {77330E},
        pages = {77330E},
          doi = {10.1117/12.859188},
       adsurl = {https://ui.adsabs.harvard.edu/abs/2010SPIE.7733E..0EK}
}

@ARTICLE{Tingay2013_MWA,
       author = {{Tingay}, S.~J. and {Goeke}, R. and {Bowman}, J.~D. and {Emrich}, D. and {Ord}, S.~M. and {Mitchell}, D.~A. and {Morales}, M.~F. and {Booler}, T. and {Crosse}, B. and {Wayth}, R.~B. and {Lonsdale}, C.~J. and {Tremblay}, S. and {Pallot}, D. and {Colegate}, T. and {Wicenec}, A. and {Kudryavtseva}, N. and {Arcus}, W. and {Barnes}, D. and {Bernardi}, G. and {Briggs}, F. and {Burns}, S. and {Bunton}, J.~D. and {Cappallo}, R.~J. and {Corey}, B.~E. and {Deshpande}, A. and {Desouza}, L. and {Gaensler}, B.~M. and {Greenhill}, L.~J. and {Hall}, P.~J. and {Hazelton}, B.~J. and {Herne}, D. and {Hewitt}, J.~N. and {Johnston-Hollitt}, M. and {Kaplan}, D.~L. and {Kasper}, J.~C. and {Kincaid}, B.~B. and {Koenig}, R. and {Kratzenberg}, E. and {Lynch}, M.~J. and {Mckinley}, B. and {Mcwhirter}, S.~R. and {Morgan}, E. and {Oberoi}, D. and {Pathikulangara}, J. and {Prabu}, T. and {Remillard}, R.~A. and {Rogers}, A.~E.~E. and {Roshi}, A. and {Salah}, J.~E. and {Sault}, R.~J. and {Udaya-Shankar}, N. and {Schlagenhaufer}, F. and {Srivani}, K.~S. and {Stevens}, J. and {Subrahmanyan}, R. and {Waterson}, M. and {Webster}, R.~L. and {Whitney}, A.~R. and {Williams}, A. and {Williams}, C.~L. and {Wyithe}, J.~S.~B.},
        title = "{The Murchison Widefield Array: The Square Kilometre Array Precursor at Low Radio Frequencies}",
      journal = {\pasa},
         year = 2013,
        month = jan,
       volume = {30},
          eid = {e007},
        pages = {e007},
          doi = {10.1017/pasa.2012.007},
archivePrefix = {arXiv},
       eprint = {1206.6945},
 primaryClass = {astro-ph.IM},
       adsurl = {https://ui.adsabs.harvard.edu/abs/2013PASA...30....7T}
}

@ARTICLE{Hotan2021_ASKAP,
       author = {{Hotan}, A.~W. and {Bunton}, J.~D. and {Chippendale}, A.~P. and {Whiting}, M. and {Tuthill}, J. and {Moss}, V.~A. and {McConnell}, D. and {Amy}, S.~W. and {Huynh}, M.~T. and {Allison}, J.~R. and {Anderson}, C.~S. and {Bannister}, K.~W. and {Bastholm}, E. and {Beresford}, R. and {Bock}, D.~C.-J. and {Bolton}, R. and {Chapman}, J.~M. and {Chow}, K. and {Collier}, J.~D. and {Cooray}, F.~R. and {Cornwell}, T.~J. and {Diamond}, P.~J. and {Edwards}, P.~G. and {Feain}, I.~J. and {Franzen}, T.~M.~O. and {George}, D. and {Gupta}, N. and {Hampson}, G.~A. and {Harvey-Smith}, L. and {Hayman}, D.~B. and {Heywood}, I. and {Jacka}, C. and {Jackson}, C.~A. and {Jackson}, S. and {Jeganathan}, K. and {Johnston}, S. and {Kesteven}, M. and {Kleiner}, D. and {Koribalski}, B.~S. and {Lee-Waddell}, K. and {Lenc}, E. and {Lensson}, E.~S. and {Mackay}, S. and {Mahony}, E.~K. and {McClure-Griffiths}, N.~M. and {McConigley}, R. and {Mirtschin}, P. and {Ng}, A.~K. and {Norris}, R.~P. and {Pearce}, S.~E. and {Phillips}, C. and {Pilawa}, M.~A. and {Raja}, W. and {Reynolds}, J.~E. and {Roberts}, P. and {Roxby}, D.~N. and {Sadler}, E.~M. and {Shields}, M. and {Schinckel}, A.~E.~T. and {Serra}, P. and {Shaw}, R.~D. and {Sweetnam}, T. and {Troup}, E.~R. and {Tzioumis}, A. and {Voronkov}, M.~A. and {Westmeier}, T.},
        title = "{Australian square kilometre array pathfinder: I. system description}",
      journal = {\pasa},
         year = 2021,
        month = mar,
       volume = {38},
          eid = {e009},
        pages = {e009},
          doi = {10.1017/pasa.2021.1},
archivePrefix = {arXiv},
       eprint = {2102.01870},
 primaryClass = {astro-ph.IM},
       adsurl = {https://ui.adsabs.harvard.edu/abs/2021PASA...38....9H}
}

@ARTICLE{Banfield2015_RGZ,
       author = {{Banfield}, J.~K. and {Wong}, O.~I. and {Willett}, K.~W. and {Norris}, R.~P. and {Rudnick}, L. and {Shabala}, S.~S. and {Simmons}, B.~D. and {Snyder}, C. and {Garon}, A. and {Seymour}, N. and {Middelberg}, E. and {Andernach}, H. and {Lintott}, C.~J. and {Jacob}, K. and {Kapi{\'n}ska}, A.~D. and {Mao}, M.~Y. and {Masters}, K.~L. and {Jarvis}, M.~J. and {Schawinski}, K. and {Paget}, E. and {Simpson}, R. and {Kl{\"o}ckner}, H.-R. and {Bamford}, S. and {Burchell}, T. and {Chow}, K.~E. and {Cotter}, G. and {Fortson}, L. and {Heywood}, I. and {Jones}, T.~W. and {Kaviraj}, S. and {L{\'o}pez-S{\'a}nchez}, {\'A}. R. and {Maksym}, W.~P. and {Polsterer}, K. and {Borden}, K. and {Hollow}, R.~P. and {Whyte}, L.},
        title = "{Radio Galaxy Zoo: host galaxies and radio morphologies derived from visual inspection}",
      journal = {\mnras},
         year = 2015,
        month = nov,
       volume = {453},
       number = {3},
        pages = {2326-2340},
          doi = {10.1093/mnras/stv1688},
archivePrefix = {arXiv},
       eprint = {1507.07272},
 primaryClass = {astro-ph.GA},
       adsurl = {https://ui.adsabs.harvard.edu/abs/2015MNRAS.453.2326B}
}

@ARTICLE{Shimwell2017_LOFAR_description,
       author = {{Shimwell}, T.~W. and {R{\"o}ttgering}, H.~J.~A. and {Best}, P.~N. and {Williams}, W.~L. and {Dijkema}, T.~J. and {de Gasperin}, F. and {Hardcastle}, M.~J. and {Heald}, G.~H. and {Hoang}, D.~N. and {Horneffer}, A. and {Intema}, H. and {Mahony}, E.~K. and {Mandal}, S. and {Mechev}, A.~P. and {Morabito}, L. and {Oonk}, J.~B.~R. and {Rafferty}, D. and {Retana-Montenegro}, E. and {Sabater}, J. and {Tasse}, C. and {van Weeren}, R.~J. and {Br{\"u}ggen}, M. and {Brunetti}, G. and {Chy{\.z}y}, K.~T. and {Conway}, J.~E. and {Haverkorn}, M. and {Jackson}, N. and {Jarvis}, M.~J. and {McKean}, J.~P. and {Miley}, G.~K. and {Morganti}, R. and {White}, G.~J. and {Wise}, M.~W. and {van Bemmel}, I.~M. and {Beck}, R. and {Brienza}, M. and {Bonafede}, A. and {Calistro Rivera}, G. and {Cassano}, R. and {Clarke}, A.~O. and {Cseh}, D. and {Deller}, A. and {Drabent}, A. and {van Driel}, W. and {Engels}, D. and {Falcke}, H. and {Ferrari}, C. and {Fr{\"o}hlich}, S. and {Garrett}, M.~A. and {Harwood}, J.~J. and {Heesen}, V. and {Hoeft}, M. and {Horellou}, C. and {Israel}, F.~P. and {Kapi{\'n}ska}, A.~D. and {Kunert-Bajraszewska}, M. and {McKay}, D.~J. and {Mohan}, N.~R. and {Orr{\'u}}, E. and {Pizzo}, R.~F. and {Prandoni}, I. and {Schwarz}, D.~J. and {Shulevski}, A. and {Sipior}, M. and {Smith}, D.~J.~B. and {Sridhar}, S.~S. and {Steinmetz}, M. and {Stroe}, A. and {Varenius}, E. and {van der Werf}, P.~P. and {Zensus}, J.~A. and {Zwart}, J.~T.~L.},
        title = "{The LOFAR Two-metre Sky Survey. I. Survey description and preliminary data release}",
      journal = {\aap},
         year = 2017,
        month = feb,
       volume = {598},
          eid = {A104},
        pages = {A104},
          doi = {10.1051/0004-6361/201629313},
archivePrefix = {arXiv},
       eprint = {1611.02700},
 primaryClass = {astro-ph.IM},
       adsurl = {https://ui.adsabs.harvard.edu/abs/2017A&A...598A.104S}
}

@ARTICLE{Shimwell2019_LoTSS_DR1,
       author = {{Shimwell}, T.~W. and {Tasse}, C. and {Hardcastle}, M.~J. and {Mechev}, A.~P. and {Williams}, W.~L. and {Best}, P.~N. and {R{\"o}ttgering}, H.~J.~A. and {Callingham}, J.~R. and {Dijkema}, T.~J. and {de Gasperin}, F. and {Hoang}, D.~N. and {Hugo}, B. and {Mirmont}, M. and {Oonk}, J.~B.~R. and {Prandoni}, I. and {Rafferty}, D. and {Sabater}, J. and {Smirnov}, O. and {van Weeren}, R.~J. and {White}, G.~J. and {Atemkeng}, M. and {Bester}, L. and {Bonnassieux}, E. and {Br{\"u}ggen}, M. and {Brunetti}, G. and {Chy{\.z}y}, K.~T. and {Cochrane}, R. and {Conway}, J.~E. and {Croston}, J.~H. and {Danezi}, A. and {Duncan}, K. and {Haverkorn}, M. and {Heald}, G.~H. and {Iacobelli}, M. and {Intema}, H.~T. and {Jackson}, N. and {Jamrozy}, M. and {Jarvis}, M.~J. and {Lakhoo}, R. and {Mevius}, M. and {Miley}, G.~K. and {Morabito}, L. and {Morganti}, R. and {Nisbet}, D. and {Orr{\'u}}, E. and {Perkins}, S. and {Pizzo}, R.~F. and {Schrijvers}, C. and {Smith}, D.~J.~B. and {Vermeulen}, R. and {Wise}, M.~W. and {Alegre}, L. and {Bacon}, D.~J. and {van Bemmel}, I.~M. and {Beswick}, R.~J. and {Bonafede}, A. and {Botteon}, A. and {Bourke}, S. and {Brienza}, M. and {Calistro Rivera}, G. and {Cassano}, R. and {Clarke}, A.~O. and {Conselice}, C.~J. and {Dettmar}, R.~J. and {Drabent}, A. and {Dumba}, C. and {Emig}, K.~L. and {En{\ss}lin}, T.~A. and {Ferrari}, C. and {Garrett}, M.~A. and {G{\'e}nova-Santos}, R.~T. and {Goyal}, A. and {G{\"u}rkan}, G. and {Hale}, C. and {Harwood}, J.~J. and {Heesen}, V. and {Hoeft}, M. and {Horellou}, C. and {Jackson}, C. and {Kokotanekov}, G. and {Kondapally}, R. and {Kunert-Bajraszewska}, M. and {Mahatma}, V. and {Mahony}, E.~K. and {Mandal}, S. and {McKean}, J.~P. and {Merloni}, A. and {Mingo}, B. and {Miskolczi}, A. and {Mooney}, S. and {Nikiel-Wroczy{\'n}ski}, B. and {O'Sullivan}, S.~P. and {Quinn}, J. and {Reich}, W. and {Roskowi{\'n}ski}, C. and {Rowlinson}, A. and {Savini}, F. and {Saxena}, A. and {Schwarz}, D.~J. and {Shulevski}, A. and {Sridhar}, S.~S. and {Stacey}, H.~R. and {Urquhart}, S. and {van der Wiel}, M.~H.~D. and {Varenius}, E. and {Webster}, B. and {Wilber}, A.},
        title = "{The LOFAR Two-metre Sky Survey. II. First data release}",
      journal = {\aap},
         year = 2019,
        month = feb,
       volume = {622},
          eid = {A1},
        pages = {A1},
          doi = {10.1051/0004-6361/201833559},
archivePrefix = {arXiv},
       eprint = {1811.07926},
 primaryClass = {astro-ph.GA},
       adsurl = {https://ui.adsabs.harvard.edu/abs/2019A&A...622A...1S}
}

@ARTICLE{Shimwell2022_LoTSS_DR2,
       author = {{Shimwell}, T.~W. and {Hardcastle}, M.~J. and {Tasse}, C. and {Best}, P.~N. and {R{\"o}ttgering}, H.~J.~A. and {Williams}, W.~L. and {Botteon}, A. and {Drabent}, A. and {Mechev}, A. and {Shulevski}, A. and {van Weeren}, R.~J. and {Bester}, L. and {Br{\"u}ggen}, M. and {Brunetti}, G. and {Callingham}, J.~R. and {Chy{\.z}y}, K.~T. and {Conway}, J.~E. and {Dijkema}, T.~J. and {Duncan}, K. and {de Gasperin}, F. and {Hale}, C.~L. and {Haverkorn}, M. and {Hugo}, B. and {Jackson}, N. and {Mevius}, M. and {Miley}, G.~K. and {Morabito}, L.~K. and {Morganti}, R. and {Offringa}, A. and {Oonk}, J.~B.~R. and {Rafferty}, D. and {Sabater}, J. and {Smith}, D.~J.~B. and {Schwarz}, D.~J. and {Smirnov}, O. and {O'Sullivan}, S.~P. and {Vedantham}, H. and {White}, G.~J. and {Albert}, J.~G. and {Alegre}, L. and {Asabere}, B. and {Bacon}, D.~J. and {Bonafede}, A. and {Bonnassieux}, E. and {Brienza}, M. and {Bilicki}, M. and {Bonato}, M. and {Calistro Rivera}, G. and {Cassano}, R. and {Cochrane}, R. and {Croston}, J.~H. and {Cuciti}, V. and {Dallacasa}, D. and {Danezi}, A. and {Dettmar}, R.~J. and {Di Gennaro}, G. and {Edler}, H.~W. and {En{\ss}lin}, T.~A. and {Emig}, K.~L. and {Franzen}, T.~M.~O. and {Garc{\'\i}a-Vergara}, C. and {Grange}, Y.~G. and {G{\"u}rkan}, G. and {Hajduk}, M. and {Heald}, G. and {Heesen}, V. and {Hoang}, D.~N. and {Hoeft}, M. and {Horellou}, C. and {Iacobelli}, M. and {Jamrozy}, M. and {Jeli{\'c}}, V. and {Kondapally}, R. and {Kukreti}, P. and {Kunert-Bajraszewska}, M. and {Magliocchetti}, M. and {Mahatma}, V. and {Ma{\l}ek}, K. and {Mandal}, S. and {Massaro}, F. and {Meyer-Zhao}, Z. and {Mingo}, B. and {Mostert}, R.~I.~J. and {Nair}, D.~G. and {Nakoneczny}, S.~J. and {Nikiel-Wroczy{\'n}ski}, B. and {Orr{\'u}}, E. and {Pajdosz-{\'S}mierciak}, U. and {Pasini}, T. and {Prandoni}, I. and {van Piggelen}, H.~E. and {Rajpurohit}, K. and {Retana-Montenegro}, E. and {Riseley}, C.~J. and {Rowlinson}, A. and {Saxena}, A. and {Schrijvers}, C. and {Sweijen}, F. and {Siewert}, T.~M. and {Timmerman}, R. and {Vaccari}, M. and {Vink}, J. and {West}, J.~L. and {Wo{\l}owska}, A. and {Zhang}, X. and {Zheng}, J.},
        title = "{The LOFAR Two-metre Sky Survey. V. Second data release}",
      journal = {\aap},
         year = 2022,
        month = mar,
       volume = {659},
          eid = {A1},
        pages = {A1},
          doi = {10.1051/0004-6361/202142484},
archivePrefix = {arXiv},
       eprint = {2202.11733},
 primaryClass = {astro-ph.GA},
       adsurl = {https://ui.adsabs.harvard.edu/abs/2022A&A...659A...1S}
}

@ARTICLE{Magliocchetti1998,
       author = {{Magliocchetti}, M. and {Maddox}, S.~J. and {Lahav}, O. and {Wall}, J.~V.},
        title = "{Variance and skewness in the FIRST survey}",
      journal = {\mnras},
         year = 1998,
        month = oct,
       volume = {300},
       number = {1},
        pages = {257-268},
          doi = {10.1046/j.1365-8711.1998.01904.x},
archivePrefix = {arXiv},
       eprint = {astro-ph/9802269},
 primaryClass = {astro-ph},
       adsurl = {https://ui.adsabs.harvard.edu/abs/1998MNRAS.300..257M}
}

@ARTICLE{Braun2019_SKA1,
       author = {{Braun}, Robert and {Bonaldi}, Anna and {Bourke}, Tyler and {Keane}, Evan and {Wagg}, Jeff},
        title = "{Anticipated Performance of the Square Kilometre Array -- Phase 1 (SKA1)}",
      journal = {arXiv e-prints},
         year = 2019,
        month = dec,
          eid = {arXiv:1912.12699},
        pages = {arXiv:1912.12699},
          doi = {10.48550/arXiv.1912.12699},
archivePrefix = {arXiv},
       eprint = {1912.12699},
 primaryClass = {astro-ph.IM},
       adsurl = {https://ui.adsabs.harvard.edu/abs/2019arXiv191212699B}
}

@ARTICLE{Bonaldi2021,
       author = {{Bonaldi}, A. and {An}, T. and {Br{\"u}ggen}, M. and {Burkutean}, S. and {Coelho}, B. and {Goodarzi}, H. and {Hartley}, P. and {Sandhu}, P.~K. and {Wu}, C. and {Yu}, L. and {Zhoolideh Haghighi}, M.~H. and {Ant{\'o}n}, S. and {Bagheri}, Z. and {Barbosa}, D. and {Barraca}, J.~P. and {Bartashevich}, D. and {Bergano}, M. and {Bonato}, M. and {Brand}, J. and {de Gasperin}, F. and {Giannetti}, A. and {Dodson}, R. and {Jain}, P. and {Jaiswal}, S. and {Lao}, B. and {Liu}, B. and {Liuzzo}, E. and {Lu}, Y. and {Lukic}, V. and {Maia}, D. and {Marchili}, N. and {Massardi}, M. and {Mohan}, P. and {Morgado}, J.~B. and {Panwar}, M. and {Prabhakar}, P. and {Ribeiro}, V.~A.~R.~M. and {Rygl}, K.~L.~J. and {Sabz Ali}, V. and {Saremi}, E. and {Schisano}, E. and {Sheikhnezami}, S. and {Vafaei Sadr}, A. and {Wong}, A. and {Wong}, O.~I.},
        title = "{Square Kilometre Array Science Data Challenge 1: analysis and results}",
      journal = {\mnras},
         year = 2021,
        month = jan,
       volume = {500},
       number = {3},
        pages = {3821-3837},
          doi = {10.1093/mnras/staa3023},
archivePrefix = {arXiv},
       eprint = {2009.13346},
 primaryClass = {astro-ph.IM},
       adsurl = {https://ui.adsabs.harvard.edu/abs/2021MNRAS.500.3821B}
}

@ARTICLE{Bonaldi2025,
       author = {{Bonaldi}, A. and {Hartley}, P. and {Braun}, R. and {Purser}, S. and {Acharya}, A. and {Ahn}, K. and {Aparicio Resco}, M. and {Bait}, O. and {Bianco}, M. and {Chakraborty}, A. and {Chapman}, E. and {Chatterjee}, S. and {Chege}, K. and {Chen}, H. and {Chen}, X. and {Chen}, Z. and {Conaboy}, L. and {Cruz}, M. and {Darriba}, L. and {De Santis}, M. and {Denzel}, P. and {Diao}, K. and {Feron}, J. and {Finlay}, C. and {Gehlot}, B. and {Ghosh}, S. and {Giri}, S.~K. and {Grumitt}, R. and {Hong}, S.~E. and {Ito}, T. and {Jiang}, M. and {Jordan}, C. and {Kim}, S. and {Kim}, M. and {Kim}, J. and {Krishna}, S.~P. and {Kulkarni}, A. and {L{\'o}pez-Caniego}, M. and {Labadie-Garc{\'\i}a}, I. and {Lee}, H. and {Lee}, D. and {Lee}, N. and {Line}, J. and {Liu}, Y. and {Mao}, Y. and {Mazumder}, A. and {Mertens}, F.~G. and {Munshi}, S. and {Nasirudin}, A. and {Ni}, S. and {Nistane}, V. and {Norregaard}, C. and {Null}, D. and {Offringa}, A. and {Oh}, M. and {Oh}, S.-H. and {Parkinson}, D. and {Pritchard}, J. and {Ruiz-Granda}, M. and {Salvador L{\'o}pez}, V. and {Shan}, H. and {Sharma}, R. and {Trott}, C. and {Yoshiura}, S. and {Zhang}, L. and {Zhang}, X. and {Zheng}, Q. and {Zhu}, Z. and {Zuo}, S. and {Akahori}, T. and {Alberto}, P. and {Allys}, E. and {An}, T. and {Anstey}, D. and {Baek}, J. and {Basavraj} and {Brackenhoff}, S. and {Browne}, P. and {Ceccotti}, E. and {Chen}, T. and {Choudhuri}, S. and {Choudhury}, M. and {Coles}, J. and {Cook}, J. and {Cornu}, D. and {Cunnington}, S. and {Das}, S. and {de Lera Acedo Acedo}, E. and {Delouis}, J.-M. and {Deng}, F. and {Ding}, J. and {Elahi}, K.~M.~A. and {Fernandez}, P. and {Fern{\'a}ndez}, C. and {Fern{\'a}ndez Alc{\'a}zar}, A. and {Galluzzi}, V. and {Gao}, L.-Y. and {Garain}, U. and {Garrido}, J. and {Gendron-Marsolais}, M.-L. and {Gessey-Jones}, T. and {Ghorbel}, H. and {Gong}, Y. and {Guo}, S. and {Hasegawa}, K. and {Hayashi}, T. and {Herranz}, D. and {Holanda}, V. and {Holloway}, A.~J. and {Hothi}, I. and {H{\"o}fer}, C. and {Jeli{\'c}}, V. and {Jiang}, Y. and {Jiang}, X. and {Kang}, H. and {Kim}, J.-Y. and {Koopmans}, L.~V. and {Lacroix}, R. and {Lee}, E. and {Leeney}, S. and {Levrier}, F. and {Li}, Y. and {Ma}, Q. and {Meriot}, R. and {Mesinger}, A. and {Mevius}, M. and {Minoda}, T. and {Miville-Desch{\^e}nes}, M.-A. and {Moldon}, J. and {Mondal}, R. and {Murmu}, C. and {Murray}, S. and {Nirmala}, Sr. and {Niu}, Q. and {Nunhokee}, C. and {O'Hara}, O. and {Pal}, S.~K. and {Pal}, S. and {Park}, J. and {Parra}, M. and {Patra}, N.~N. and {Pindor}, B. and {Remazeilles}, M. and {Rey}, P. and {Rubino-Martin}, J.~A. and {Saha}, S. and {Selvaraj}, A. and {Semelin}, B. and {Shah}, R. and {Shao}, Y. and {Shaw}, A.~K. and {Shi}, F. and {Shimabukuro}, H. and {Singh}, G. and {Sohn}, B.~W. and {Stagni}, M. and {Starck}, J.-L. and {Sui}, C. and {Swinbank}, J.~D. and {S{\'a}nchez}, J. and {S{\'a}nchez-Exp{\'o}sito}, S. and {Takahashi}, K. and {Takeuchi}, T. and {Tripathi}, A. and {Verdes-Montenegro}, L. and {Vielva}, P. and {Vitello}, F.~R. and {Wang}, G.-J. and {Wang}, Q. and {Wang}, X. and {Wang}, Y. and {Wang}, Y.-X. and {Wiegert}, T. and {Wild}, A. and {Williams}, W.~L. and {Wolz}, L. and {Wu}, X. and {Wu}, P. and {Xia}, J.-Q. and {Xu}, Y. and {Yan}, R. and {Yan}, Y.-P. and {Yin}, Z. and {You}, Z. and {Yu}, X. and {Yu}, K. and {Yue}, B. and {Zhao}, Z. and {Zhao}, X. and {Zhou}, X.},
        title = "{Square Kilometre Array Science Data Challenge 3a: foreground removal for an EoR experiment}",
      journal = {\mnras},
         year = 2025,
        month = oct,
       volume = {543},
       number = {2},
        pages = {1092-1119},
          doi = {10.1093/mnras/staf1466},
archivePrefix = {arXiv},
       eprint = {2503.11740},
 primaryClass = {astro-ph.IM},
       adsurl = {https://ui.adsabs.harvard.edu/abs/2025MNRAS.543.1092B}
}

@misc{Mohan2015_PyBDSF,
       author = {{Mohan}, Niruj and {Rafferty}, David},
        title = "{PyBDSF: Python Blob Detection and Source Finder}",
 howpublished = {Astrophysics Source Code Library, record ascl:1502.007},
         year = 2015,
        month = feb,
          eid = {ascl:1502.007},
       adsurl = {https://ui.adsabs.harvard.edu/abs/2015ascl.soft02007M}
}

@ARTICLE{Hancock2012_Aegean,
       author = {{Hancock}, P.~J. and {Murphy}, T. and {Gaensler}, B.~M. and {Hopkins}, A. and {Curran}, J.~R.},
        title = "{Compact continuum source finding for next generation radio surveys}",
      journal = {\mnras},
         year = 2012,
        month = may,
       volume = {422},
       number = {2},
        pages = {1812-1824},
          doi = {10.1111/j.1365-2966.2012.20768.x},
archivePrefix = {arXiv},
       eprint = {1202.4500},
 primaryClass = {astro-ph.IM},
       adsurl = {https://ui.adsabs.harvard.edu/abs/2012MNRAS.422.1812H}
}

@ARTICLE{Hancock2018_Aegean,
       author = {{Hancock}, Paul J. and {Trott}, Cathryn M. and {Hurley-Walker}, Natasha},
        title = "{Source Finding in the Era of the SKA (Precursors): Aegean 2.0}",
      journal = {\pasa},
         year = 2018,
        month = mar,
       volume = {35},
          eid = {e011},
        pages = {e011},
          doi = {10.1017/pasa.2018.3},
archivePrefix = {arXiv},
       eprint = {1801.05548},
 primaryClass = {astro-ph.IM},
       adsurl = {https://ui.adsabs.harvard.edu/abs/2018PASA...35...11H}
}

@ARTICLE{Lao2023_HeTu,
       author = {{Lao}, B. and {Jaiswal}, S. and {Zhao}, Z. and {Lin}, L. and {Wang}, J. and {Sun}, X. and {Qin}, S. -L.},
        title = "{Radio sources segmentation and classification with deep learning}",
      journal = {Astronomy and Computing},
         year = 2023,
        month = jul,
       volume = {44},
          eid = {100728},
        pages = {100728},
          doi = {10.1016/j.ascom.2023.100728},
archivePrefix = {arXiv},
       eprint = {2306.01426},
 primaryClass = {astro-ph.IM},
       adsurl = {https://ui.adsabs.harvard.edu/abs/2023A&C....4400728L}
}

@InProceedings{Carion2020_DETR,
    author="Carion, Nicolas
    and Massa, Francisco
    and Synnaeve, Gabriel
    and Usunier, Nicolas
    and Kirillov, Alexander
    and Zagoruyko, Sergey",
    editor="Vedaldi, Andrea
    and Bischof, Horst
    and Brox, Thomas
    and Frahm, Jan-Michael",
    title="End-to-End Object Detection with Transformers",
    booktitle="Computer Vision -- ECCV 2020",
    year="2020",
    publisher="Springer International Publishing",
    address="Cham",
    pages="213--229",
    isbn="978-3-030-58452-8"
}

@InProceedings{Meng2021_ConditionalDETR,
    author    = {Meng, Depu and Chen, Xiaokang and Fan, Zejia and Zeng, Gang and Li, Houqiang and Yuan, Yuhui and Sun, Lei and Wang, Jingdong},
    title     = {Conditional DETR for Fast Training Convergence},
    booktitle = {Proceedings of the IEEE/CVF International Conference on Computer Vision (ICCV)},
    month     = {October},
    year      = {2021},
    pages     = {3651-3660}
}

@ARTICLE{Chen2024_LWDETR,
       author = {{Chen}, Qiang and {Su}, Xiangbo and {Zhang}, Xinyu and {Wang}, Jian and {Chen}, Jiahui and {Shen}, Yunpeng and {Han}, Chuchu and {Chen}, Ziliang and {Xu}, Weixiang and {Li}, Fanrong and {Zhang}, Shan and {Yao}, Kun and {Ding}, Errui and {Zhang}, Gang and {Wang}, Jingdong},
        title = "{LW-DETR: A Transformer Replacement to YOLO for Real-Time Detection}",
      journal = {arXiv e-prints},
         year = 2024,
        month = jun,
          eid = {arXiv:2406.03459},
        pages = {arXiv:2406.03459},
          doi = {10.48550/arXiv.2406.03459},
archivePrefix = {arXiv},
       eprint = {2406.03459},
 primaryClass = {cs.CV},
       adsurl = {https://ui.adsabs.harvard.edu/abs/2024arXiv240603459C}
}

@article{
    Oquab2023_DINOv2,
    title={{DINO}v2: Learning Robust Visual Features without Supervision},
    author={Maxime Oquab and Timoth{\'e}e Darcet and Th{\'e}o Moutakanni and Huy V. Vo and Marc Szafraniec and Vasil Khalidov and Pierre Fernandez and Daniel HAZIZA and Francisco Massa and Alaaeldin El-Nouby and Mido Assran and Nicolas Ballas and Wojciech Galuba and Russell Howes and Po-Yao Huang and Shang-Wen Li and Ishan Misra and Michael Rabbat and Vasu Sharma and Gabriel Synnaeve and Hu Xu and Herve Jegou and Julien Mairal and Patrick Labatut and Armand Joulin and Piotr Bojanowski},
    journal={Transactions on Machine Learning Research},
    issn={2835-8856},
    year={2024},
    url={https://openreview.net/forum?id=a68SUt6zFt},
    note={Featured Certification}
}

@inproceedings{
    Zhu2021_DeformableDETR,
    title={Deformable {\{}DETR{\}}: Deformable Transformers for End-to-End Object Detection},
    author={Xizhou Zhu and Weijie Su and Lewei Lu and Bin Li and Xiaogang Wang and Jifeng Dai},
    booktitle={International Conference on Learning Representations (ICLR)},
    year={2021},
}

@ARTICLE{Sapkota2025,
       author = {{Sapkota}, Ranjan and {Harsha Cheppally}, Rahul and {Sharda}, Ajay and {Karkee}, Manoj},
        title = "{RF-DETR Object Detection vs YOLOv12 : A Study of Transformer-based and CNN-based Architectures for Single-Class and Multi-Class Greenfruit Detection in Complex Orchard Environments Under Label Ambiguity}",
      journal = {arXiv e-prints},
         year = 2025,
        month = apr,
          eid = {arXiv:2504.13099},
        pages = {arXiv:2504.13099},
          doi = {10.48550/arXiv.2504.13099},
archivePrefix = {arXiv},
       eprint = {2504.13099},
 primaryClass = {cs.CV},
       adsurl = {https://ui.adsabs.harvard.edu/abs/2025arXiv250413099S}
}

@ARTICLE{Dahiya2025,
       author = {{Dahiya}, Neeraj and {Prakash}, Deo and {Kundu}, Shakti and {Kuttan}, Shanu Rakesh and {Suwalka}, Isha and {Ayadi}, Manel and {Dubale}, Mitiku and {Hashmi}, Arshad},
        title = "{Optimised RFO tuned RF-DETR model for precision urine microscopy for renal and systemic disease diagnosis}",
      journal = {Scientific Reports},
         year = 2025,
        month = jul,
       volume = {15},
       number = {1},
          eid = {25842},
        pages = {25842},
          doi = {10.1038/s41598-025-11725-0},
       adsurl = {https://ui.adsabs.harvard.edu/abs/2025NatSR..1525842D}
}

@INPROCEEDINGS{Jarvis2015,
       author = {{Jarvis}, M. and {Bacon}, D. and {Blake}, C. and {Brown}, M. and {Lindsay}, S. and {Raccanelli}, A. and {Santos}, M. and {Schwarz}, D.~J.},
        title = "{Cosmology with SKA Radio Continuum Surveys}",
    booktitle = {Advancing Astrophysics with the Square Kilometre Array (AASKA14)},
         year = 2015,
        month = apr,
          eid = {18},
        pages = {18},
          doi = {10.22323/1.215.0018},
 primaryClass = {astro-ph.CO},
       adsurl = {https://ui.adsabs.harvard.edu/abs/2015aska.confE..18J}
}

@ARTICLE{Alonso2021,
       author = {{Alonso}, David and {Bellini}, Emilio and {Hale}, Catherine and {Jarvis}, Matt J. and {Schwarz}, Dominik J.},
        title = "{Cross-correlating radio continuum surveys and CMB lensing: constraining redshift distributions, galaxy bias, and cosmology}",
      journal = {\mnras},
         year = 2021,
        month = mar,
       volume = {502},
       number = {1},
        pages = {876-887},
          doi = {10.1093/mnras/stab046},
archivePrefix = {arXiv},
       eprint = {2009.01817},
 primaryClass = {astro-ph.CO},
       adsurl = {https://ui.adsabs.harvard.edu/abs/2021MNRAS.502..876A}
}

@ARTICLE{Hale2024,
       author = {{Hale}, C.~L. and {Schwarz}, D.~J. and {Best}, P.~N. and {Nakoneczny}, S.~J. and {Alonso}, D. and {Bacon}, D. and {B{\"o}hme}, L. and {Bhardwaj}, N. and {Bilicki}, M. and {Camera}, S. and {Heneka}, C.~S. and {Pashapour-Ahmadabadi}, M. and {Tiwari}, P. and {Zheng}, J. and {Duncan}, K.~J. and {Jarvis}, M.~J. and {Kondapally}, R. and {Magliocchetti}, M. and {Rottgering}, H.~J.~A. and {Shimwell}, T.~W.},
        title = "{Cosmology from LOFAR Two-metre Sky Survey Data Release 2: angular clustering of radio sources}",
      journal = {\mnras},
         year = 2024,
        month = jan,
       volume = {527},
       number = {3},
        pages = {6540-6568},
          doi = {10.1093/mnras/stad3088},
archivePrefix = {arXiv},
       eprint = {2310.07627},
 primaryClass = {astro-ph.CO},
       adsurl = {https://ui.adsabs.harvard.edu/abs/2024MNRAS.527.6540H}
}

@ARTICLE{Hale2025,
       author = {{Hale}, C.~L. and {Best}, P.~N. and {Duncan}, K.~J. and {Kondapally}, R. and {Jarvis}, M.~J. and {Magliocchetti}, M. and {R{\"o}ttgering}, H.~J.~A. and {Schwarz}, D.~J. and {Smith}, D.~J.~B. and {Zheng}, J.},
        title = "{The clustering of active galactic nuclei and star-forming galaxies in the LoTSS Deep Fields}",
      journal = {\mnras},
         year = 2025,
        month = dec,
       volume = {544},
       number = {2},
        pages = {1323-1348},
          doi = {10.1093/mnras/staf1626},
archivePrefix = {arXiv},
       eprint = {2510.01029},
 primaryClass = {astro-ph.CO},
       adsurl = {https://ui.adsabs.harvard.edu/abs/2025MNRAS.544.1323H}
}

@ARTICLE{Tanidis2025,
       author = {{Tanidis}, Konstantinos and {Asorey}, Jacobo and {Saraf}, Chandra Shekhar and {Hale}, Catherine Laura and {Bahr-Kalus}, Benedict and {Parkinson}, David and {Camera}, Stefano and {Norris}, Ray and {Hopkins}, Andrew and {Bilicki}, Maciej and {Gupta}, Nikhel},
        title = "{Cross-correlating the EMU Pilot Survey 1 with CMB lensing: Constraints on cosmology and galaxy bias with harmonic-space power spectra}",
      journal = {\pasa},
         year = 2025,
        month = may,
       volume = {42},
          eid = {e062},
        pages = {e062},
          doi = {10.1017/pasa.2025.10033},
archivePrefix = {arXiv},
       eprint = {2411.05913},
 primaryClass = {astro-ph.CO},
       adsurl = {https://ui.adsabs.harvard.edu/abs/2025PASA...42...62T}
}

@ARTICLE{Best2012,
       author = {{Best}, P.~N. and {Heckman}, T.~M.},
        title = "{On the fundamental dichotomy in the local radio-AGN population: accretion, evolution and host galaxy properties}",
      journal = {\mnras},
         year = 2012,
        month = apr,
       volume = {421},
       number = {2},
        pages = {1569-1582},
          doi = {10.1111/j.1365-2966.2012.20414.x},
archivePrefix = {arXiv},
       eprint = {1201.2397},
 primaryClass = {astro-ph.CO},
       adsurl = {https://ui.adsabs.harvard.edu/abs/2012MNRAS.421.1569B}
}

@ARTICLE{Mingo2019,
       author = {{Mingo}, B. and {Croston}, J.~H. and {Hardcastle}, M.~J. and {Best}, P.~N. and {Duncan}, K.~J. and {Morganti}, R. and {Rottgering}, H.~J.~A. and {Sabater}, J. and {Shimwell}, T.~W. and {Williams}, W.~L. and {Brienza}, M. and {Gurkan}, G. and {Mahatma}, V.~H. and {Morabito}, L.~K. and {Prandoni}, I. and {Bondi}, M. and {Ineson}, J. and {Mooney}, S.},
        title = "{Revisiting the Fanaroff-Riley dichotomy and radio-galaxy morphology with the LOFAR Two-Metre Sky Survey (LoTSS)}",
      journal = {\mnras},
         year = 2019,
        month = sep,
       volume = {488},
       number = {2},
        pages = {2701-2721},
          doi = {10.1093/mnras/stz1901},
archivePrefix = {arXiv},
       eprint = {1907.03726},
 primaryClass = {astro-ph.GA},
       adsurl = {https://ui.adsabs.harvard.edu/abs/2019MNRAS.488.2701M}
}

@ARTICLE{Mingo2022,
       author = {{Mingo}, B. and {Croston}, J.~H. and {Best}, P.~N. and {Duncan}, K.~J. and {Hardcastle}, M.~J. and {Kondapally}, R. and {Prandoni}, I. and {Sabater}, J. and {Shimwell}, T.~W. and {Williams}, W.~L. and {Baldi}, R.~D. and {Bonato}, M. and {Bondi}, M. and {Dabhade}, P. and {G{\"u}rkan}, G. and {Ineson}, J. and {Magliocchetti}, M. and {Miley}, G. and {Pierce}, J.~C.~S. and {R{\"o}ttgering}, H.~J.~A.},
        title = "{Accretion mode versus radio morphology in the LOFAR Deep Fields}",
      journal = {\mnras},
         year = 2022,
        month = apr,
       volume = {511},
       number = {3},
        pages = {3250-3271},
          doi = {10.1093/mnras/stac140},
archivePrefix = {arXiv},
       eprint = {2201.04433},
 primaryClass = {astro-ph.GA},
       adsurl = {https://ui.adsabs.harvard.edu/abs/2022MNRAS.511.3250M}
}

@ARTICLE{Horton2025,
       author = {{Horton}, M.~A. and {Hardcastle}, M.~J. and {Miley}, G.~K. and {Tasse}, C. and {Shimwell}, T.},
        title = "{Complex morphology and precession indicators of active galactic nuclei jets in LoTSS DR2}",
      journal = {\aap},
         year = 2025,
        month = jul,
       volume = {699},
          eid = {A338},
        pages = {A338},
          doi = {10.1051/0004-6361/202453559},
archivePrefix = {arXiv},
       eprint = {2504.18518},
 primaryClass = {astro-ph.GA},
       adsurl = {https://ui.adsabs.harvard.edu/abs/2025A&A...699A.338H}
}

@ARTICLE{Hardcastle2020,
       author = {{Hardcastle}, M.~J. and {Croston}, J.~H.},
        title = "{Radio galaxies and feedback from AGN jets}",
      journal = {\nar},
         year = 2020,
        month = jun,
       volume = {88},
          eid = {101539},
        pages = {101539},
          doi = {10.1016/j.newar.2020.101539},
archivePrefix = {arXiv},
       eprint = {2003.06137},
 primaryClass = {astro-ph.HE},
       adsurl = {https://ui.adsabs.harvard.edu/abs/2020NewAR..8801539H}
}

@ARTICLE{Cochrane2023,
       author = {{Cochrane}, R.~K. and {Kondapally}, R. and {Best}, P.~N. and {Sabater}, J. and {Duncan}, K.~J. and {Smith}, D.~J.~B. and {Hardcastle}, M.~J. and {R{\"o}ttgering}, H.~J.~A. and {Prandoni}, I. and {Haskell}, P. and {G{\"u}rkan}, G. and {Miley}, G.~K.},
        title = "{The LOFAR Two-metre Sky Survey: the radio view of the cosmic star formation history}",
      journal = {\mnras},
         year = 2023,
        month = aug,
       volume = {523},
       number = {4},
        pages = {6082-6102},
          doi = {10.1093/mnras/stad1602},
archivePrefix = {arXiv},
       eprint = {2305.15510},
 primaryClass = {astro-ph.GA},
       adsurl = {https://ui.adsabs.harvard.edu/abs/2023MNRAS.523.6082C}
}

@ARTICLE{Simonte2024,
       author = {{Simonte}, M. and {Andernach}, H. and {Br{\"u}ggen}, M. and {Miley}, G.~K. and {Barthel}, P.},
        title = "{Giant radio galaxies in the LOFAR deep fields}",
      journal = {\aap},
         year = 2024,
        month = jun,
       volume = {686},
          eid = {A21},
        pages = {A21},
          doi = {10.1051/0004-6361/202348904},
archivePrefix = {arXiv},
       eprint = {2403.08037},
 primaryClass = {astro-ph.GA},
       adsurl = {https://ui.adsabs.harvard.edu/abs/2024A&A...686A..21S}
}

@ARTICLE{Morabito2025,
       author = {{Morabito}, Leah K. and {Kondapally}, R. and {Best}, P.~N. and {Yue}, B.-H. and {de Jong}, J.~M.~G.~H.~J. and {Sweijen}, F. and {Bondi}, Marco and {Schwarz}, Dominik J. and {Smith}, D.~J.~B. and {van Weeren}, R.~J. and {R{\"o}ttgering}, H.~J.~A. and {Shimwell}, T.~W. and {Prandoni}, Isabella},
        title = "{A hidden active galactic nucleus population: the first radio luminosity functions constructed by physical process}",
      journal = {\mnras},
         year = 2025,
        month = jan,
       volume = {536},
       number = {1},
        pages = {L32-L37},
          doi = {10.1093/mnrasl/slae104},
archivePrefix = {arXiv},
       eprint = {2411.05069},
 primaryClass = {astro-ph.GA},
       adsurl = {https://ui.adsabs.harvard.edu/abs/2025MNRAS.536L..32M}
}




\appendix

\section{Mask R-CNN baseline: \texttt{caesar-mrcnn}}
\label{subsec:mrcnn}

Mask R-CNN combines object detection, classification, and instance segmentation in a two-stage framework \citep{He2017_MaskRCNN}. We adopt \texttt{caesar-mrcnn}\footnote{\url{https://github.com/SKA-INAF/caesar-mrcnn-tf2}} as a radio-oriented Mask R-CNN baseline against the transformer-based RF-DETR. \texttt{Caesar-mrcnn} is a refactored TensorFlow~2 implementation of the system presented by \citet{Riggi2023}, featuring an improved preprocessing pipeline and additional backbone options. In this work, we use \texttt{caesar-mrcnn} as a detection-oriented baseline: for each radio cutout, it returns an instance mask, an associated bounding box, and a class label, while detailed source characterisation is deferred to downstream analysis tools.

For comparison, \texttt{caesar-mrcnn} uses the same cutouts, data splits, and augmentation policy as RF-DETR (Section~\ref{subsec:data-prep}), namely \(132\times132\)~pixel single-channel LOFAR images at a scale of \(1.5\arcsec\,\mathrm{pixel}^{-1}\). As Mask R-CNN requires per-instance binary masks, we derive these masks via threshold-based segmentation on the ZScale-remapped images using a \(3\sigma\) threshold. The segmentation is recomputed after each geometric augmentation to avoid WCS misalignment and to preserve consistency between masks, bounding boxes, and class labels.

Algorithmically, \texttt{caesar-mrcnn} follows the standard two-stage Mask R--CNN pipeline: a ResNet--FPN backbone feeds a Region Proposal Network (RPN), whose proposals are refined via ROIAlign and passed to (i) a classification and bounding-box regression head and (ii) a fully convolutional mask head producing \(28\times28\) per-instance masks. Training optimises the standard multi-task loss $L = L_{\rm class} + L_{\rm box} + L_{\rm mask}.$
Relative to a vanilla Mask R-CNN configuration, the radio-oriented implementation extends the RPN anchor scales (to $\sim$8--128\,px) and broadens the anchor aspect-ratio priors (0.2--5.0) to better capture elongated and multi-island radio emission. In addition, the bounding-box and mask loss terms are down-weighted relative to the classification loss to balance the segmentation-dominated objective, following the recommendations of \citet{Riggi2023}.

Preprocessing is radio-specific and handled internally by the \texttt{caesar-mrcnn} package, including single-channel FITS ingestion, NaN sanitisation, and ZScale intensity remapping. We therefore ingest LOFAR cutouts directly as single-channel FITS images, with conversion to network tensors performed internally. While the input representation differs from that of RF-DETR, the training/validation split and augmentation strategy are kept identical across both models for consistency. We largely follow the default configuration recommended by \citet{Riggi2023}, adopting a ResNet-101 backbone with FPN, optimisation with Adam at a learning rate of $5\times10^{-4}$, and wide RPN anchor ratios and scales together with reduced bounding-box loss weights to accommodate both compact and extended radio sources. The full model configuration is summarised in Table~\ref{tab:mrcnn-config}. We evaluate \texttt{caesar-mrcnn} on the same ELAIS-N1 validation set using the identical matching criteria and thresholds adopted for RF-DETR. The resulting detection and classification metrics are summarised in Table~\ref{tab:mrcnn-evaluation}, providing a useful point of reference for comparison with RF-DETR.

\begin{table}
\centering
\caption{Main parameters of \texttt{caesar-mrcnn} used in this work.}
\label{tab:mrcnn-config}
\begin{tabular}{ll}
\hline
Component & Value \\
\hline
\multicolumn{2}{l}{\emph{Architecture \& Optimisation}}\\
Backbone & ResNet-101 \\
Backbone strides & [4, 8, 16, 32, 64] \\
Optimiser & Adam \\
Learning rate & $5\times10^{-4}$ \\
Epochs & 250 \\
Batch size & 2 \\
\hline
\multicolumn{2}{l}{\emph{RPN / ROI configuration}}\\
RPN anchor ratios & $[0.2,0.3,0.5,1,2,3,4,5]$ \\
RPN anchor scales & $[8,16,32,64,128]$ \\
RPN NMS threshold & 0.7 \\
Train anchors per image & 256 \\
Train ROIs per image & 256 \\
Max GT instances & 100 \\
\hline
\multicolumn{2}{l}{\emph{Loss weights}}\\
RPN class & 1.0 \\
RPN bbox & 0.1 \\
MRCNN class & 1.0 \\
MRCNN bbox & 0.1 \\
MRCNN mask & 0.1 \\
\hline
\end{tabular}
\end{table}

\begin{table*}
\centering
\small
\caption{Validation results for \texttt{caesar-mrcnn} on the same ELAIS-N1 validation set at IoU$\ge 0.5$ and score$\ge 0.5$.}
\label{tab:mrcnn-evaluation}
\begin{tabular}{lccccccc}
\toprule
\multirow{2}{*}{Class} & \multicolumn{3}{c}{Detection} & \multicolumn{3}{c}{Classification} \\
\cmidrule(lr){2-4}\cmidrule(lr){5-7}
 & Completeness (\%) & Reliability (\%) & F1 score (\%) & Recall (\%) & Precision (\%) & F1 score (\%) \\
\midrule
All                      & 85.6 & 82.1 & 83.8 & 86.4 & 86.4 & 86.4 \\
\texttt{Compact}         & 90.0 & 85.7 & 87.8 & 89.9 & 98.4 & 94.0 \\
\texttt{Single-Extended} & 80.4 & 86.5 & 83.3 & 41.1 & 95.7 & 57.5 \\
\texttt{Multi-Extended}  & 65.2 & 35.6 & 46.1 & 52.2 & 32.4 & 40.0 \\
\texttt{Flagged}         & 48.4 & 87.5 & 62.2 & 22.6 & 100.0 & 36.8 \\
\texttt{Spurious}        & 18.6 & 22.1 & 20.2 & 12.1 & 88.2 & 21.3 \\
\bottomrule
\end{tabular}
\end{table*}

\section{Computational runtimes}
\label{subsec:computation}

All computations were conducted on a high-performance computing cluster using a single NVIDIA Tesla V100 GPU. Training RF-DETR on the ELAIS-N1 dataset for 200 epochs, using the configuration listed in Table~\ref{tab:rfdetr-config}, required approximately 70 hours of wall-clock time on a single GPU. 
Inference on the full LoTSS Deep Field continuum images was performed using the tiled sliding-window strategy described in Section~\ref{subsec:inference}. For RF-DETR, wall-clock runtimes for the detection and consolidation stage ranged from approximately 0.4 to 1.2 hours per field on a single GPU, depending on mosaic size and source density. Additional photometry and post-processing steps may extend the total end-to-end pipeline runtime.



\bsp	
\label{lastpage}
\end{document}